\documentclass[aps,pre,amsmath,amssymb,superscriptaddress,showpacs,showkeys]{revtex4-2}

\usepackage{graphicx}
\usepackage{dcolumn}
\usepackage{bm}
\usepackage{CJKutf8}
\usepackage{xcolor}

\begin{document}



\title{Simulation Method of Microscale Fluid-Structure Interactions: Diffuse-Resistance-Domain Approach
} 

\author{Min Gao}
\affiliation{School of General Studies, Guangzhou College of Technology and Business, 5 Guangming Road, Guangzhou, Guangdong 510850, China}

\author{Zhihao Li} 
\affiliation{Department of Civil Engineering and Smart Cities, College of Engineering, Shantou University, Shantou, 515063, Guangdong, China.}
\affiliation{Department of Physics and MATEC Key Lab, Guangdong Technion - Israel Institute of Technology, 241 Daxue Road, Shantou, Guangdong 515063, China}

\author{Xinpeng Xu}\thanks{Corresponding authors}
\email{xu.xinpeng@gtiit.edu.cn}
\affiliation{Department of Physics and MATEC Key Lab, Guangdong Technion - Israel Institute of Technology, 241 Daxue Road, Shantou, Guangdong 515063, China}
\affiliation{Technion – Israel Institute of Technology, Haifa 3200003, Israel}


\date{\today}

\begin{abstract}
Direct numerical simulations (DNS) of microscale fluid-structure interactions (mFSI) in multicomponent multiphase flows pose many challenges, including the thermodynamic consistency of multiphysics couplings, tracking of moving interfaces, dynamics of moving triple-phase contact lines, and the coupling of multiphase hydrodynamics with phase transition dynamics. We propose and validate a generic DNS approach: Diffuse-Resistance-Domain (DRD) approach. It overcomes the above challenges by employing Onsager's variational principle (OVP) to formulate dynamic models and combining traditional diffuse-interface models for fluid-fluid interfacial dynamics with a novel implementation of complex fluid-solid interfacial conditions via smooth interpolations of dynamic-resistance coefficients across interfaces. After careful validation by numerous benchmark simulations, we simulated several cutting-edge, challenging mFSI problems across diverse fields. This generic DNS approach offers a promising tool for elucidating physical mechanisms, manipulating microscale fluid dynamics, and optimizing engineering processes across diverse fields. 
\end{abstract}


\maketitle


\section{Introduction} \label{sec:introduction}

Microscale fluid-structure interactions (mFSI) refer to the complex dynamic interplay between multicomponent multiphase flows and moving, deformable, or evolving solid structures at length scales that typically range from $1$ to $100 \, \mathrm{\mu m}$~\cite{Kirby2010Book,Subramaniam2020}. mFSI differs from the fluid-structure interactions at macroscales in many aspects, and physical effects such as viscous dissipation, interfacial tension, wettability, and fluid slip start to dominate the system. Understanding mFSI is crucial for various applications, including microfluidics, additive manufacturing, active matter physics, biomedical engineering, geophysics, and civil engineering, \emph{etc}~\cite{Kirby2010Book}. However, a comprehensive study of mFSI remains a significant challenge, largely due to the complexities of multiphysics, multi-field couplings, and the high surface-area-to-volume ratio~\cite{Hou2012,Dowell2001Rev}. For most mFSI problems, deriving analytical solutions to the governing equations remains unattainable. Consequently, direct numerical simulations (DNS) are commonly employed to investigate the fundamental physics underlying mFSI phenomena~\cite{Hou2012,Dowell2001Rev,Subramaniam2020,Griffith2020IBM}.

\begin{figure}[htbp] 
\centering
\includegraphics[width=0.7\columnwidth]{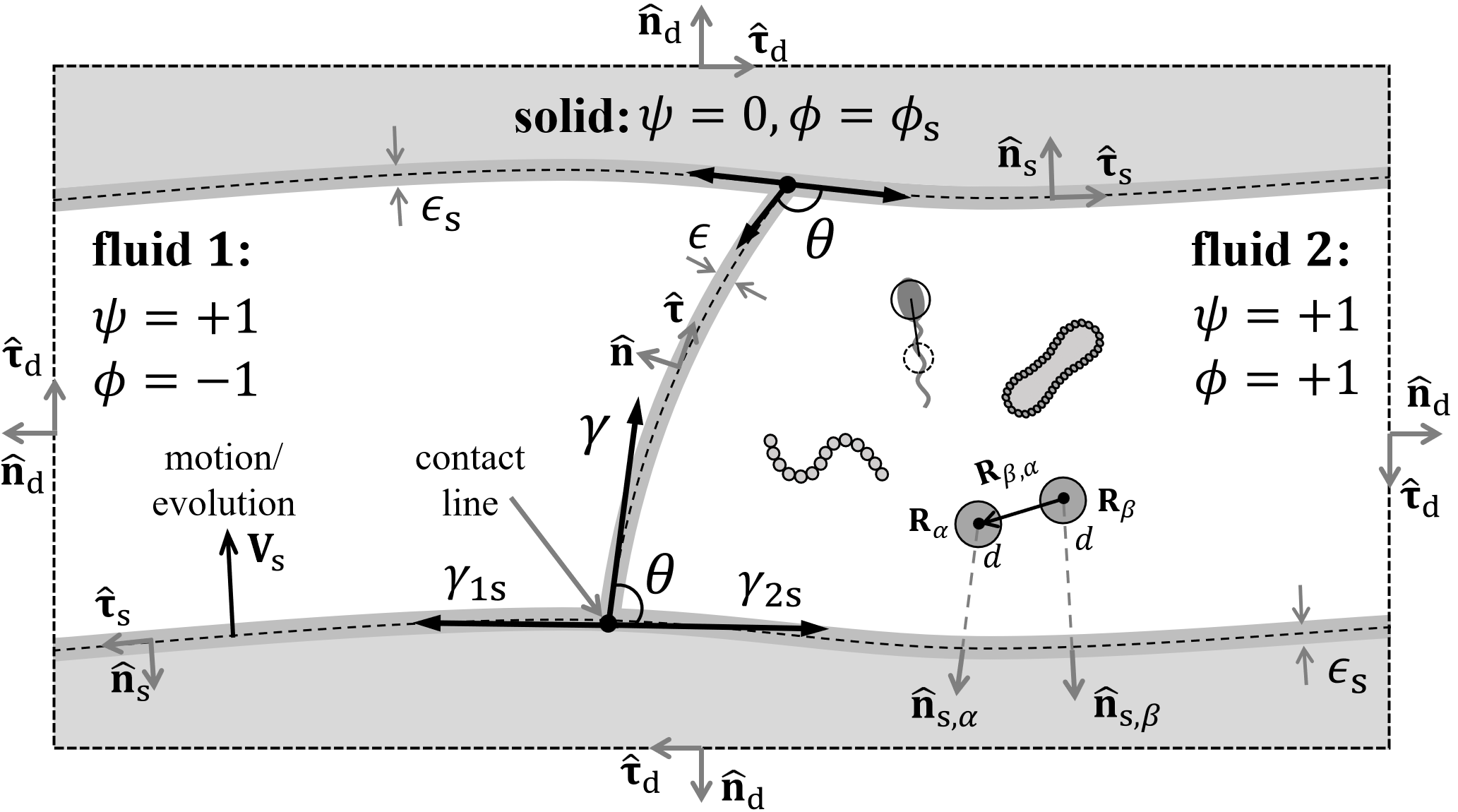}
\caption {Schematic illustration of typical mFSI scenarios simulated using the DRD approach. This approach integrates traditional diffuse-interface models for fluid-fluid interfacial dynamics with a novel implementation of complex fluid-solid interfacial conditions, achieved through smooth interpolations of dynamic-resistance coefficients across different domains. These scenarios typically involve multiphase flows where fluid-fluid interfacial dynamics are intricately coupled with wetting behavior and contact line motion on complex, corrugated solid surfaces. The DRD approach also captures the dynamics of suspended, actively deformable objects (including rigid particles, flexible fibers, microswimmers, and elastic capsules) as well as the evolution of solid surfaces driven by interfacial processes such as precipitation and dissolution. Here, immiscible two-phase fluid regions are distinguished by different values of phase parameter $\phi$, while sharing the same value $\psi=+1$. The solid domain is represented by $\psi=0$ and $\phi=\phi_{\mathrm{s}}$. 
\label{Fig:Schematic}}
\end{figure}

Numerous DNS methods have been developed to tackle the complexities of mFSI. These methods can be broadly classified into two types~\cite{Hou2012} -- partitioned and monolithic approaches, each with its strengths and limitations. The partitioned approach treats fluids and solids as two computational domains solved separately and supplemented by proper interfacial conditions at fluid-solid interfaces. 
Its major advantage is that it integrates available models and numerical methods that have been validated to solve complicated fluid or solid problems. Examples are the immersed boundary methods~\cite{Griffith2020IBM}, arbitrary Lagrangian-Eulerian method~\cite{Feng2009ALE}, and some particle-based simulation methods, \emph{e.g.}, the smoothed particle hydrodynamics method~\cite{Shadloo2016SPH}, the dissipative particle dynamics method~\cite{Hoogerbrugge1992DPD}, and the multi-particle collision dynamics method~\cite{Padding2004PRLMPC}, \emph{etc}. 
However, this approach requires either explicit tracking of interfaces or constructing proper particle-particle interactions at interfaces, which can present significant practical challenges, particularly for mFSI in multiphase flows with moving contact lines. In contrast, the monolithic approach integrates fluids and solids into a unified regular computational domain, employing a single mathematical framework and solving a coupled set of equations with a unified algorithm. The interfacial conditions are imposed implicitly, and there is no need to track the interface explicitly. 
Examples include multi-phase-field method~\cite{Chun2015,Mokbel2018,Kim2020}, fictitious domain method~\cite{Glowinski2001FDM}, diffuse domain method~\cite{Lowengrub2009DD,Lowengrub2021DD}, smoothed profile method~\cite{Yamamoto2005SPM,Yamamoto2021SPM}, and fluid particle dynamics (FPD) method~\cite{tanaka2000simulation,tanaka2018physical}, \emph{etc}. Some other popular methods exist, such as the lattice Boltzmann method~\cite{De2013LBM} that can be partitioned or monolithic depending on the solved model. 
Each of the above methods has its advantages and limitations, and the choice of method depends on the specific problem. Researchers continue to develop and refine these methods to improve their accuracy, efficiency, and applicability to real-world scenarios. 

Despite the development of these powerful DNS methods for mFSI, two main challenges persist. (i) \emph{mFSI in multicomponent multiphase flows}~\cite{Hou2012}. Most DNS methods discussed above primarily address mFSI between immersed solid structures and single-component, single-phase flows. In contrast, mFSI in multicomponent multiphase flows remains largely unexplored, posing significant challenges in both model construction (\emph{e.g.}, ensuring the thermodynamic consistency of the model) and algorithm development. These challenges arise from the intricate interplay among factors such as multicomponent fluid composition, fluid viscoelasticity, fluid-fluid interfacial tension, solid wettability, fluid slip, contact line dynamics, and external fields (\emph{e.g.}, electrical, magnetic, or acoustic fields). 
(ii) \emph{Boundary conditions at multiple dynamic interfaces and near triple-phase contact lines}~\cite{Kirby2010Book,Subramaniam2020,Ladd2021}. The high surface-area-to-volume ratio at microscales makes it crucial to define and implement precise boundary conditions that accurately capture interfacial or boundary physics for reliable simulations. However, this task becomes particularly challenging and computationally demanding when dealing with complex solid geometries or dynamic interfaces that move, deform, or evolve due to reactions or phase transitions, such as solidification and evaporation. 
In this work, we propose a generic monolithic DNS approach, the Diffuse-Resistance-Domain (DRD) approach, to address these challenges and enable the simulation of mFSI in multicomponent multiphase flows.

The remainder of this paper is organized as follows. Section~\ref{sec:theory} introduces the main idea of the DRD approach, illustrating its application to simulate one-dimensional (1D) diffusion and the dynamic coupling between viscous fluids and rigid solids. We then extend the application of DRD to the simulation of mFSI in two-phase flows, where we employ Onsager's variational principle (OVP) to derive the general governing equations. Sec.~\ref{sec:Benchmark} presents a series of benchmark simulations that address various classical mFSI problems in single-phase and two-phase fluid flows, validating the DRD approach as a robust direct numerical simulation (DNS) method for mFSI in multiphase flows. In addition, this section highlights cutting-edge applications of the DRD approach in areas such as microfluidics and active matter hydrodynamics. Section~\ref{sec:Apps-PorousFlows} delves into more complex mFSI scenarios, including cases where the solid structure evolves dynamically due to interfacial processes like precipitation and dissolution in porous media. Finally, concluding remarks are provided in Section~\ref{sec:summary}.

\section{Diffuse-resistance-domain (DRD) approach}\label{sec:theory} 

\subsection{Key features of the DRD approach}\label{sec:theory-KeyFeaturers} 

We propose a generic, monolithic DNS framework for simulating mFSI in multicomponent multiphase flows that we call \emph{diffuse-resistance-domain (DRD) approach}. It has several key features as follows.

(i) \textbf{Monolithic and diffuse-interface characterization.} The DRD approach is monolithic and all the dynamic equations apply to the whole extended fluid-solid domain with regular boundaries. Herein, all the interfaces are treated as smooth \emph{diffuse} interfaces (of small but non-zero thickness) described by phase parameter fields. For example, in the typical mFSI scenarios shown in Fig.~\ref{Fig:Schematic}, the phase field $\phi$ describes the fluid-fluid interface and $\psi$ describes the fluid-solid interface. 

(ii) \textbf{Thermodynamic consistency}. The thermodynamic consistency of the governing equations and boundary conditions is ensured by the application of Onsager's variational principle (OVP)~\cite{Doi2013,Doi2021,Xu2021,Xu2022,Arroyo2017} and linear response theory for small deviations near equilibrium~\cite{Onsager1931a,deGroot1984}, where the linear dynamic constitutive relations are assumed: $X_{\mathrm{i}}=\mathcal{R}_{ij} \dot{\alpha}_j$ with $X_{\mathrm{i}}$ and $\dot{\alpha}_{\mathrm{i}}$ being the conjugate thermodynamic force-flux pairs, and $\mathcal{R}_{ij}$ being the \emph{resistance} coefficient matrix that is symmetric and positive definite. Off-diagonal entries of $\mathcal{R}_{ij}$ are referred to as cross-coupling coefficients between different irreversible processes (with $i\neq j$). Typical linear kinetic laws of Onsager's form~\cite{deGroot1984} include Fick's law of diffusion, Fourier's law of heat conduction, Newton's law of viscosity, and Ohm's law of electrical conduction, \emph{etc}. For a more comprehensive introduction to Onsager's variational principle (OVP) and its applications in modeling complex fluids, we refer the reader to our review paper~\cite{Xu2017}, Masao Doi's textbook~\cite{Doi2013}, and the detailed review articles~\cite{Doi2011,Doi2021}. We emphasize that employing the OVP ensures thermodynamic consistency in our DRD method, while inherently constraining its applicability due to OVP's foundational assumptions. Like OVP, DRD is thus limited to systems satisfying Onsager’s linear relations, valid only near equilibrium. Nevertheless, our formulation requires local equilibrium—-even under globally far-away from equilibrium conditions—-by assuming the relaxation timescale of material elements to local equilibrium is negligible compared to that of the global dynamics~\cite{deGroot1984,Doi2013}. Under this well-separated timescale condition, all equilibrium thermodynamic quantities (\emph{e.g.}, entropy and restricted free energy) remain well-defined locally, thereby justifying the use of linear response theory, the Onsager relations, and OVP in our model construction. We acknowledge that in cases where the relaxation time is comparable to the global timescale (\emph{e.g.}, in glass or in active colloidal suspensions), these approximations would break down and the current formulation would no longer be applicable. That is, the proposed DNS method exhibits limitations in regimes characterized by strongly non-equilibrium dynamics or high-Reynolds-number flows, where local equilibrium assumptions become invalid and the equilibrium free energy and quadratic dissipation functionals are not well-defined.

(iii) \textbf{Prescribed fluid-solid interfacial profiles.} In contrast to multi-phase-field methods for mFSI, the DRD approach delineates fluid-solid interfaces through phase parameter fields that do not evolve by dynamic equations (such as Cahn-Hilliard or Allen-Cahn equations) but are prescribed by functions changing smoothly across the boundary from the fluid to the solid domain. This simplifies or avoids the energy construction for solids—especially in the rigid limit—and substantially reduces computational costs, particularly when the solids move, deform, or evolve.
For example, the $\psi$ in Fig.~\ref{Fig:Schematic} is prescribed by
\begin{equation}\label{Eq:theory-DRD-psi}
\psi(\boldsymbol{r}, t)=\frac{1}{2}\left[1-\tanh \left(\frac{\mathcal{D}(\boldsymbol{r}, t)}{\sqrt{2} \epsilon_{\mathrm {s}}}\right)\right],
\end{equation}
changing smoothly across the boundary from the fluid domains ($\psi=1$) to the solid domain ($\psi=0$). Here $\epsilon_{\mathrm{s}}$ is the characteristic fluid-solid interfacial thickness and $\mathcal{D}(\boldsymbol{r}, t)$ is the signed distance function away from the solid surface with ${\mathcal{D}}>0$ ($<0$) in the solid (fluid) domain. Note that $\mathcal{D}(\boldsymbol{r}, t)$ encodes the complicated dynamic solid geometries (including the shape and the instantaneous position) of the solid structure that moves, deforms, or evolves with known or separately-solved local velocities. The fluid-solid phase field $\psi$ is prescribed \emph{a priori} by the form of Eq.~(\ref{Eq:theory-DRD-psi}), mainly because this choice provides a smooth transition between the fluid and solid domains and it makes the sharp-interface limit analysis (see Sec.~\ref{sec:App3-SharpLimit}) more self-consistent. By doing so, we simplify the numerical implementation of fluid-solid interfacial dynamics. We avoid the need to construct a free-energy function to characterize the coexisting fluid and solid phases and circumvent the computational expense associated with solving the full dynamic equations for $\psi$. This is particularly advantageous for mFSI systems where solids are rigid of complex geometries or exhibit complex motions, deformations, or evolutions.

(iv) \textbf{Solids treated as special fluids}. The solid domains are treated as special fluid domains where the resistance coefficients $\mathcal{R}_{ij}$ have a significant contrast compared to the resistance coefficients of other ``real'' fluid domains. 
In practice, we employ the fluid-solid phase-field parameter (\emph{e.g.}, $\psi(\boldsymbol{r}, t)$) to smoothly interpolate $\mathcal{R}_{ij}$ between fluid and solid domains, \emph{e.g.}, in the form of Eq.~(\ref{Eq:theory-Dprofile12}) or Eq.~(\ref{Eq:theory-Dprofile3}). By this, we can then obtain the required governing equations applicable to the whole fluid-solid system easily from those already developed in sharp (fluid-solid) interface models of mFSI~\cite{Xu2012,Xu2012b}. Moreover, in comparison to the diffuse-domain method~\cite{Lowengrub2009DD}, the governing equations in the DRD approach applicable to both multiphase fluids and solid structures take the same form as those for multiphase fluids alone. This can significantly simplify the development of numerical algorithms.

In the following subsections, we will further elucidate the key features of the DRD approach. First, we apply it to simulate one-dimensional (1D) diffusion and investigate the dynamic coupling between viscous fluids and rigid solids. Subsequently, we will further demonstrate the application of DRD in modeling mFSI within two-phase flows.

\begin{figure}[htbp] 
  \centering
  \includegraphics[width=0.95\columnwidth]{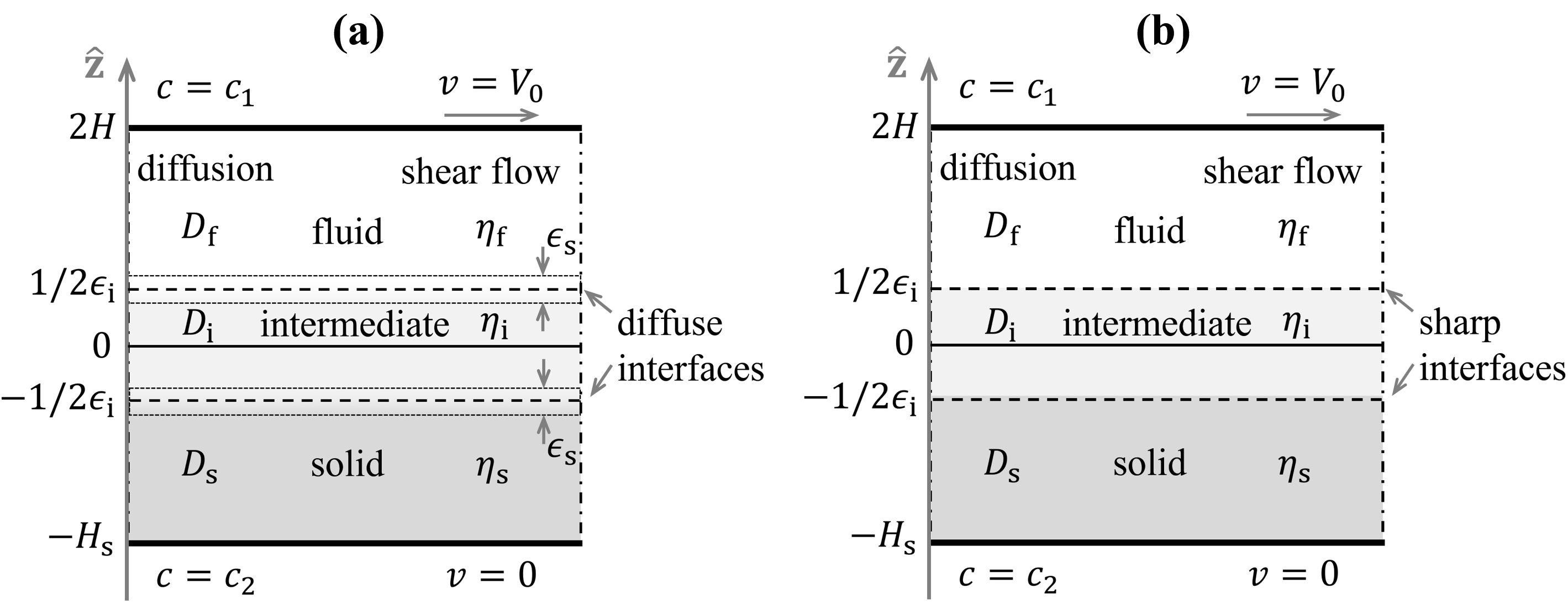}
  \caption {Schematic illustrations of the 1D boundary value problem in a system including three (fluid, intermediate, and solid) subdomains with (a) diffuse interfaces (of small but non-zero thickness $\epsilon_{\mathrm{s}}$) and (b) sharp interfaces (discussed mainly in Appendix~\ref{sec:App-1DDiffusion} to elucidate the validity of DRD approach), respectively. The discussions and conclusions about the diffusion equation for 1D diffusion apply equally to the Navier-Stokes equation for laminar simple shear flow. } 
  \label{Fig:theory-1DSchematic}
\end{figure}

\subsection{Simulating 1D diffusion: DRD approach in a nutshell}\label{sec:theory-nutshell} 

We begin by explaining the main components of the DRD approach and demonstrate its application by solving the one-dimensional (1D) diffusion equation with three representative types of boundary conditions. Consider a 1D diffusion process within a fluid domain defined by $0\leqslant z \leqslant 2H$ that is in contact with a solid domain defined by $-H_{\mathrm{s}}\leqslant z < 0$, as illustrated in Fig.~\ref{Fig:theory-1DSchematic}. The governing equation for the concentration $c=c(z,t)$ is the 1D diffusion equation
\begin{equation}\label{Eq:theory-ct}
\partial_t c=\partial_{\mathrm z}\left(D\partial_{\mathrm z}c \right),
\end{equation}
subject to the boundary condition $c(H,t)=c_1$ and a suitable boundary condition at the fluid-solid interface $z=0$. Three common types of boundary conditions are typically considered: (1) Dirichlet boundary condition, $c(0,t)=c_2$; (2) Neumann boundary condition, $\partial_z c(0,t)=0$; (3) Robin boundary condition, $\ell_{\mathrm{s}} \partial_z c(0,t)= c(0,t)-c_2$. While the numerical solution to such a 1D problem is relatively straightforward, various algorithms have been developed to address it. In contrast, solving this problem in a real three-dimensional (3D) system with complex, curved solid geometries presents a more significant challenge. To overcome this complexity, we apply the DRD approach to solve the 1D diffusion equation under different boundary conditions.

\begin{figure}[htbp] 
  \centering
  \includegraphics[width=0.73\columnwidth]{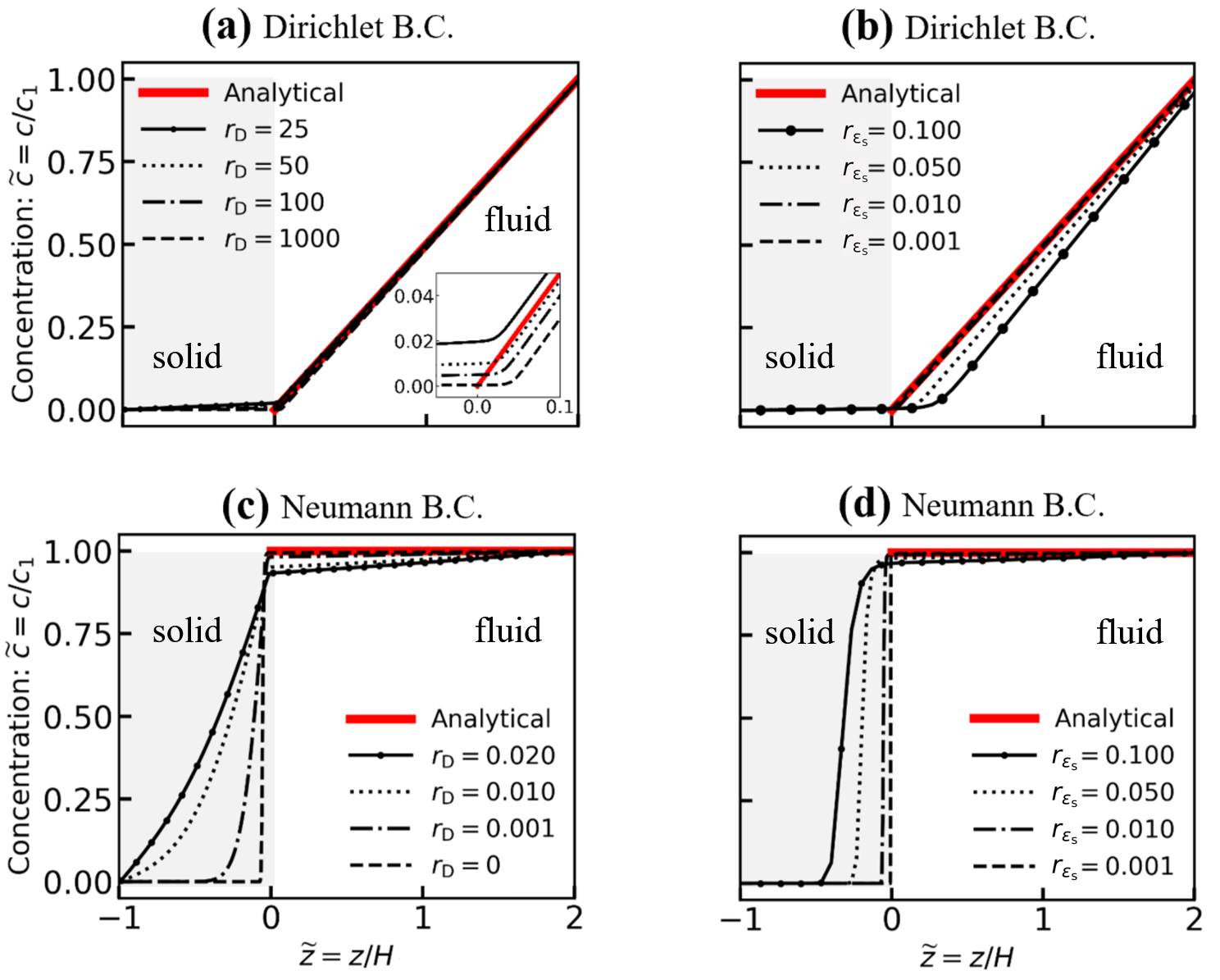}
  \caption {Steady-state solutions $\tilde{c}(z)=c/c_1$ of the 1D diffusion equation (\ref{Eq:theory-ct}) in the fluid domain $0\leqslant z \leqslant 2H$ under (a,b) Dirichlet ($c(0,t)=0$) and (c,d) Neumann ($\partial_z c(0,t)=0$) boundary conditions. Numerical solutions using the DRD approach for different diffusivity ratios $r_{\mathrm D}=D_{\mathrm{s}}/D_{\mathrm{f}}$ and relative interface thicknesses $r_{\epsilon_{\mathrm s}}=\epsilon_{\mathrm s}/H$ are compared to the analytical solution (red solid curves). For sufficiently small $r_{\epsilon_{\mathrm s}}\leqslant 0.01$, (a,b) if $r_{\mathrm D}\gg 1$, we reproduce Dirichlet boundary condition, while (c,d) if $r_{\mathrm D}\to 0$, we reproduce Neumann boundary condition. 
  Here, we take the physics parameters to be: ${H^2}/{\tau_0 D_\mathrm{f}} = 1.0$, $c_2/c_1=0$, and $H_s/H=1.0$. There are two ``free'' parameters: $r_{\epsilon_{\mathrm s}} =\epsilon_{\mathrm s}/H$ and $r_{\mathrm D}=D_{\mathrm{s}}/D_{\mathrm{f}}$. 
  In (a,c), we vary $r_{\mathrm D}$ and fix $r_{\epsilon_{s}}=0.01$; in (b), we vary $r_{\epsilon_{\mathrm s}}$ and fix the diffusivity ratios $r_{\mathrm D}=100$; in (d), we vary $r_{\epsilon_{\mathrm s}}$ and fix $r_{\mathrm D}=0$.}
  \label{Fig:theory-BC12}
\end{figure}

According to the DRD approach, we treat the fluid-solid interface at $z=0$ as a diffuse interface of small finite thickness $\epsilon_{\mathrm{s}}$ (see Fig.~\ref{Fig:theory-1DSchematic}(a)), characterized by a smooth (time-independent) interfacial profile function $\psi(z)$ (also called phase parameter as in the phase-field approach of multiphase flows~\cite{Qian2006JFM}):
\begin{equation}\label{Eq:theory-psi}
\psi(z)=\frac{1}{2}\left[1+\tanh \left(\frac{z}{\sqrt{2} \epsilon_{\mathrm {s}}}\right)\right],
\end{equation}
which changes smoothly from $\psi=1$ in the fluid domain to $\psi=0$ in the solid domain. 
Furthermore, to be more accurate near the fluid-solid interface, we assume the combined phase parameter $\psi(z) c(z,t)$ (instead of $c(z,t)$) is locally conserved following
\begin{equation} \label{Eq:theory-ct2}
\partial_t(\psi c) = -\partial_z J,
\end{equation}
with $J$ being the diffusion flux. Then, the governing dynamic equation can be derived using Onsager's variational principle (OVP)~\cite{Qian2006JFM,Doi2011,Doi2013,Doi2021,Xu2017,Xu2021,Xu2022}. 

Firstly, the free energy for the fluid-solid two-phase system is given by
\begin{equation}\label{Eq:theory-F}
\mathcal{F}\left[c(z,t), \psi(z)\right]=\int d\boldsymbol {r} \left[\psi k_{\mathrm B}Tc\ln(c/c_0)\right], 
\end{equation}  
where $k_{\mathrm B}$ is Boltzmann constant, $T$ is the temperature, and $c_0$ is the some reference concentration. Using Eq.~(\ref{Eq:theory-ct2}), we obtain the rate of change of the free energy as $\dot{\mathcal{F}}[J]=\int d\boldsymbol{r}\left(J \partial_z \mu\right)$ with the chemical potential $\mu =k_{\mathrm B}T(1+\ln(c/c_0))$.
Secondly, the dissipation function is given by~\cite{Doi2011,Doi2013,Doi2021}
\begin{equation}\label{Eq:theory-Phi}
\Phi[J]=\int d \boldsymbol{r}\left[\frac{k_{\mathrm B}T}{2cD(\psi)}  J^2\right].
\end{equation}
Physically, the quadratic form of $\Phi$ is equivalent to Onsager's linear force-flux relations~\cite{Onsager1931a}. 
Thirdly, minimizing the Rayleighian $\mathcal {R}[J]=\dot{\mathcal{F}} +\Phi$ with respect to $J$ gives the dynamic equation:  
\begin{equation}\label{Eq:theory-ct3}
\partial_t (\psi c)=\partial_{\mathrm z}\left[D(\psi)\partial_{\mathrm z}c \right],
\end{equation}
supplemented with Dirichlet (essential) boundary conditions, $c=c_1$ and $c=c_2$ at $z=2H$ and $z=-H_{\mathrm{s}}$, respectively.

\begin{figure}[htbp]
  \centering
  \includegraphics[width=1.0\columnwidth]{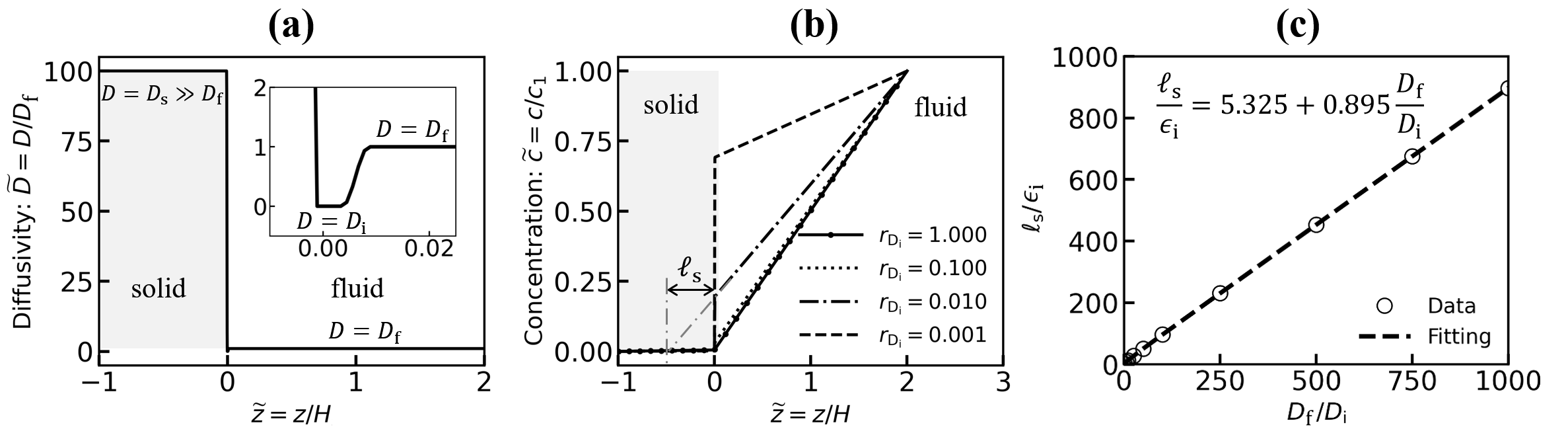} 
  \caption { Steady-state solutions $\tilde{c}(z)=c/c_1$ of the 1D diffusion equation (\ref{Eq:theory-ct}) in the fluid domain $0\leqslant z \leqslant 2H$ under the Robin boundary condition, $\ell_{\mathrm{s}} \partial_z c(0,t)= c(0,t)$. (a) The interpolation profile of the diffusivity $D(z)$ takes the form of Eq.~(\ref{Eq:theory-Dprofile3}). (b) ``Slip'' solution for the Robin boundary condition is reproduced using the DRD approach for different relative slip lengths $\ell_{\mathrm{s}}/H$, prescribed by different diffusivity ratios $r_{{\mathrm D}_{\mathrm{i}}}=D_{\mathrm{i}}/D_{\mathrm{f}}$. Here, we take the physics parameters to be: ${H^2}/{\tau_0 D_\mathrm{f}} = 1.0$, $c_2/c_1=0$, and $H_s/H = 1.0$. We vary $r_{{\mathrm D}_{\mathrm{i}}}$ and fix the three ``free'' parameters: $r_{\epsilon_{\mathrm s}} =\epsilon_{\mathrm s}/H=0.01$, $r_{\epsilon_{\mathrm i}} =\epsilon_{\mathrm i}/H=0.005$, and $r_{\mathrm D}=D_{\mathrm{s}}/D_{\mathrm{f}}=100$. (c) The relative ``slip" length $\ell_{\mathrm{s}}/\epsilon_{\mathrm{i}}$ is found to be linearly proportional to $r_{{\mathrm D}_{\mathrm{i}}}^{-1}=D_{\mathrm{f}}/D_{\mathrm{i}}$, as predicted by sharp-interface calculations in the triple-domain system (see Appendix~\ref{sec:App-1DDiffusion}). }
  \label{Fig:theory-BC3}
\end{figure}

According to the DRD approach, to reproduce the three types of boundary conditions mentioned above at $z=0$, we need to take proper smooth interpolations of the diffusivity $D$ across the fluid and the solid domains. For the Dirichlet and the Neumann boundary conditions, we choose a smooth interpolation of $D(z)$ by taking it as a simple linear function of $\psi$:
\begin{subequations}\label{Eq:theory-Dprofile}
\begin{equation}\label{Eq:theory-Dprofile12}
D(\psi)= D_{\mathrm{s}} +\left(D_{\mathrm{f}}-D_{\mathrm{s}}\right)\psi, 
\end{equation}
with $D(\psi=1)=D_{\mathrm{f}}$ and $D(\psi=0)=D_{\mathrm{s}}$ in the fluid and solid domains, respectively. In the sharp interface limit ($\epsilon_{\mathrm{s}} \to 0$), if $D_{\mathrm{s}} \gg D_{\mathrm{f}}$, we reproduce the Dirichlet boundary condition at $z=0$ effectively as shown in Fig.~\ref{Fig:theory-BC12}(a) and Fig.~\ref{Fig:theory-BC12}(b), whereas, if $D_{\mathrm{s}} \ll D_{\mathrm{f}}$, we would instead reproduce the Neumann boundary condition at $z=0$ effectively as shown in Fig.~\ref{Fig:theory-BC12}(c) and Fig.~\ref{Fig:theory-BC12}(d). Note that in the case of Neumann boundary condition with $D_{\mathrm{s}} \ll D_{\mathrm{f}}$ and hence $D(\psi)\approx  D_{\mathrm{f}} \psi$, the diffusion equation becomes $\partial_t (\psi c)=\partial_{\mathrm z}\left(\psi D_{\mathrm{f}}\partial_{\mathrm z}c \right)$, which takes the same form of that in diffuse domain method~\cite{Lowengrub2009DD}. 

In contrast, to reproduce the Robin boundary condition, we have to introduce the third intermediate domain of small thickness $\epsilon_{\mathrm{i}}$ with a very small diffusivity $D_{\mathrm i}$ ($<D_{\mathrm{f}}\ll D_{\mathrm{s}}$), as shown in Fig.~\ref{Fig:theory-1DSchematic}(a), and take the following interpolation form of $D$ over the three domains
\begin{equation}\label{Eq:theory-Dprofile3}
D(\psi)= 
\begin{cases}
D_{\mathrm{f}}+ 2(1-\psi)(D_{\mathrm{i}}-D_{\mathrm{f}}) & \text { if } \quad \psi \geqslant 0.5 \quad \text{and}\quad \psi=\psi_1, \\
2D_{\mathrm{i}}-D_{\mathrm{s}}+ 2(1-\psi)(D_{\mathrm{s}}-D_{\mathrm{i}}) & \text { if } \quad \psi < 0.5\quad \text{and}\quad \psi=\psi_2,
\end{cases}
\end{equation}    
\end{subequations}
as schematically shown in Fig.~\ref{Fig:theory-1DSchematic}(a) and plotted in Fig.~\ref{Fig:theory-BC3}(a), with $\psi_1(z)=\frac{1}{4}\left[3+\tanh \left((z-\epsilon_{\mathrm{i}}/2)/{\sqrt{2} \epsilon_{\mathrm {s}}}\right)\right]$ and $\psi_2(z)=\frac{1}{4}\left[1+\tanh \left((z+\epsilon_{\mathrm{i}}/2)/{\sqrt{2} \epsilon_{\mathrm {s}}}\right)\right]$ both taking a form similar to that of Eq.~(\ref{Eq:theory-psi}). In the limits of sharp interface (as $\epsilon_{\mathrm{s}} \to 0$) and $D_{\mathrm{s}} \gg D_{\mathrm{f}}$, we reproduce the Robin boundary condition at $z=0$ effectively as shown in Fig.~\ref{Fig:theory-BC3}(b), and Fig.~\ref{Fig:theory-BC3}(c) shows the effective ``slip'' length $\ell_{\mathrm{s}}/\epsilon_{\mathrm{i}}$ is linearly proportional to the diffusivity ratio $D_{\mathrm{f}}/D_{\mathrm{i}}$.

Note that alternative interpolation forms to those in Eqs.~(\ref{Eq:theory-Dprofile}) are also possible; however, our primary concern is to ensure that the interpolation function preserves the essential characteristics—whether monotonic or non-monotonic (with an special intermediate domain)—with respect to the fluid–solid phase parameter, $\psi$. For example, we employ the linear, monotonic interpolation in Eq.(\ref{Eq:theory-Dprofile12}) since our objective is not to impose any specialized structure to enforce Dirichlet or Neumann boundary conditions. We have also tested a Heaviside function for $\psi$ and observed no significant differences in the simulation outcomes. However, to reproduce Robin boundary conditions, we adopted the non-monotonic interpolation in Eq.~(\ref{Eq:theory-Dprofile3}) and compared its performance with that of a piecewise cubic polynomial interpolation for $\psi$ as used in our previous work~\cite{zhang2015anisotropic}. Both approaches yielded only quantitative differences in the relationship between the effective slip length, $\ell_{\mathrm{s}}/\epsilon_{\mathrm{i}}$, and the diffusivity ratio, $r_{D_{\mathrm{i}}}=D_{\mathrm{i}}/D_{\mathrm{f}}$. In contrast, the piecewise linear interpolation in Eq.~(\ref{Eq:theory-Dprofile3}) is simpler to implement and provides a larger range of slip lengths, as shown in Fig.\ref{Fig:theory-BC3}(c).

 Before carrying out simulations, we first non-dimensionalize the 1D diffusion equation by taking the length unit to be $H$, time by some characteristic measurement time $\tau_0$, and concentration by $c_1$. Eight dimensionless parameters appear, as shown in the following comprehensive list along with their physical interpretations. Moreover, we classify these dimensionless parameters into two general categories:  
\begin{itemize}
\item (i) Measurable physics parameters defined based on measurable quantities. (1) $H^2/\tau_0 D_{\mathrm{f}}$: the ratio of the diffusion time $H^2/D_{\mathrm{f}}$ to the time unit $\tau_0$. (2) $c_2/c_1$: the concentration ratio. (3) $H_{\mathrm{s}}/H$: the dimension ratio of the solid and the fluid region. (4) $\ell_{\mathrm{s}}/H$: the relative ``slip'' length in the Robin boundary condition.  
\item (ii) Parameters arising from the DRD treatment of fluid-solid interfaces. Diffusivity ratios: (1) $r_D=D_{\mathrm{s}}/D_{\mathrm{f}}$ and (2) $r_{D_{\mathrm{i}}}=D_{\mathrm{i}}/D_{\mathrm{f}}$. Relative interfacial thicknesses: (3) $r_{\epsilon_{\mathrm{s}}}=\epsilon_{\mathrm{s}}/H$ and (4) $r_{\epsilon_{\mathrm{i}}}=\epsilon_{\mathrm{i}}/H$. 
\end{itemize}
In the appendix~\ref{sec:App-1DDiffusion}, we analyze the sharp-interface models for the 1D diffusion problem. It reveals that some parameters from the DRD treatment are related to measurable quantities. Consequently, the effective number of free parameters is reduced. For example, the slip length $\ell_{\mathrm{s}}$ in the Robin boundary condition is found to scale as $\ell_{\mathrm{s}}/\epsilon_{\mathrm{i}} \sim D_{\mathrm{f}}/D_{\mathrm{i}}=r_{D_{\mathrm{i}}}^{-1}$. Therefore, only \emph{three} free parameters remain: $r_{\epsilon_{\mathrm{s}}}=\epsilon_{\mathrm{s}}/H$, $r_{\epsilon_{\mathrm{i}}}=\epsilon_{\mathrm{i}}/H$, and $r_D=D_{\mathrm{s}}/D_{\mathrm{f}}$.

 The DRD simulation results for 1D diffusion are presented in Figs.~\ref{Fig:theory-BC12} and \ref{Fig:theory-BC3} with careful convergence analysis. They provide practical guidelines for the value selection of the three ``free'' parameters: sufficiently small interface thicknesses, $r_{\epsilon_{\mathrm{s}}}\sim r_{\epsilon_{\mathrm{i}}} \leqslant 0.01$; proper values of diffusivity ratios, $r_D \geqslant 100$ for Dirichlet and Robin boundary condition, and $r_D=0$ for Neumann boundary condition. In this way, the steady-state numerical solutions at $0\leqslant z \leqslant 2H$ obtained from the DRD approach agree very well with the analytical steady-state solutions for the 1D diffusion equation under the three different types of boundary conditions. Such validity of the DRD approach in reproducing the solutions of 1D diffusion equation with different boundary conditions can be easily understood by considering the analytical steady-state solutions of 1D diffusion equation (\ref{Eq:theory-ct}) in a triple-domain system (for $-H_{\mathrm{s}}\leqslant z\leqslant 2H$) separated by two ``sharp" interfaces located at $z=\epsilon_{\mathrm i}/2$ and $z=-\epsilon_{\mathrm i}/2$ as shown in Fig.~\ref{Fig:theory-1DSchematic}(b), and taking proper limits of the diffusivity ratio $r_D=D_{\mathrm{s}}/D_{\mathrm{f}}$ in different cases (see Appendix~\ref{sec:App-1DDiffusion} for a detailed discussion).

\subsection{Simulating dynamic coupling between viscous fluids and rigid solids}

The idea of the DRD approach mentioned above can also be used to numerically solve the dynamic equations for other transport phenomena such as the heat conduction equation, Navier-Stokes equation, Cahn-Hilliard and Allen-Cahn equations, electrokinetic equation, and so on. Here, we provide an additional example demonstrating the application of the DRD approach to solving the Navier-Stokes equations.

To be specific, we consider the 1D laminar flow field (with velocity $\boldsymbol{v}=v(z,t) \boldsymbol{\hat{x}}$ and pressure $p=\mathrm{const}$), which is induced by simple shear in a viscous fluid (with $0\leqslant z \leqslant 2H$) as shown in Fig.~\ref{Fig:theory-1DSchematic}. The governing equation for the 1D velocity field is the following 1D Navier-Stokes equation 
\begin{equation}\label{Eq:theory-vt}
\partial_t v = \partial_{\mathrm z}\left(\nu \partial_{\mathrm z}v \right),
\end{equation}
which takes the same form of the 1D diffusion equation (\ref{Eq:theory-ct}) with $\nu$ being the kinematic viscosity. When Eq.~(\ref{Eq:theory-vt}) is subject to the boundary condition $v(2H,t)=V_0$ and one of the three types of boundary conditions for $v$ at the fluid-solid interface $z=0$, we can then use the DRD approach to solve it numerically as discussed above. All the above discussions for the solution of the 1D diffusion equation also apply to the solution of 1D Navier-Stokes equation here. Here, the only relevant resistance coefficient is the viscosity $\nu$, analogous to diffusivity $D$ for the 1D diffusion. The three types of boundary conditions at $z=0$ can be reproduced effectively if we take proper smooth interpolations of $\nu$ over different domains. For the interpolation form in Eq.~(\ref{Eq:theory-Dprofile12}), if $\nu_2 \gg \nu_1$ (and for sufficiently small interfacial thickness $\epsilon_{\mathrm{s}}$), we reproduce the Dirichlet or no-slip boundary condition at $z=0$ effectively: $v|_{z=0}=0$. Whereas, if $\nu_2 \ll \nu_1$ (and for sufficiently small interfacial thickness $\epsilon_{\mathrm{s}}$), we would reproduce the Neumann or free-surface boundary condition at $z=0$ effectively: $\partial_z v|_{z=0}=0$.
In contrast, for the interpolation form in Eq.~(\ref{Eq:theory-Dprofile3}), if $D_{\mathrm i}<D_{\mathrm{f}}\ll D_{\mathrm{s}}$, we reproduce the Robin boundary condition -- Navier's slip boundary condition: $\ell_{\mathrm s}\partial_z v|_{z=0}=v(z=0)$ with $\ell_{\mathrm s}$ being the fluid slip length. More generally, when employing the DRD approach to simulate 3D viscous flows over solid surfaces with complex geometries, it suffices to solve the Navier-Stokes equations with spatially varying viscosity. Numerous efficient algorithms are already available and can be directly utilized for this purpose~\cite{gao_efficient_2014}.

 


Finally, we would like to point out that similar ideas of using smoothed profiles of viscosity across the interfacial region have been proposed before to facilitate the numerical study of colloidal dynamics~\cite{tanaka2000simulation,tanaka2018physical} (so-called fluid particle dynamics method, or FPD method), particle dispersion in nematic liquid-crystal solvents~\cite{Yamamoto2001}, the impact of solid objects on free surfaces~\cite{Mirzaii2012,Chun2015}, two-phase flows in or on stationary solid domains with complex geometries~\cite{Kim2020}, the melting and flowing of solids~\cite{Carlson2002}, and the fluid-structure interaction in presence of a hyperelastic body~\cite{Esmailzadeh2014}, \emph{etc}. Particularly, the FPD method has been shown by careful numerical~\cite{tanaka2018physical,Xu2023PoF} and analytical studies~\cite{Fujitani2007} to be capable of reproducing arbitrary flow fields derived from conventional sharp-interface methods as the viscosity ratio tends to be large enough and the interface thickness tends to be sufficiently small.  

\subsection{Simulating mFSI in two-phase flows: General equations}\label{sec:theory-Eqns} 

To further demonstrate the capability of the DRD approach in simulating mFSI, we consider an immiscible two-phase flow including some suspending (active deformable) objects composed of $N$ rigid spherical particles over rigid corrugated solid boundary surfaces, as shown in Fig.~\ref{Fig:Schematic}. In the DRD approach, we use the interfacial profile functions or phase parameters $\phi(\boldsymbol{r},t)$, $\psi_{\alpha}(\boldsymbol{r},t)$ (with $\alpha=1,..., N$), and $\psi(\boldsymbol{r})$ to describe the immiscible fluid-fluid interface, the surface of $\alpha$-th particle, and the corrugated solid boundary surface, respectively. 
Particularly, for the $\alpha$-th particle of diameter $d$ centered at $\boldsymbol {R}_{\alpha}(t)$, 
the interfacial profile function $\psi_{\alpha}$ takes the form of Eq.~(\ref{Eq:theory-DRD-psi}) with the signed distance function given by $\mathcal{D}_{\alpha}(\boldsymbol{r}, t)= d/2-\left| \boldsymbol{r}-\boldsymbol{R}_{\alpha} (t)\right|$, that is, 
\begin{equation}\label{Eq:theory-DRD-Dalpha}
\psi_{\alpha}(\boldsymbol{r}, t)=\frac{1}{2}\left[1-\tanh \left(\frac{d/2-\left| \boldsymbol{r}-\boldsymbol{R}_{\alpha} (t)\right|}{\sqrt{2} \epsilon_{\mathrm {s}}}\right)\right],
\end{equation}
and the velocity of the particle is given by 
\begin{equation}\label{Eq:theory-DRD-Valpha}
\dot{\boldsymbol{R}}_{\alpha}=\boldsymbol{V}_{\alpha}(t)\equiv \frac{\int d\boldsymbol{r}^{\prime} (1-\psi_{\alpha}) \boldsymbol{v}(\boldsymbol{r}^{\prime},t) }{\int  d\boldsymbol{r}^{\prime} (1-\psi_{\alpha})},
\end{equation}
with the superscript dot hereafter denoting the material time derivative, \emph{i.e.,} $\dot{\boldsymbol{R}}_{\alpha}={d{\boldsymbol{R}}_{\alpha}}/{dt}$.
 Note that Eq.~(\ref{Eq:theory-DRD-Valpha}) does not impose force-free conditions; rather, it is applicable even when external forces are present. In fact, Eq.~(\ref{Eq:theory-DRD-Valpha}) originates from one of the key features of our DRD approach—namely, the treatment of solids as special fluids—which is outlined at the beginning of Sec.~\ref{sec:theory-KeyFeaturers}. Herein, rigid particles are treated as highly viscous fluids. Then by numerically solving the dynamic equations (\ref{Eq:theory-DRD-Dyn}), we obtain an effective ``flow'' field within the particle that primarily reflects its translation and rotation. Hence, it is natural that the particle’s translational velocity is computed by Eq.~(\ref{Eq:theory-DRD-Valpha}) as the average flow velocity inside the particle.

For the corrugated solid boundary surfaces, the interfacial profile function $\psi$ also takes the form of Eq.~(\ref{Eq:theory-DRD-psi}): $\psi(\boldsymbol{r}, t)=\frac{1}{2}\left[1-\tanh \left({\mathcal{D}(\boldsymbol{r}, t)}/{\sqrt{2} \epsilon_{\mathrm {s}}}\right)\right]$ with $\mathcal{D}(\boldsymbol{r}, t)$ being the signed distance away from the solid boundary surfaces. 
Furthermore, to improve the conservation of the phase parameter $\phi$ within the fluid domain, we assume the combined phase parameter $\Psi(\boldsymbol{r},t)\phi(\boldsymbol{r},t)$ (instead of $\phi(\boldsymbol{r},t)$) is locally conserved following
\begin{equation} \label{Eq:theory-DRD-phi}
\partial_t(\Psi\phi) =-\nabla \cdot(\Psi\phi\boldsymbol{v})-\nabla \cdot \boldsymbol{J},
\end{equation}
where $\Psi(\boldsymbol{r},t) \equiv \psi\prod_{\alpha=1}^N \psi_\alpha$ with $\Psi=0$ inside solid boundary or inside particles and $\Psi=1$ inside fluids, $\boldsymbol{J}$ is the diffusion flux, and $\boldsymbol{v}(\boldsymbol{r},t)$ denotes the flow velocity, following the incompressibility condition $\nabla\cdot \boldsymbol{v}=0$ (alternatively, one can also take $\nabla\cdot (\psi \boldsymbol{v})=0$ as in the diffuse domain method~\cite{Lowengrub2009DD,Lowengrub2021DD}). Then, the governing dynamic equations can be derived using Onsager's variational principle (OVP)~\cite{Qian2006JFM,Doi2011,Doi2013,Doi2021,Xu2017,Xu2021,Xu2022}. 

Firstly, the free energy for the fluid-fluid-solid triple-phase system includes three contributions:
\begin{equation}\label{Eq:theory-DRD-F}
\mathcal{F}\left[\phi(\boldsymbol{r},t), \psi (\boldsymbol{r}),  \psi_{\alpha}\left(\boldsymbol{r}; \boldsymbol{R}_{\alpha}\right)\right]=\mathcal{F}_{\mathrm b}[\phi(\boldsymbol{r},t),\psi(\boldsymbol{r},t)]+\mathcal{F}_{\mathrm s}\left[\phi(\boldsymbol{r},t), \psi (\boldsymbol{r}), \psi_{\alpha}\left(\boldsymbol{r}; \boldsymbol{R}_{\alpha}\right)\right]. 
\end{equation} 
Here the fluid-bulk energy $\mathcal{F}_{\mathrm b}=\int d\boldsymbol {r} \hat{f}_{\mathrm b}$ takes the form of Cahn-Hilliard free energy as
\begin{subequations}\label{Eq:theory-DRD-FbFs}
\begin{equation} \label{Eq:theory-DRD-FbFs-Fb}
\mathcal{F}_{\mathrm b}[\phi(\boldsymbol r,t),\psi(\boldsymbol{r},t)]=\int d\boldsymbol {r} \left[\Psi\left(f_{\mathrm b}(\phi)+\frac{1}{2}K|\nabla \phi|^2 \right)\right],
\end{equation} 
with $\hat{f}_{\mathrm b}\equiv f_{\mathrm b}+\frac{1}{2}K|\nabla \phi|^2$ being the bulk free energy density and $f_{\mathrm b}(\phi)=\frac{1}{4}a\left(\phi^2-1\right)^2$ being the double-well potential and $\phi=\pm 1$ representing the two fluid phases. The energy parameter $a$ and the interfacial stiffness $K$ are positive constants.
The thickness and interfacial tension of the fluid-fluid interface are given by $\epsilon=\sqrt{K/a}$ and $\gamma=2\sqrt{2} a\epsilon /3$, respectively.
The fluid-solid interfacial energy $\mathcal{F}_{\mathrm s}$ is given by
\begin{equation} \label{Eq:theory-DRD-FbFs-Fs}
\mathcal{F}_{\mathrm s}\left[\phi(\boldsymbol{r},t), \psi (\boldsymbol{r}),  \psi_{\alpha}\left(\boldsymbol{r}; \boldsymbol{R}_{\alpha}\right)\right]=  \int d \boldsymbol{r} \left[\epsilon_{\mathrm s} f_{\mathrm{s}}(\phi) \left(\sum_{{\alpha}=1}^N\left|\nabla \psi_{\alpha}\right|^2 + \left|\nabla\psi\right|^2\right)\right],
\end{equation}
\end{subequations}
with $f_{\mathrm s}(\phi)=-\frac{1}{4} \gamma \cos \theta_{\mathrm{s}} (\phi^3-3\phi)$ being the surface energy density~\cite{xia2009flow,Lowengrub2021DD} and $\theta_{\mathrm{s}}$ being the static contact angle at the solid surfaces. 
Using Eqs.~(\ref{Eq:theory-DRD-Valpha})--(\ref{Eq:theory-DRD-FbFs}), we obtain the rate of change of the total free energy $\dot{\mathcal{F}}_{\mathrm T}=\dot{\mathcal{F}}-\dot{\mathcal{W}}_{\mathrm{ext}}$ as (for a detailed derivation, refer to the appendix~\ref{sec:App-DRD-theory})
\begin{equation}\label{Eq:theory-DRD-FTdot}
\dot{\mathcal{F}}_{\mathrm T} [\boldsymbol{v},\boldsymbol{J}]=\int d\boldsymbol{r}\left(-\boldsymbol{v}\cdot \hat{\mu}\nabla(\Psi\phi) +\boldsymbol{J}\cdot \nabla\hat{\mu}  -\boldsymbol{v} \cdot\boldsymbol{f}_{\mathrm{tot}}\right),
\end{equation} 
where the total chemical potential $\hat{\mu}$ is given by
\begin{equation}\label{Eq:theory-DRD-muT} 
\Psi\hat{\mu}\equiv \Psi\mu_{\mathrm b}-\nabla \cdot (K\Psi \nabla \phi)+{\epsilon_{\mathrm s}}(\sum_{\alpha=1}^{N} \left|\nabla\psi_{\alpha}\right|^{2}+ \left|\nabla\psi \right|^{2})\mu_{\mathrm{s}},
\end{equation} 
with $\mu_{\mathrm b}=\partial f_{\mathrm b}(\phi)/\partial \phi=a\phi \left(\phi^2-1\right)$, 
and $\mu_{\mathrm{s}}=\partial f_{\mathrm{s}}(\phi)/\partial \phi=-\frac{3}{4} \gamma \cos \theta_{\mathrm{s}} (\phi^2-1)$, and $\boldsymbol{f}_{\mathrm{tot}}=\sum_{\alpha=1}^{N}(1-\psi_{\alpha})(\boldsymbol{F}_{\mathrm{ext},\alpha}-\hat{p}_\alpha\nabla\psi_\alpha)/{\int d\boldsymbol{r}^{\prime}(1-\psi_{\alpha})}$ is the total force density applied on each particle with the pressure $\hat{p}_\alpha$ defined by $\hat{p}_{\alpha} \equiv \hat{p}\psi \prod_{\beta\neq \alpha} \psi_\beta+\nabla \cdot (2\epsilon_{\mathrm{s}}f_{\mathrm{s}}\nabla\psi_\alpha)$ and the generalized pressure 
\begin{equation}\label{Eq:theory-DRD-phat} 
\hat{p}\equiv -\hat{f}_{\mathrm b} +\hat{\mu}\phi.   
\end{equation}
Moreover, here, we have neglected the motion of solid boundary, and employed the natural boundary condition: $\hat{\boldsymbol{n}}_{\mathrm{d}} \cdot K\Psi\nabla \phi=0$, the impermeability conditions: $\hat{\boldsymbol{n}}_{\mathrm{d}} \cdot\boldsymbol{v}=0$ and $\hat{\boldsymbol{n}}_{\mathrm{d}} \cdot\boldsymbol{J}=0$, with $\hat{\boldsymbol{n}}_{\mathrm{d}}$ being the outward unit vector of the whole computational domain as shown in Fig.~\ref{Fig:Schematic}. Note that the force density due to $-\hat{p}_\alpha\nabla\psi_\alpha$ (small and non-zero only in the fluid-solid interfacial region) arises from the assumption of the conservation of $\Psi \phi$ in Eq.~(\ref{Eq:theory-DRD-phi}).

The work power $\dot{\mathcal{W}}_{\mathrm{ext}}=\sum_{\alpha}\boldsymbol{F}_{\mathrm{ext},\alpha}\cdot\dot{\boldsymbol{R}}_{\alpha}$ is the work power done by ``external forces'' $\boldsymbol{F}_{\mathrm{ext},\alpha}$ and the particle velocity $\dot{\boldsymbol{R}}_{\alpha}$ is a function of velocity field $\boldsymbol{v}(\boldsymbol{r},t)$ given in Eq.~(\ref{Eq:theory-DRD-Valpha}). 
The total external force, $\boldsymbol{F}_{\mathrm{ext},\alpha}$, acting on the $\alpha$-th particle within the suspended active deformable object can be generally expressed as $\boldsymbol{F}_{\mathrm{ext},\alpha} \equiv \boldsymbol{F}_{\mathrm{act},\alpha} - \partial U/\partial \boldsymbol{R}_{\alpha}$ with $\boldsymbol{F}_{\mathrm{act},\alpha}$ being the active or self-propelled forces (of biological or chemical origin) applied on the $\alpha$-th particle and $U\left(\left\{\boldsymbol{R}_{\alpha}\right\} \right)$ being the interaction potential. For the system shown in Fig.~\ref{Fig:Schematic}, the potential $U\left(\left\{\boldsymbol{R}_{\alpha}\right\} \right)$ is given by 
\begin{subequations}\label{Eq:theory-DRD-U}
\begin{equation}\label{Eq:theory-DRD-U1}
U\left(\left\{\boldsymbol{R}_{\alpha} \right\} \right)=\sum_{\zeta > \beta} U_{\mathrm {pp}}\left(|\boldsymbol{{R}}_{\zeta}-\boldsymbol{{R}}_{\beta} |\right)+\sum_{\beta}U_{\mathrm {ps}}\left(|\mathcal{D}(\boldsymbol{{R}}_{\beta})|\right),
\end{equation}
where $U_{\mathrm{pp}}$ and $U_{\mathrm ps}$ denote the two-body particle-particle interactions and the particle-solid boundary interactions, respectively, and $|\mathcal{D}(\boldsymbol{{R}}_{\beta})|$ denotes the minimal distance of the $\beta$-th particle at $\boldsymbol{r}=\boldsymbol{{R}}_{\beta}$ away from the solid boundary. In this work, the following harmonic potentials are employed to take into account the excluded volume interactions between particles and between particles and solid boundaries: 
\begin{align}\label{Eq:theory-DRD-Upp}
U_{\mathrm{pp}}\left(|\boldsymbol{{R}}_{\zeta}-\boldsymbol{{R}}_{\beta} |\right)= 
\begin{cases} 
\frac{1}{2} k_{\mathrm{s}}\left(R_{\beta\zeta}-d\right)^2, & {R_{\beta\zeta}\le d}, \\ 
0, & {R_{\beta\zeta}>d},
\end{cases}
\end{align}
\begin{align}\label{Eq:theory-DRD-Ups}
U_{\mathrm{ps}}\left(|\mathcal{D}(\boldsymbol{{R}}_{\beta})|\right)= 
\begin{cases} 
\frac{1}{2}k_{\mathrm{s}} \left[|\mathcal{D}(\boldsymbol{{R}}_{\beta})|-d/2\right]^2, & {|\mathcal{D}(\boldsymbol{{R}}_{\beta})|\le d/2}, \\ 
0, & {|\mathcal{D}(\boldsymbol{{R}}_{\beta})|>d/2},
\end{cases}
\end{align}
\end{subequations}
with $R_{\beta\zeta}\equiv {|\boldsymbol{R}_{\beta\zeta}|}$ and $\boldsymbol{R}_{\beta\zeta}\equiv \boldsymbol {R}_{\zeta}-\boldsymbol{R}_{\beta}$. 
From Eqs.~(\ref{Eq:theory-DRD-U}), we then obtain 
\begin{align} \label{Eq:theory-DRD-Fext}
\boldsymbol{F}_{\mathrm{ext},\alpha}&= \boldsymbol{F}_{\mathrm{act},\alpha}
-\sum_{\beta \neq \alpha}\frac{\partial U_{\mathrm{pp}}}{\partial R_{\beta \alpha}}\boldsymbol{\hat{R}}_{\beta\alpha}
-\frac{\partial U_{\mathrm{ps}}}{\partial |\mathcal{D}(\boldsymbol{{R}}_{\alpha})|}(-\boldsymbol{\hat{n}}_{\mathrm{s},\alpha}), \nonumber \\
&= \boldsymbol{F}_{\mathrm{act},\alpha}
-\sum_{\beta \neq \alpha} k_{\mathrm{s}} \left(R_{\beta\alpha}-d\right)\boldsymbol{\hat{R}}_{\beta\alpha}
-k_{\mathrm{s}} \left(|\mathcal{D}(\boldsymbol{{R}}_{\alpha})|-d/2\right) (-\boldsymbol{\hat{n}}_{\mathrm{s},\alpha}),
\end{align}
with $\boldsymbol{\hat{R}}_{\beta\alpha}\equiv \boldsymbol{R}_{\beta\alpha}/R_{\beta\alpha}$ and $\boldsymbol{\hat{n}}_{\mathrm{s},\alpha}$ being the inward unit normal vector of the solid boundary passing through the center (at $\boldsymbol{r}=\boldsymbol{{R}}_{\alpha}$) of the $\alpha$-th particle (see Fig.~\ref{Fig:Schematic}).

Secondly, the dissipation function of the system is given by
\begin{equation}\label{Eq:theory-DRD-Phi}
\Phi[\boldsymbol{v},\boldsymbol{J}]=\int d \boldsymbol{r}\left[\frac{1}{4}\eta\left(\phi,\psi,  \psi_{\alpha}\right)(\nabla {\boldsymbol{v}}+\nabla {\boldsymbol{v}}^{\mathrm{T}})^{2}+ \frac{1}{2}M^{-1}\left(\phi,\psi,\psi_{\alpha}\right)\boldsymbol{J}^2\right],
\end{equation}
with the superscript $\mathrm T$ denoting the tensor transpose. Again, physically, the quadratic form of $\Phi$ is equivalent to Onsager's linear force-flux relations~\cite{Onsager1931a}. The two resistance coefficients $\eta\left(\phi, \psi, \psi_{\alpha}\right)$ and $M^{-1}\left(\phi, \psi, \psi_{\alpha}\right)$ are the viscosity and the inverse of mobility, respectively. In contrast to diffusion domain method~\cite{Lowengrub2009DD,Lowengrub2021DD}, we follow the idea of DRD approach to incorporate the fluid-solid boundary conditions by interpolating the two resistance coefficients smoothly over different domains using either form of Eq.~(\ref{Eq:theory-Dprofile12}) and Eq.~(\ref{Eq:theory-Dprofile3}), where the solid viscosity $\eta_{\mathrm{s}}$ is very large, and the solid mobility $M_{\mathrm{s}}$ is very small in comparison to the viscosity $\eta_{\mathrm{f}}$ and the mobility $M_{\mathrm{f}}$ in fluids, \emph{i.e.}, $\eta_{\mathrm{s}}\gg \eta_{\mathrm{f}}$ and $M_{\mathrm{s}}\ll M_{\mathrm{f}}$.

Thirdly, imposing the incompressibility condition $\nabla\cdot\boldsymbol{v}=0$, we obtain the Rayleighian 
\begin{equation}\label{Eq:theory-DRD-R}
\mathcal {R}[\boldsymbol{v},\boldsymbol{J}]=\dot{\mathcal{F}}+\Phi-\dot{\mathcal{W}}_{\mathrm{ext}}-\int d \boldsymbol{r}P\nabla\cdot \boldsymbol{v},
\end{equation} 
with $P$ being a local Lagrange multiplier and representing the pressure. Minimizing $\mathcal{R}$ with respect to $\boldsymbol{v}$ and $\boldsymbol{J}$ gives a set of thermodynamically-consistent dynamic equations:
\begin{subequations}\label{Eq:theory-DRD-Dyn}
\begin{equation} \label{Eq:theory-DRD-Dyn-Stokes}
\rho\left({\partial_t \boldsymbol{v}}+\boldsymbol{v} \cdot \nabla \boldsymbol{v}\right)=-\nabla P+\nabla \cdot\left[\eta\left(\phi,\psi,\psi_{\alpha}\right) \left(\nabla {\boldsymbol{v}}+\nabla {\boldsymbol{v}}^{\mathrm T} \right)\right]+\hat{\mu}\nabla(\Psi\phi) +\boldsymbol{f}_{\mathrm{tot}},
\end{equation}
\begin{equation}\label{Eq:theory-DRD-Dyn-CH}
\partial_t(\Psi\phi)+\nabla \cdot(\Psi\phi\boldsymbol{v})=\nabla \cdot\left(M\left(\phi,\psi,\psi_{\alpha}\right) \nabla \hat{\mu}\right),
\end{equation}
\end{subequations}
where we have added the inertial terms in the Navier-Stokes equation with $\rho$ being the fluid density. 

In summary, the governing dynamic equations for the immiscible two-phase flows including $N$ suspending particles over a stationary rigid corrugated solid surface include Eqs.~\eqref{Eq:theory-DRD-Dalpha}, \eqref{Eq:theory-DRD-Valpha}, (\ref{Eq:theory-DRD-muT}), (\ref{Eq:theory-DRD-Fext}), (\ref{Eq:theory-DRD-Dyn}), and the incompressibility condition $\nabla\cdot\boldsymbol{v}=0$, supplemented with proper boundary conditions at the boundary of the whole regular computational domain: $\boldsymbol{v}=0$, $\Psi \hat{\boldsymbol{n}}_{\mathrm{d}} \cdot \nabla \phi=0$, and $\hat{\boldsymbol{n}}_{\mathrm{d}} \cdot \nabla \hat{\mu}=0$. 

Finally, before ending this subsection, we would like to give some remarks as follows.

(i) When we derive the dynamic equations by minimizing the Rayleighian in Eq.~\eqref{Eq:theory-DRD-R}, we have used the fact that the particle velocity $\boldsymbol{V}_{\alpha}(t)=\dot{\boldsymbol{R}}_{\alpha}$ is not an independent rate~\cite{tanaka2018physical} but is slaved by the flow field $\boldsymbol{v}(\boldsymbol{r},t)$ via Eq.~(\ref{Eq:theory-DRD-Valpha}).

(ii) The inertial terms in the Navier-Stokes equation~\eqref{Eq:theory-DRD-Dyn-Stokes} can also be included into OVP directly, as discussed in Onsager's early work with Machlup~\cite{Onsager1953b}, by introducing $\beta$-type state variables $\dot{\alpha}_i$ (the change rate of $\alpha$-type state variables, $\alpha_i$) and adding the change rate of kinetic energy $\dot{E}_{\mathrm{K}}$ to the Rayleighian in Eq.~\eqref{Eq:theory-DRD-R}~\cite{Chun2015,Chun2022}.

(iii) The free energy functional $\mathcal{F}$ has only included the information about the interactions (energy and entropy) among constituents (particles or molecules) of the system fluids; external forces applied at the system boundaries or those arising from additional internal (hidden) degrees of freedom (\emph{e.g.}, ATP-driven biochemistry, biophysical, and chemophysical processes) have not been included and have to be appended as supplementary terms (through external work power $\dot{\mathcal{W}}_{\mathrm{ext}}$) in the OVP.

(iv) Here, we have only considered neutrally buoyant rigid particles. That is, we have neglected the difference between the fluid density and that of particles. More generally, mass density $\rho$ should be a function of $\psi_\alpha$, \emph{i.e.}, $\rho(\psi_\alpha,\psi,\phi)$, and we should add gravitational force density $\rho \boldsymbol{g}$ in the Navier-Stokes equation~\eqref{Eq:theory-DRD-Dyn-Stokes}. Moreover, external torques can also be applied to each particle.

(v) The DRD approach can also describe the inertial motion of massive rigid particles (as how the FPD method can do~\cite{Xu2023PoF,tanaka2018physical}). To show this, we can multiply both sides of Eq.~\eqref{Eq:theory-DRD-Dyn-Stokes} by $1-\psi_\alpha(\boldsymbol{r})$, integrate over space, and obtain an approximate equation for the translational motion of rigid particles as~\cite{tanaka2018physical}
\begin{equation}\label{Eq:Method-ParticleEqn}
M_{\alpha} {d \boldsymbol{V}_{\alpha}}/{d t}=\boldsymbol{F}_{{\mathrm{ext}},\alpha}+\boldsymbol{F}_{{\rm v},\alpha}+\boldsymbol{F}_{{\rm c},\alpha},
\end{equation} 
with $M_\alpha$ and $\boldsymbol{V}_\alpha$ being the mass and the velocity of $\alpha$-th particle, given via $M_\alpha\boldsymbol{V}_\alpha=\int d\boldsymbol{r} \rho(1-\psi_\alpha)\boldsymbol{v}$. Here, $\boldsymbol{F}_{{\mathrm{ext}},\alpha}$ is the external force applied on the $\alpha$-th particle; $\boldsymbol{F}_{{\rm v},\alpha}$ and $\boldsymbol{F}_{{\rm c},\alpha}$ are the viscous force and the capillary force exerted by the surrounding two-phase fluids on the rigid particle~\cite{goto2015purely}:
\begin{equation}\label{Eq:Method-Fv}
\boldsymbol{F}_{{\rm v},\alpha}= \int d A_\alpha \hat{\boldsymbol{n}}_{{\rm p},\alpha} \cdot \left [ -P \boldsymbol{I}+ \eta_{\rm f} \left(\nabla \boldsymbol{v}+\nabla \boldsymbol{v}^{T}\right)\right], \quad
\boldsymbol{F}_{{\rm c},\alpha}\approx -\int d A_\alpha \hat{\boldsymbol{n}}_{{\rm p},\alpha} \cdot \boldsymbol{\Pi}-\mathbf{F}_{\mathrm{cs},\alpha} .    
\end{equation}
in which $\boldsymbol{I}$ is the unit tensor, the integral is over the particle surface, $dA_\alpha$ and $\hat{\boldsymbol{n}}_{{\rm p},\alpha}$ are the surface element and the unit outward normal vector of the particle, Korteweg capillary pressure tensor $\boldsymbol{\Pi}$ and the surface capillary force, $\mathbf{F}_{\mathrm{cs},\alpha}$, are given by
\begin{equation}\label{Eq:DRD-Pi}
\boldsymbol{\Pi}\equiv \hat{p}+K\nabla \phi \nabla \phi,\quad
\mathbf{F}_{\mathrm{cs},\alpha} 
\equiv -\frac{\partial}{\partial \mathbf{R}_\alpha}\int d\mathbf{r} \epsilon_{\mathrm s} f_{\mathrm{s}} |\nabla \psi_\alpha|^2,
\end{equation}
respectively. We have used the identity $\nabla \left|\nabla\psi_\alpha\right|^{2}=-\partial \left|\nabla\psi_\alpha\right|^{2}/\partial \mathbf{R}_\alpha$. Furthermore, for non-spherical particles, one also needs to consider their rotational motion and the angular acceleration induced by torques~\cite{Xu2023PoF,tanaka2018physical}. 
Particularly, in the overdamped limit, the characteristic time scaling as $d/V_0$ is much larger than the viscous relaxation time $M_\alpha/\eta_{\rm f} d\approx \rho d^2/\eta_{\rm f}$ with $V_0$ being the characteristic velocity of the particle. That is, the particle Reynolds number, defined by ${\rm Re}_{\rm p}\equiv \rho d V_0/\eta_{\rm f}$, is very small: ${\rm Re}_{\rm p} \ll 1$. In this limit, we have $\boldsymbol{F}_{\mathrm{ext},\alpha}+\boldsymbol{F}_{{\rm v},\alpha}+\boldsymbol{F}_{{\rm c},\alpha}=0$.




\subsection{Numerical implementations} \label{sec:Numerics}

The DRD approach proposed for mFSI problems demonstrates its strength by employing simple models to achieve accurate simulations while maintaining intrinsic thermodynamic consistency by smoothly interpolating dynamic-resistance coefficients across interfaces. In particular, for the mFSI problems in two-phase flows discussed in Sec.~\ref{sec:theory-Eqns}, the main governing equations~\eqref{Eq:theory-DRD-muT}, \eqref{Eq:theory-DRD-Dyn-Stokes}, and \eqref{Eq:theory-DRD-Dyn-CH} differ from those of the standard sharp fluid-solid interface model~\cite{gao_efficient_2014, yang_efficient_2018, zhu_fully_2020} in their treatment of the fluid-solid interactions and the smooth interpolation of dynamic-resistance coefficients across the fluid-solid interfaces, \emph{i.e.}, the viscosity coefficient in the Navier-Stokes equation~\eqref{Eq:theory-DRD-Dyn-Stokes} and the mobility coefficient in the Cahn-Hilliard equation~\eqref{Eq:theory-DRD-Dyn-CH}. Additional low-order terms in Eq.~\eqref{Eq:theory-DRD-muT} arise from wettability (contact angle) conditions, which are significant only in the vicinity of the fluid-solid interfaces.

The complete equation system can be solved efficiently using established two-phase flow solvers~\cite{gao_efficient_2014, yang_efficient_2018, zhu_fully_2020}, wherein pressure and velocity are decoupled in the Navier-Stokes equations, and a stabilization scheme is applied to the Cahn-Hilliard equation. 

The low-order terms are discretized explicitly, which does not impose additional severe time constraints. The resulting system of discretized equations is efficiently solved using a GMRES (Generalized Minimal RESidual) solver, preconditioned with an FFT(Fast-Fourier-Transform)-based method~\cite{gao_efficient_2014}. The smooth interpolation of dynamic-resistance coefficients across the diffuse fluid–solid interface is treated implicitly, which eliminates the time-step constraints associated with explicit schemes for stiff coefficients. While our DRD approach, focused on transient dynamics, incurs a higher computational cost than standard sharp fluid-solid interface methods (approximately $5$ times slower~\cite{gao_efficient_2014}) for planar solid surfaces, its primary advantage manifests for complex solid-surface geometries. For such geometries, where traditional finite difference methods are poorly suited and finite element methods (FEM) are typically required. In comparison, the DRD method extends the computational domain to a rectangular region, enabling the use of structured grids. This allows highly efficient structured-grid solvers (\emph{e.g.}, FFT, multigrid), which are typically $10$ times faster than FEM solvers for comparable problems. Furthermore, these structured-grid solvers benefit from lower communication overhead and exhibit superior performance on modern GPU architectures. Consequently, for mFSI problems with complex, moving, or evolving solid surfaces, our DRD method offers a considerably simpler and computationally more efficient alternative. 

Specifically, the dynamic system is discretized as follows. Given a time step $\Delta t$, at the time $t^n=n\Delta t$, we denote particle phase-parameter by $\psi_\alpha^n$, particle positions by ${\boldsymbol{R}}_{\alpha}^n$ (for $\alpha=1,...,N$), flow velocity by $\boldsymbol{v}^n$, pressure by $p^n$, and two-fluid phase-parameter by $\phi^n$. At the next timestep $t^{n+1}=(n+1)\Delta t$, we update the above variables $\psi_\alpha^{n+1}$, ${\boldsymbol{R}}_{\alpha}^{n+1}$, $\boldsymbol{v}^{n+1}$, $p^{n+1}$ and $\phi^{n+1}$ by the following three steps.

\textbf{Step 1:} We discretize the Cahn-Hilliard equation using a stabilization method \cite{gao_efficient_2014}. The phase parameter $\phi^{n+1}$ is updated by discretizing Eqs.~\eqref{Eq:theory-DRD-muT} and \eqref{Eq:theory-DRD-Dyn-CH} as follows:
\begin{subequations}\label{Eq:Numerics-DRD-Dyn-discrete}
\begin{equation}\label{Eq:Numerics-DRD-Dyn-CH-discete}
\frac{1}{\Delta t}(\Psi^n\phi^{n+1} - \Psi^n\phi^{n}) + \nabla \cdot(\Psi^n\phi^n\boldsymbol{v}^n) = \nabla \cdot\left[M\left(\phi^n,\psi^n,\psi_{\alpha}^n\right) \nabla \hat{\mu}^{n+1}\right],
\end{equation}
\begin{equation}\label{Eq:Numerics-DRD-muT-discrete} 
\Psi^n\hat{\mu}^{n+1} = \Psi^n \left[\mu_{\mathrm b}^n + S(\phi^{n+1} - \phi^n) \right]- \nabla \cdot (K\Psi^n \nabla \phi^{n+1})+{\epsilon_{\mathrm s}}\left(\sum_{\alpha=1}^{N} \left|\nabla\psi_{\alpha}^n\right|^{2}+ \left|\nabla\psi\right|^{2}\right)\mu_{\mathrm{s}}^n,
\end{equation} 
\end{subequations}
with boundary conditions $\hat{\boldsymbol{n}}_{\mathrm{d}} \cdot \nabla \phi^{n+1}=0$,  $\hat{\boldsymbol{n}}_{\mathrm{d}} \cdot \nabla \hat{\mu}^{n+1} = 0$. Here $\psi_\alpha^n$ is defined in Eq.~\eqref{Eq:theory-DRD-Dalpha} with $\boldsymbol{R}_{\alpha} (t)$ replaced with ${\boldsymbol{R}}_{\alpha}^n$; $\Psi^n= \psi\prod_{\alpha=1}^N \psi_\alpha^n$. The higher-order terms $\nabla \cdot (K\Psi^n \nabla \phi^{n+1})$ and $\nabla \cdot\left[M\left(\phi^n,\psi^n,\psi_{\alpha}^n\right) \nabla \hat{\mu}^{n+1}\right]$ are treated implicitly to alleviate time constraints, and the extra low-order terms are handled explicitly. The term $S(\phi^{n+1}-\phi^n)$ is a stabilization term introduced to remove instabilities from the nonlinear bulk chemical potential $\mu_b^n$, where $S$ is a stabilization coefficient and $S=1.5$ is sufficient to ensure the stability in our simulations.

\textbf{Step 2:} We discretize the incompressible Navier-Stokes equations using a pressure stabilization scheme~\cite{guermond_splitting_2009}. The velocity is first updated as
\begin{equation} \label{Eq:Numerics-DRD-Dyn-Stokes}
\rho^n \left( \frac{\boldsymbol{v}^{n+1}-\boldsymbol{v}^n}{\Delta t} + \boldsymbol{v}^n \cdot \nabla \boldsymbol{v}^n \right) =-\nabla (2p^n-p^{n-1}) + \nabla \cdot\left[\eta^n \left(\nabla {\boldsymbol{v}}+\nabla {\boldsymbol{v}}^{\mathrm T} \right)^{n+1}\right]+\hat{\mu}^{n+1}\nabla(\Psi^n\phi^{n+1}) +\boldsymbol{f}_{\mathrm{tot}}^n,
\end{equation}
with boundary conditions $\boldsymbol{v}^{n+1}=0$,  
where density $\rho^n$ and viscosity $\eta^n$ are interpolated by phase parameters. The viscous term is treated implicitly to remove time constraints. Then, pressure is updated by solving the following Poisson equation with constant coefficient:
\begin{equation}\label{ns_discrete2}
\Delta (p^{n+1}-p^n) = \frac{\bar{\rho}}{\Delta t} \nabla\cdot\boldsymbol{v}^{n+1},     
\end{equation}
with boundary condition $\hat{\boldsymbol{n}}_{\mathrm{d}} \cdot \nabla  p^{n+1} = 0$, 
where $\bar{\rho}$ denotes the minimum value of density.

\textbf{Step 3:} Lastly, we update particles positions according to Eq.\eqref{Eq:theory-DRD-Valpha}:
\begin{equation}\label{Eq:Numerics-DRD-Valpha-discrete}
\boldsymbol{R}_{\alpha}^{n+1} = \boldsymbol{R}_{\alpha}^{n} + \Delta t \frac{\int d\boldsymbol{r}^{\prime} (1-\psi_{\alpha}^n) \boldsymbol{v}^{n+1} }{\int  d\boldsymbol{r}^{\prime} (1-\psi_{\alpha}^n)},
\end{equation}
and then $\psi_\alpha^{n+1}$ is updated using Eq.~\eqref{Eq:theory-DRD-Dalpha}, weher $\boldsymbol{R}_{\alpha} (t)$ is replaced by ${\boldsymbol{R}}_{\alpha}^{n+1}$.

A key advantage of the proposed DRD approach is that it allows discretization of the entire space using a regular rectangular mesh (or grid) of dimensions $L \times H$ or $L_x \times L_z$, thereby greatly simplifying numerical implementation. This setup enables the direct use of simple, structured, and adaptive block-structured grids with well-established numerical solvers. In contrast to traditional DNS methods for mFSI problems that require boundary-fitted meshes or involve complex geometric treatments, the DRD framework offers a significantly more efficient and straightforward alternative. In our simulations, uniform grids are employed, and the domain is discretized using a finite-difference method on staggered grids, with velocity defined at cell faces and pressure, the two-fluid phase parameter $\phi$, and the fluid-solid interfacial parameter $\psi$ defined at cell centers~\cite{zhu_fully_2020}.

Moreover, one key tuning parameter in the proposed DRD method is the fluid-solid interfacial thickness $\epsilon_{s}$ used in various interpolations to enforce different boundary conditions. As a result, the grid resolution (\emph{i.e.}, mesh size $\Delta x, \Delta z$) must be carefully selected to ensure accurate results. For single-phase flows with Dirichlet or Neumann boundary conditions, a mesh size satisfying $\max(\Delta x, \Delta z) \leq \epsilon_{\mathrm{s}}$ is sufficient to achieve mesh-independent results. In the case of the Robin boundary condition, a finer resolution is required to resolve the intermediate interpolation layer, with numerical simulations indicating that a mesh size smaller than $0.5\epsilon_{\mathrm{s}}$ is adequate. For two-phase flows on solid surfaces, the interfacial thickness $\epsilon_{\mathrm{s}}$ must be refined further to accurately resolve the wettability (contact angle) condition. In our implementation, we typically set $\epsilon_{\mathrm{s}} = 0.5\epsilon$, where $\epsilon$ denotes the interfacial thickness associated with two-phase flows. Accordingly, in the hydrodynamic simulations presented in this work, we choose $\epsilon_{\mathrm{s}} = \epsilon = 0.01$ for single-phase flows, and $\epsilon_{\mathrm{s}} = 0.005$, $\epsilon = 0.01$ for two-phase flows. Unless otherwise specified, the mesh sizes are set to $\Delta x = \Delta z = \epsilon = 0.01$ in both cases.

\quad

\section{Benchmark validations and applications} \label{sec:Benchmark}

\subsection{Benchmark simulation 1: Complicated (moving) solid geometries }\label{sec:Benchmark-1}

\subsubsection{Governing dynamic equations \label{sec:App1-DynEqns}}

In this section, we highlight the first advantage of our DRD approach—its capability to simulate viscous flows over moving or geometrically complex solid surfaces. Specifically, we investigate the inertial focusing of a single passive colloidal particle suspended in a two-dimensional (2D), single-phase, pressure-driven flow within stationary microchannels, both without and with surface corrugations, as shown in Fig.~\ref{Fig:App1-Focusing1}(a) and Fig.~\ref{Fig:App1-Focusing2}(a). In this fluid-solid two-phase system, no fluid-fluid interface is present, eliminating the need to consider the phase parameter $\phi$. The only relevant interfacial phase parameters are $\psi$ for the solid boundaries and $\psi_1$ for the suspended colloidal particle. Then, from the general dynamic equations presented in Sec.~\ref{sec:theory-Eqns}, we obtain the following Navier-Stokes equation:
\begin{equation} \label{Eq:App1-Dyn-NS}
\rho\left({\partial_t \boldsymbol{v}}+\boldsymbol{v} \cdot \nabla \boldsymbol{v}\right)=-\nabla P+\nabla \cdot\left[\eta\left(\psi,\psi_1\right) \left(\nabla {\boldsymbol{v}}+\nabla {\boldsymbol{v}}^{\mathrm T} \right)\right]+\frac{(1-\psi_1)\boldsymbol{F}_{\mathrm{ext},1}}{\int d\boldsymbol{r}^{\prime}(1-\psi_1)},
\end{equation} 
supplemented with the incompressibility condition, $\nabla \cdot \boldsymbol{v}=0$, and the boundary condition at the boundary of the whole regular computational domain: $\boldsymbol{v}=0$. 
Here, the total force $\boldsymbol{F}_{\mathrm{ext},1}$ on the particle is applied by the solid boundary and is given from Eq.~(\ref{Eq:theory-DRD-Fext}) by 
\begin{equation}\label{Eq:App1-Fext}
\boldsymbol{F}_{\mathrm{ext},1} =
-\frac{\partial U_{\mathrm{ps}}}{\partial |\mathcal{D}(\boldsymbol{{R}}_1)|}(-\boldsymbol{\hat{n}}_{\mathrm{s},1})= 
- k_{\mathrm{s}} \left(|\mathcal{D}(\boldsymbol{{R}}_1)|-d/2\right) (-\boldsymbol{\hat{n}}_{\mathrm{s},1}), 
\end{equation}
where $|\mathcal{D}(\boldsymbol{{R}}_1)|$ denotes the minimal distance of the particle at $\boldsymbol{r}=\boldsymbol{{R}}_1$ away from the solid boundary and it includes the information of complicated solid geometries. $\boldsymbol{\hat{n}}_{\mathrm{s},1}$ is the inward unit normal vector of the solid boundary passing through the center of the particle (see Fig.~\ref{Fig:Schematic}). 
Fluid slip is neglected at all solid boundaries and hence we take the spatially-varying viscosity $\eta({\boldsymbol r})$ in the form of Eq.~(\ref{Eq:theory-Dprofile12}) as 
\begin{equation} \label{Eq:App1-etaM-eta}
\eta(\psi,\psi_{1})=\eta_{\mathrm{s}}+\left(\eta_{\mathrm{f}}-\eta_{\mathrm{s}} \right)\psi\psi_1,
\end{equation}
where the interfacial profile functions $\psi$ (of the solid boundary) and $\psi_1$ (of the particle) both take the form of Eq.~(\ref{Eq:theory-DRD-psi}):
\begin{equation}\label{Eq:App1-psipsi1}
\psi(\boldsymbol{r})=\frac{1}{2}\left[1-\tanh \left(\frac{\mathcal{D}(\boldsymbol{r})}{\sqrt{2} \epsilon_{\mathrm {s}}}\right)\right], \quad
\psi_1(\boldsymbol{r}, t)=\frac{1}{2}\left[1-\tanh \left(\frac{d/2-\left| \boldsymbol{r}-\boldsymbol{R}_1 (t)\right|}{\sqrt{2} \epsilon_{\mathrm {s}}}\right)\right],
\end{equation}
with the particle center $\boldsymbol{R}_1 (t)$ evolving as Eq.~(\ref{Eq:theory-DRD-Valpha}) by 
\begin{equation}\label{Eq:App1-Valpha}
\dot{\boldsymbol{R}}_1=\boldsymbol {V}_1(t)=\frac{\int d\boldsymbol{r}^{\prime} (1-\psi_1) \boldsymbol{v}(\boldsymbol{r}^{\prime},t) }{\int  d\boldsymbol{r}^{\prime} (1-\psi_1)}.
\end{equation}
In addition, we employ the parabolic profile of the inlet velocity at $x=0$ to induce the pressure-driven flows inside the microchannel:
\begin{equation}\label{Eq:App1-inlet}
\boldsymbol{v}|_{x=0}= V_0 \left[z(H-z)/(H/2)^2\right].
\end{equation}

\subsubsection{Particle focusing in 2D microchannels \label{sec:App1-Benchmark}} 

Before carrying out simulations, as done in Sec.~\ref{sec:App1-Benchmark}, we first non-dimensionalize the above dynamic equations by scaling the length by the channel width $H$, velocity by the maximal inlet velocity $V_0$, time by $\tau_0 \equiv H/V_0$, viscosity by $\eta_{\mathrm{f}}$, pressure and stress by $\eta_{\mathrm{f}}V_0/H$, and force by $\eta_{\mathrm f} V_0$.
Five dimensionless parameters appear, as shown in the following comprehensive list along with their physical interpretations. Moreover, we classify them into two general categories: 

\begin{figure}[htbp] 
  \centering
  \includegraphics[width=0.83\columnwidth]{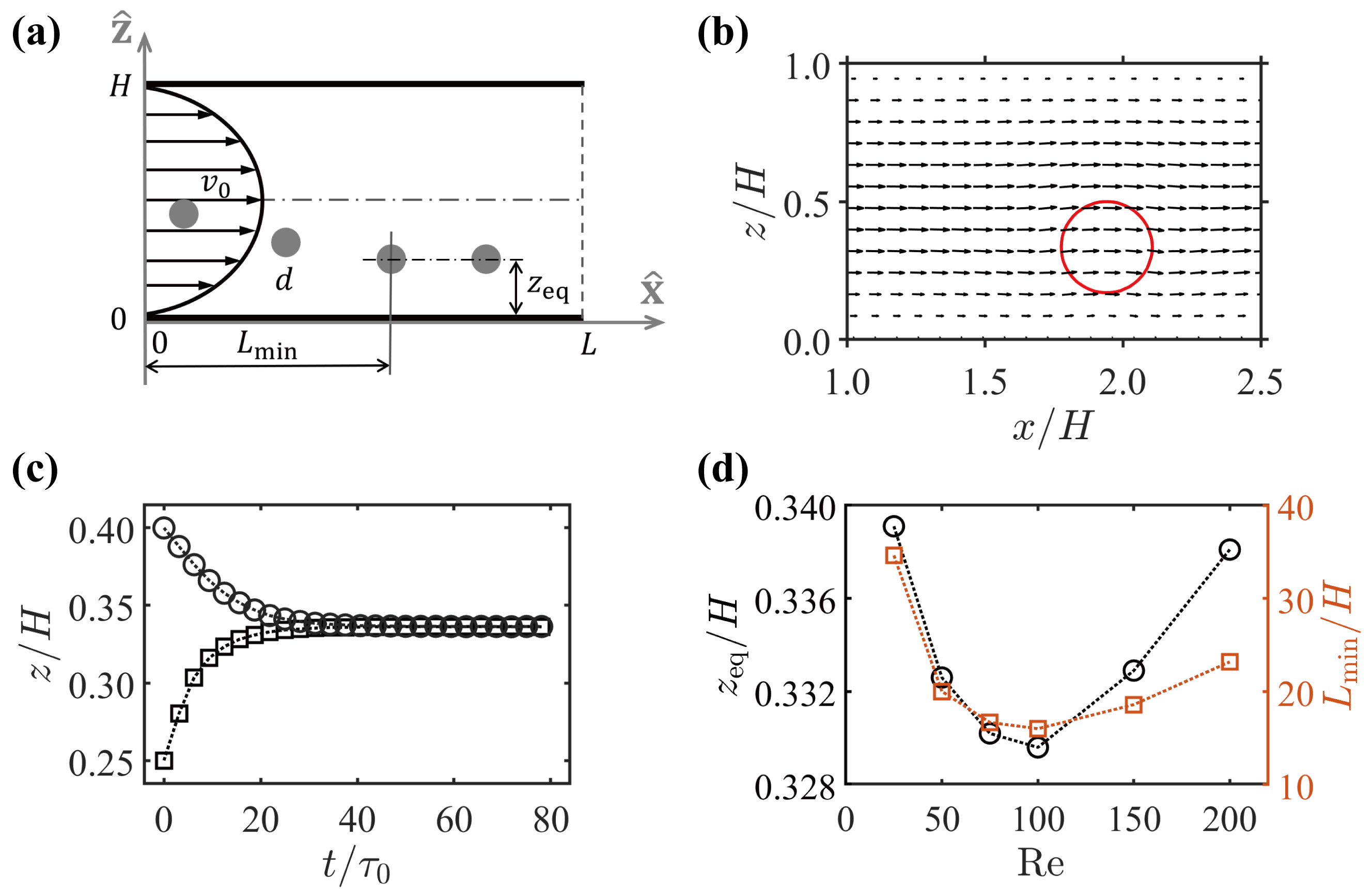}
  \caption {Inertial focusing of one (passive, suspended) circular colloidal particle by 2D single-phase pressure-driven flows in a microchannel with planar surfaces. (a) Schematic illustration of the simulation setup of dimensions $L\times H$. Here, the equilibration distance $L_{\mathrm{min}}$ refers to the minimum length in the $x$-direction at which the particle becomes focused from the initial center position of channel inlet at $x=0$, and the equilibrium position $z_{\mathrm{eq}}$ denotes (after the particle being focused) the distance of the particle center away from the bottom surface of the channel at $z=0$.  (b) Flow fields near the particle that has already been focused at equilibrium position $z_{\mathrm{eq}}$. (c) Equilibration (after $\sim 30 \tau_0$) of two individual particles starting from two different initial positions ($z$-coordinates) at the inlet boundary $x=0$. (d) Dependence of equilibrium position $z_{\mathrm{eq}}$ and equilibration distance $L_{\mathrm{min}}$ on Reynolds number $\mathrm{Re}=\rho V_0 H/\eta_{\mathrm f}$. Here, we take the physics parameters to be: $\mathrm{Re}=\rho V_0 H/\eta_{\mathrm f}=32.0$, $\mathrm{Re_p}=\rho V_0 d/\eta_{\mathrm f}=10.67$, $r_{\mathrm d}=d/H=1/3$, and $L/H=6.0$. We take the two ``free'' parameters to be: $r_{\epsilon_{\mathrm{s}}}=\epsilon_{\mathrm{s}}/H=0.01$ and $r_\eta = r_{\mathrm{s}}/r_{\mathrm{f}}=100.0$. Periodic boundary conditions have been used in the $x$-direction.}
  \label{Fig:App1-Focusing1}
\end{figure} 

\begin{itemize}
\item (i) Measurable physics parameters defined based on measurable quantities. (1) Reynolds number: ${\mathrm{Re}} \equiv \rho V_0 H/\eta_{\mathrm f}$, or, the particle Reynolds number: ${\mathrm{Re}}_{\mathrm p}\equiv \rho V_0d/\eta_{\mathrm f}$. Alternatively, we can define another particle Reynolds number: ${ \mathrm{Re}_{\mathrm {ps}}}= {\rho V_0 d^2}/{{\eta_{\mathrm f}H}}$. It is the characteristic particle Reynolds number in shear flows~\cite{zettner2001moderate,huang2012rotation}, which is the ratio of inertial forces $\rho (2V_0d/H)^2 (d/2)$ to hydrodynamic shear force $2\eta_{\mathrm f}V_0d/H$.  
(2) Stiffness parameter for the particle-solid boundary interaction: $\mathcal{K}={k}/{H \eta_{\mathrm f} V_0}$. In our simulations, we take $\mathcal{K}_{\mathrm s}$ to be a large constant to avoid overlapping between the particle and the solid boundary. (3) Confinement strength or particle-diameter to channel width ratio: $r_{\mathrm{d}}=d/H$ with $0 \leqslant r_{\mathrm{d}} \leqslant 1 $. Particularly, $r_{\mathrm{d}} \to 0$ corresponds to the limit of free (no confinement) particle dynamics in shear flows. $r_{\mathrm{d}} \to 1$ or $d\to H$ corresponds to the strongest (stuck) confinement case. 
\item (ii) ``Free'' parameters arising from the DRD treatment of fluid-solid interfaces. (1) Viscosity ratios: $r_\eta=\eta_{\mathrm{s}} / \eta_{\mathrm{f}}$. 
(2) Relative interfacial thicknesses: $r_{\epsilon_{\mathrm{s}}}=\epsilon_{\mathrm{s}}/H$. 
\end{itemize}

\begin{figure}[htbp] 
  \centering
  \includegraphics[width=0.98\columnwidth]{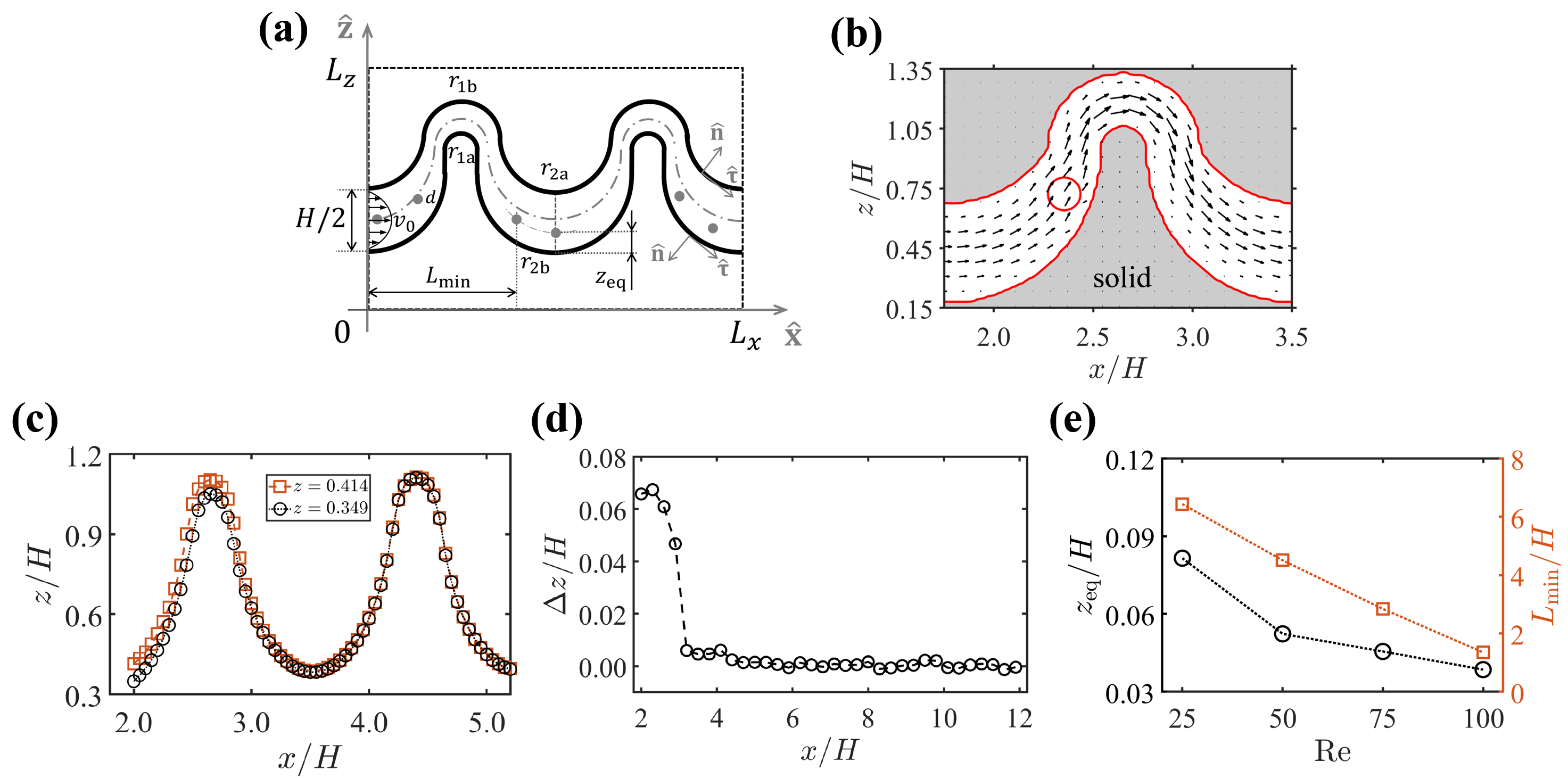}
  \caption {Inertial focusing of one (passive, suspended) circular colloidal particle of diameter $d$ by 2D single-phase pressure-driven flows in a corrugated (an asymmetrically curved) microchannel of inlet width $H/2$. (a) Schematic illustration of the 2D simulation setup of dimensions $L_{\mathrm x}\times L_{\mathrm z}$. Here, the equilibration distance $L_{\mathrm{min}}$ refers to the minimum length in the $x$-direction at which the particle becomes focused from the initial center position of channel inlet at $x=0$, and the equilibrium position $z_{\mathrm{eq}}$ denotes (after the particle being focused) the distance of the particle center away from the bottom surface of the channel as it passes through the middle of the channel at $x=L_{\rm x}/2$. (b) Flow fields near the particle that has already been focused at equilibrium position $z_{\mathrm{eq}}$. (c) Equilibration (after $\sim 30 \tau_0$) of two individual particles starting from two different initial positions ($z$-coordinate) at the inlet boundary $x=0$. The trajectories of the two individual particles gradually converge or coincide. (d) A further illustration of the equilibration of the two individual particles by calculating the difference $\Delta z$ in the $z$-coordinates of two particles when they are passing the same $x$ position. (e) Dependence of $z_{\mathrm{eq}}$ and $L_{\mathrm{min}}$ on Reynolds number $\mathrm{Re}=\rho V_0 H/\eta_{\mathrm f}$. Here, we take the physics parameters to be: particle-solid boundary interaction stiffness, $\mathcal{K}=k/H\eta_{\mathrm{f}V_0} = 50.0$, $r_{\mathrm d}=d/H=1/6$, $L_{\rm x}/H=3.5$, $L_{\rm z}/H=1.5$; in the simulations of (b-d), we take $\mathrm{Re}=\rho V_0 H/\eta_{\mathrm f}=100.0$ and hence $\mathrm{Re_p}=\rho V_0 d/\eta_{\mathrm f}=16.67$. We take the two ``free'' parameters to be: $r_{\epsilon_{\mathrm{s}}}=\epsilon_{\mathrm{s}}/H=0.005$ and $r_\eta = \eta_{\mathrm{s}}/\eta_{\mathrm{f}}=100.0$. Periodic boundary conditions have been used in the $x$-direction.} 
 
  \label{Fig:App1-Focusing2}
\end{figure}



In this study, we perform only two-dimensional (2D) simulations in the $x$-$z$ plane, as shown in Figs.~\ref{Fig:App1-Focusing1} and \ref{Fig:App1-Focusing2}. The microchannel has a length $L$ and a characteristic width $H$. These simulations serve as standard benchmarks for assessing the coupling between rigid particles and confined fluid flow. Firstly, for particles focusing in planar microchannels, we observe that particles reach an equilibrium position $z_{\mathrm{eq}}$ after traveling a minimum distance $L_{\mathrm{min}}$. This equilibrium position is independent of the initial particle location but exhibits a non-monotonic dependence on the Reynolds number, $\mathrm{Re} = \rho V_0 H / \eta_{\mathrm{f}}$ (see Fig.~\ref{Fig:App1-Focusing1}(d)). Our results show qualitative agreement with previous direct numerical simulation (DNS) studies~\cite{Esipov2020Focusing,Bazaz2020Focusing}; more quantitative comparison and the exploration of focusing mechanisms leave to another undergoing work.

Secondly, for inertial particles focusing in 2D corrugated (asymmetrically curved) microchannels, we find that particles also focus onto equilibrium positions after traveling a distance $L_{\mathrm{min}}$, which is again independent of the initial particle position. However, in contrast to planar microchannels, $L_{\mathrm{min}}$ exhibits a monotonic dependence on $\mathrm{Re}$ (see Fig.~\ref{Fig:App1-Focusing2}(e)). Notably, the focusing distance in corrugated microchannels is approximately an order of magnitude smaller than in planar channels, indicating that channel corrugation significantly enhances focusing efficiency, consistent with experimental findings~\cite{Di2007Focusing}.

These benchmark simulations provide practical guidelines for selecting the two key ``free'' parameters: (i) a sufficiently small interface thickness, $r_{\epsilon_{\mathrm{s}}} \leqslant 0.01$, and (ii) an appropriate viscosity ratio, $r_\eta = \eta_{\mathrm{s}} / \eta_{\mathrm{f}} \geqslant 100$.

\subsection{Benchmark simulation 2: Deformable (active) objects}\label{sec:Benchmark-2}

\subsubsection{Governing dynamic equations}\label{sec:App2-Eqn}

In this section, we demonstrate the second advantage of our DRD approach -- the capability of simulating hydrodynamic interactions between viscous fluids and (active) deformable objects. We combine the DRD approach with particle-based models of (active) deformable objects that have already been widely used in particle-based molecular simulations in soft matter and biological physics~\cite{Xu2023PoF,MarkD2018deformableparticle}. Specifically, we consider the following fluid-solid two-phase flow: dynamics of active or passive, deformable or rigid objects in two-dimensional (single-phase) viscous shear flows, as shown in Fig.~\ref{Fig:App2-microswimmer}. In this fluid-solid two-phase flow, the fluid-fluid interface also does not exist, so the phase parameter $\phi$ is not needed, and the only interfacial phase parameters are $\psi$ for solid boundaries and $\psi_{\alpha}$ (with $\alpha =1,...,N$) for the rigid particles that compose the deformable object. Then, from the general dynamic equations (\ref{Eq:theory-DRD-Dyn}) presented in Sec.~\ref{sec:theory-Eqns}, we obtain the following Navier-Stokes equation: 
\begin{equation} \label{Eq:App2-Dyn-NS}
\rho\left({\partial_t \boldsymbol{v}}+\boldsymbol{v} \cdot \nabla \boldsymbol{v}\right)=-\nabla P+\nabla \cdot\left[\eta({\psi}_\alpha) \left(\nabla {\boldsymbol{v}}+\nabla {\boldsymbol{v}}^{\mathrm T} \right)\right]+\boldsymbol{f}_{\mathrm{tot}},
\end{equation} 
supplemented with the incompressibility condition, $\nabla \cdot \boldsymbol{v}=0$. 
Here, the total force density $\boldsymbol{f}_{\mathrm{tot}}$ is given by
\begin{equation}\label{Eq:App2-Dyn-NS-Fext}
\boldsymbol{f}_{\mathrm{tot}}=\sum_{\alpha=1}^{N}(1-\psi_{\alpha})\boldsymbol{F}_{\mathrm{ext},\alpha}/{\int d\boldsymbol{r}^{\prime}(1-\psi_{\alpha})}
\end{equation}
with the total external force $\boldsymbol{F}_{\mathrm{ext},\alpha}$ acting on the $\alpha$-th particle given by Eq.~(\ref{Eq:theory-DRD-Fext}). 
In the simulations of active deformable objects, we also take the simple harmonic form of the particle-solid boundary interaction potential $U_{\mathrm{ps}}$ in Eq.~(\ref{Eq:theory-DRD-Ups}). However, the specific form of $U_{\mathrm{pp}}$ depends on the specific particle-based model of the specific active deformable objects that will be discussed in the following subsections. 



\begin{figure}[htbp]
  \centering
  \includegraphics[width=0.6\columnwidth]{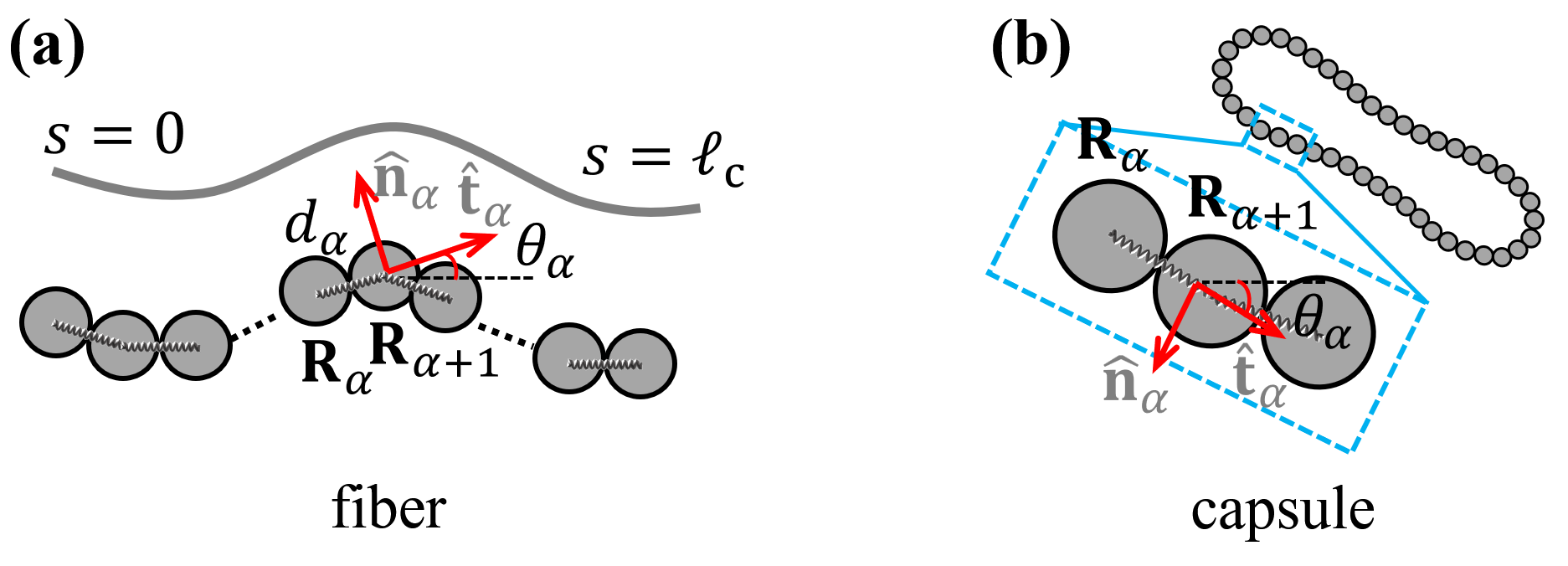}
  \caption {Schematic illustration of deformable objects composed of rigid particles. (a) A flexible fiber and (b) an elastic capsule composed of connected rigid particles using bead-spring model. } 
  \label{Fig:App2-fiber}
\end{figure}

\subsubsection{Particle-based models for deformable (active) objects}\label{sec:App2-Fiber}

\emph{Flexible fibers: Bead-spring model~\cite{Xu2023PoF} -- }
A passive flexible fiber can be simulated using the bead-spring model~\cite{yamamoto1993method,Xu2023PoF}. As shown schematically in Fig.~\ref{Fig:App2-fiber}(a), the fiber consists of $N$ identical rigid circular beads (of diameter $d$) that are connected by ($N-1$) identical springs (of equilibrium length $\ell_0$). Flexible fibers show elastic resistance to both compression and bending. In this case, the interactions between fiber bead particles are represented by the following form of energy (per length in 2D):
\begin{equation}\label{Eq:App2-Upp}
U_{\mathrm{pp}}\left(\boldsymbol{R}_\alpha\right)=\sum_{\alpha=1}^{N-1}\frac{1}{2}k_{\mathrm s}\left(R_{\alpha,\alpha+1}-\ell_0\right)^{2} -\sum_{\alpha=1}^{N-1}k_{\mathrm b}\left(\boldsymbol{\hat{t}}_{\alpha+1}\cdot \boldsymbol{\hat{t}}_{\alpha}\right)^{2}.
\end{equation}
Here $R_{\alpha,\alpha+1}=|\boldsymbol{R}_{\alpha,\alpha+1}|$ is the distance between the center-of-mass of the two neighboring beads with $\boldsymbol{R}_{\alpha,\alpha+1}=\boldsymbol{R}_{\alpha+1}-\boldsymbol{R}_{\alpha}$ and $\boldsymbol{R}_\alpha$ being the center-of-mass position of each bead. $\hat{\boldsymbol{t}}_{\alpha}=\boldsymbol{R}_{\alpha,\alpha+1}/R_{\alpha,\alpha+1}$ is tangential unit vector of the fiber. $k_{\mathrm {s}}$ is the spring constant, and for inextensible fibers, a very large $k_{\mathrm {s}}$ should be used such that the spring length is close to its equilibrium length $\ell_0$, and hence the contour length of the fiber is also almost constant to be $\ell_{\mathrm c}=N\ell_0$. $k_{\mathrm b}$ is the bending constant that characterizes the bending stiffness of the fiber. 

Substituting the energy $U_{\mathrm{pp}}$ in Eq.~(\ref{Eq:App2-Upp}) and $U_{\mathrm{ps}}$ in Eq.~(\ref{Eq:theory-DRD-Ups}) into Eq.~(\ref{Eq:theory-DRD-Fext}), we obtain the total force applied on each particle bead of the flexible fiber to be
\begin{equation}\label{Eq:App2-Ftot-Fiber}
\boldsymbol{F}_{\mathrm{ext},\alpha}=\boldsymbol{F}_{\mathrm{s},\alpha}+\boldsymbol{F}_{\mathrm{b},\alpha}
-k_{\mathrm{s}}\left(|\mathcal{D}(\boldsymbol{{R}}_{\alpha})|-d/2\right) (-\boldsymbol{\hat{n}}_{\mathrm{s},\alpha}),
\end{equation}
in which the lateral spring force $\boldsymbol{F}_{\mathrm{s},\alpha}$ is given by
\begin{subequations}\label{Eq:Method-BeadSpringFs}
\begin{equation}\label{Eq:Method-BeadSpringFsi}
\boldsymbol{F}_{\mathrm{s},\alpha}=-k_{\mathrm{s}}\left[\left(R_{\alpha-1, \alpha}-\ell_0\right) \hat{\boldsymbol{t}}_{\alpha-1}-\left(R_{\alpha, \alpha+1}-\ell_0\right) \hat{\boldsymbol{t}}_{\alpha}\right], \quad \text{for}\quad 2\leqslant \alpha\leqslant N-1,
\end{equation} 
\begin{equation}\label{Eq:Method-BeadSpringFs1N}
\boldsymbol{F}_{\mathrm{s},1}=k_{\mathrm{s}}\left(R_{1,2}-\ell_0\right) \hat{\boldsymbol{t}}_{1},\quad \boldsymbol{F}_{\mathrm{s},N}=-k_{\mathrm{s}}\left(R_{N-1, N}-\ell_0\right) \hat{\boldsymbol{t}}_{N-1},
\end{equation}
\end{subequations}
and the bending force $\boldsymbol{F}_{\mathrm{b},\alpha}$ is given by
\begin{subequations}\label{Eq:Method-BeadSpringFb}
\begin{align}\label{Eq:Method-BeadSpringFbi}
\boldsymbol{F}_{\mathrm{b},\alpha}= k_{\mathrm{b}}\left[\left(\hat{\boldsymbol{t}}_{\alpha}+\hat{\boldsymbol{t}}_{\alpha-2}\right) \cdot \frac{\left(\boldsymbol{I}-\hat{\boldsymbol{t}}_{\alpha-1} \hat{\boldsymbol{t}}_{\alpha-1}\right)}{R_{\alpha-1, \alpha}}   -\left(\hat{\boldsymbol{t}}_{\alpha+1}+\hat{\boldsymbol{t}}_{\alpha-1}\right) \cdot \frac{\left(\boldsymbol{I}-\hat{\boldsymbol{t}}_{\alpha} \hat{\boldsymbol{t}}_{\alpha}\right)}{R_{\alpha, \alpha+1}}\right], \quad \text{for}\quad 3\leqslant \alpha \leqslant N-2,
\end{align} 
\begin{equation}\label{Eq:Method-BeadSpringFb1N}
\boldsymbol{F}_{\mathrm{b},1}=-k_{\mathrm{b}} \hat{\boldsymbol{t}}_{2} \cdot \frac{\left(\boldsymbol{I}-\hat{\boldsymbol{t}}_{1} \hat{\boldsymbol{t}}_{1}\right)}{R_{1,2}}, \quad  
\boldsymbol{F}_{\mathrm{b},N}=k_{\mathrm{b}} \hat{\boldsymbol{t}}_{N-2} \cdot \frac{\left(\boldsymbol{I}-\hat{\boldsymbol{t}}_{N-1} \hat{\boldsymbol{t}}_{N-1}\right)}{R_{N-1, N}}. 
\end{equation}  
\begin{equation}\label{Eq:Method-BeadSpringFb1N2}
\boldsymbol{F}_{\mathrm{b},2}=k_{\mathrm{b}}\left[\hat{\boldsymbol{t}}_{2} \cdot \frac{\left(\boldsymbol{I}-\hat{\boldsymbol{t}}_{1} \hat{\boldsymbol{t}}_{1}\right)}{R_{1,2}}-\left(\hat{\boldsymbol{t}}_{3}+\hat{\boldsymbol{t}}_{1}\right) \cdot \frac{\left(\boldsymbol{I}-\hat{\boldsymbol{t}}_{2} \hat{\boldsymbol{t}}_{2}\right)}{R_{2,3}}\right],
\end{equation}
\begin{align}\label{Eq:Method-BeadSpringFbN1N3}
 \boldsymbol{F}_{\mathrm{b},N-1}=k_{\mathrm{b}}\left[\left(\hat{\boldsymbol{t}}_{N-1}+\hat{\boldsymbol{t}}_{N-3}\right) \cdot \frac{\left(\boldsymbol{I}-\hat{\boldsymbol{t}}_{N-2} \hat{\boldsymbol{t}}_{\boldsymbol{N}-2}\right)}{R_{N-2, N-1}} -\hat{\boldsymbol{t}}_{N-2} \cdot \frac{\left(\boldsymbol{I}-\hat{\boldsymbol{t}}_{N-1} \hat{\boldsymbol{t}}_{N-1}\right)}{R_{N-1, N}}\right],   
\end{align}
with $\boldsymbol{I}$ being the 2D unit tensor. 
\end{subequations}

\emph{Active (self-propelled) microswimmers: Dumbbell model ---}
Next, to model self-propelling microswimmers, we employ the dumbbell force-dipole model proposed by Furukawa \emph{et al.}~\cite{Furukawa2014PRE}, where a swimming microorganism comprises a dumbbell with a pair of prescribed dipolar force (see Fig.~\ref{Fig:App2-microswimmer}(a)). Each dumbbell microswimmer consists of one spherical body particle with diameter $d_{\mathrm B}$ and one ``phantom" particle with diameter $d_{\mathrm P}$. The phantom particle can be thought of as a far-field modeling of the effect of a thin flagellar bundle, whereby a force is exerted on the fluid at the position with a distance $d_0$ away from the microswimmer's body. The propulsion force $-\boldsymbol{F}_{\mathrm {act},\alpha}={F}_0 \boldsymbol{\hat{u}}_\alpha$ is exerted on the fluid via \& at the phantom particle; an equal and opposite force $\boldsymbol{F}_{\mathrm {act},\alpha}={F}_0 \boldsymbol{\hat{u}}_\alpha$ is exerted on the body particle, which ensures the dipolar character of the swimming mechanism. Here the active propulsion force is assumed to be applied along the direction of unit vector $\hat{\boldsymbol{u}}_{\alpha}=\left(\boldsymbol{R}^{\mathrm{B}}_{\alpha}-\boldsymbol{R}^{\mathrm{P}}_{\alpha}\right) /\left|\boldsymbol{R}^{\mathrm{B}}_{\alpha}-\boldsymbol{R}^{\mathrm{P}}_{\alpha}\right|$, where $\boldsymbol{R}^m_{\alpha}$ denotes and the diameter $d_m$ of the particles and the distances of separation between them determine the shape of the microswimmer, where the subscripts $m={\mathrm{B}}$ and ${\mathrm{P}}$ denote the body particles and the phantom particle of the dumbbell, respectively. Particularly, in 2D, we can represent $\hat{\boldsymbol{u}}_{\alpha}$ by $\hat{\boldsymbol{u}}_{\alpha}= (\cos\theta_{\alpha}, \sin\theta_{\alpha})$ with $\theta_{\alpha}$ denoting the angle of $\hat{\boldsymbol{u}}_{\alpha}$ relative to the $+\boldsymbol{\hat x}$ direction.  


Since each microswimmer is composed of two particles, it is more convenient to express the total force density $\boldsymbol{f}_{\mathrm{tot}}$ in the dynamic equation~(\ref{Eq:App2-Dyn-NS}) into an alternative form as 
\begin{equation}\label{Eq:App2-Dyn-NS-Fext2}
\boldsymbol{f}_{\mathrm{tot}}=\sum_{\alpha=1}^N \sum_{m={\mathrm {B,P}}}  
\frac{(1-\psi^{m}_{\alpha}) \boldsymbol{F}^{m}_{\alpha}}{\int d\boldsymbol{r} (1-\psi^{m}_{\alpha})}, 
\end{equation}
where the external force $\boldsymbol{F}^{m}_{\alpha}$ (with $m=\mathrm{B},\mathrm{P}$ for body and phantom particles, respectively) acting on the center-of-mass of each microswimmer particle is given (from Eq.~(\ref{Eq:theory-DRD-Fext})) by
\begin{equation}\label{Eq:Method-force}
\boldsymbol{F}^{{\mathrm B}}_{\alpha}=\boldsymbol{F}_{\mathrm {act},\alpha}-\sum_{\beta \neq \alpha}\frac{\partial U_{\mathrm {pp}}}{\partial{R}^{\mathrm{B}}_{\beta\alpha}}{\boldsymbol{\hat{R}}^{\mathrm{B}}_{\beta\alpha}}
-\frac{\partial U_{\mathrm{ps}}}{\partial |\mathcal{D}(\boldsymbol{{R}}^B_{\alpha})|}(-\boldsymbol{\hat{n}}^B_{\mathrm{s},\alpha}), \quad \boldsymbol{F}^{{\mathrm P}}_{\alpha} = -\boldsymbol{F}_{\mathrm {act},\alpha}, 
\end{equation} 
with $\boldsymbol{\hat{n}}^{\mathrm B}_{\mathrm{s},\alpha}$ being the outward unit vector of the solid surface closest to the body of the $\alpha$-th microswimmer (pointing toward the solid domain), $\boldsymbol{R}^{\mathrm B}_{\beta\alpha}\equiv \boldsymbol {R}^{\mathrm B}_{\alpha}-\boldsymbol{R}^{\mathrm B}_{\beta}$, $R^{\mathrm B}_{\beta\alpha}\equiv {|\boldsymbol{R}^{\mathrm B}_{\beta\alpha}|}$, and $\boldsymbol{\hat{R}}^{\mathrm B}_{\beta\alpha}\equiv \boldsymbol{R}^{\mathrm B}_{\beta\alpha}/R^{\mathrm B}_{\beta\alpha}$.
The particle-solid boundary interaction potential $U_{\mathrm{ps}}$ is taken to be the simple harmonic form as given in Eq. (\ref{Eq:theory-DRD-Ups}). 

For the active force $\boldsymbol{F}_{\mathrm {act},\alpha}=F_0\boldsymbol{\hat{u}}_{\alpha}$ that is stochastic and drives the self-propulsion of the microswimmer, we assume it follows run-and-tumble (RTP) dynamics where the magnitude $F_0$ is constant but the direction of $\boldsymbol{F}_{\mathrm {act},\alpha}$ or $\boldsymbol{\hat{u}}_{\alpha}$ randomly undergoes complete reorientations (``tumbles") at a certain rate~\cite{solon2015pressure}. The time interval $\Delta t$ between two subsequent tumbling is chosen using an exponential distribution $P(\Delta t)=\tau_{\mathrm p}^{-1} e^{- \Delta t/{\tau_{\mathrm p}}}$ with $\tau_{\mathrm p}$ being the persistent time. When this time is reached, the microswimmer tumbles, a new orientation $\boldsymbol{\hat{u}}_\alpha$ is chosen uniformly in all 3D directions, or in 2D the orientation angle $\theta_{\alpha}$ is chosen uniformly in $[0,2 \pi]$, and then a new time interval $\Delta t$ is generated again from the same exponential distribution. In addition, we assume there are no interactions between phantom particles of different microswimmers, which therefore can overlap. However, an overlap between the body and phantom particles of different microswimmers leads to an unphysical effect~\cite{Furukawa2014PRE}. Therefore, we assume that if the phantom particle of the $\alpha$-th microswimmer overlaps with the ``body" of another $\beta$-th microswimmer, that is, when $\left|\boldsymbol{R}^{\mathrm P}_\alpha-\boldsymbol{R}^{\mathrm{B}}_{\beta}\right|<(d_{\mathrm B}+d_{\mathrm P})/2$, the active force $\boldsymbol{F}_{\mathrm {act},\alpha}$ applied on the $\alpha$-th microswimmer will be switched off by setting ${F}_0=0$ until its phantom particle once again lies in a purely fluid region~\cite{Furukawa2014PRE}. Similar switch rules also apply to the overlap of the phantom particle with the solid boundary. On the other hand, the body particles of different microswimmers should not overlap and exclude each other, and we take the (excluded volume) interactions $U_{\mathrm{pp}}$ between the body particles of different microswimmers to be a simple harmonic form as given in Eq.~(\ref{Eq:theory-DRD-Upp}): 
\begin{align}\label{Eq:App2-Up0}
U_{\mathrm {pp}}\left(\left\{{R}_{\beta\alpha}^{\mathrm B}\right\}\right)= 
\sum_{\beta\neq \alpha} 
\begin{cases} 
\frac{1}{2} k\left({R}_{\beta\alpha}^{\mathrm B}-d\right)^2, & {R}_{\beta\alpha}^{\mathrm B} \le d, \\ 
0, & {{R}_{\beta\alpha}^{\mathrm B}>d}.
\end{cases}
\end{align}  
with $k$ being the interaction stiffness. 
Fluid slip is neglected at all particle surfaces and at solid boundaries and hence we take the spatially-varying viscosity $\eta({\boldsymbol r})$ in the form of Eq.~(\ref{Eq:theory-Dprofile12}) as   
\begin{equation} \label{Eq:App2-eta}
\eta(\psi_{\alpha})=\eta_{\mathrm{s}}+\left(\eta_{\mathrm{f}}-\eta_{\mathrm{s}} \right)\prod_{\alpha=1}^N \psi^{B}_\alpha,
\end{equation} 
where 
the interfacial profile function $\psi^m_{\alpha}$ of the particle takes the form of Eq.~(\ref{Eq:theory-DRD-Dalpha}) as
\begin{equation}\label{Eq:App2-psi1}
\psi_\alpha^m(\boldsymbol{r}, t)=\frac{1}{2}\left[1-\tanh \left(\frac{d_m/2-\left| \boldsymbol{r}-\boldsymbol{R}_\alpha^m (t)\right|}{\sqrt{2} \epsilon_{\mathrm {s}}}\right)\right],
\end{equation}
with $\boldsymbol{R}_\alpha^m (t)$ being the center position of the body particle ($m={\mathrm B}$) and the phantom particle ($m={\mathrm P}$) of $\alpha$-th microswimmer. The positions $\boldsymbol{R}_\alpha^m$ of the microswimmer's body and phantom particles are updated using the body-particle velocity ${\boldsymbol{V}}_{\alpha}^{\mathrm B} (t)$ and the orientation $\hat{\boldsymbol{u}}_\alpha(t)$ of the microswimmer
\begin{equation}\label{Eq:App2-Valpha}
\boldsymbol{\dot{R}}_\alpha^{\mathrm B}={\boldsymbol{V}}_{\alpha}^{\mathrm B} (t)\equiv \frac{\int d\boldsymbol{r}^{\prime}(1-{\psi_{\alpha}^{\mathrm B}})\boldsymbol{v}(\boldsymbol{r}^{\prime},t)}{\int d\boldsymbol{r}^{\prime} (1-\psi_{\alpha}^{\mathrm B})}, \quad
\boldsymbol{R}^{\mathrm{P}}_\alpha(t)=\boldsymbol{R}^{\mathrm{B}}_\alpha(t)-d_0 \hat{\boldsymbol{u}}_\alpha(t),
\end{equation}
respectively. 

\subsubsection{Far-field dipolar flow field surrounding dumbbell microswimmers\label{sec:App2-Benchmark}}


As done in Sec.~\ref{sec:App1-Benchmark}, we non-dimensionalize the above dynamic equations. In addition to the six dimensionless parameters, ${\mathrm{Re}} \equiv \rho V_0 H/\eta_{\mathrm f}$ or ${\mathrm{Re}}_{\mathrm p}\equiv \rho V_0d/\eta_{\mathrm f}$ or ${ \mathrm{Re}_{\mathrm {ps}}}= {\rho V_0 d^2}/{{\eta_{\mathrm f}H}}$ with $V_0$ being the characteristic velocity; P\'eclet number, ${\mathrm{Pe}_{\mathrm{s}}} \equiv \tau_{\mathrm{p}}/\tau_0$ with $\tau_{\mathrm{p}}$ being the persistence time of the run-and-tumble motion of the microswimmer and $\tau_0=H/V_0$ being the time unit; $\mathcal{K}_{\mathrm s}={k_{\mathrm{s}}}/{H \eta_{\mathrm f} V_0}$; $r_{{\eta}}\equiv \eta_ {\mathrm s}/\eta_ {\mathrm f}\gg 1$; $r_{\epsilon}\equiv {\epsilon}_{\mathrm s}/H$; $r_{\mathrm{d}}=d/H$, as given in Sec.~\ref{sec:App1-Benchmark}, one more dimensionless parameter arises: $\mathcal{K}_{\mathrm b}={k_{\mathrm{b}}}/{H \eta_{\mathrm f} V_0}$, the stiffness parameter for the fiber bending resistance.   

\begin{figure}[htbp]
  \centering
  \includegraphics[width=0.72\columnwidth]{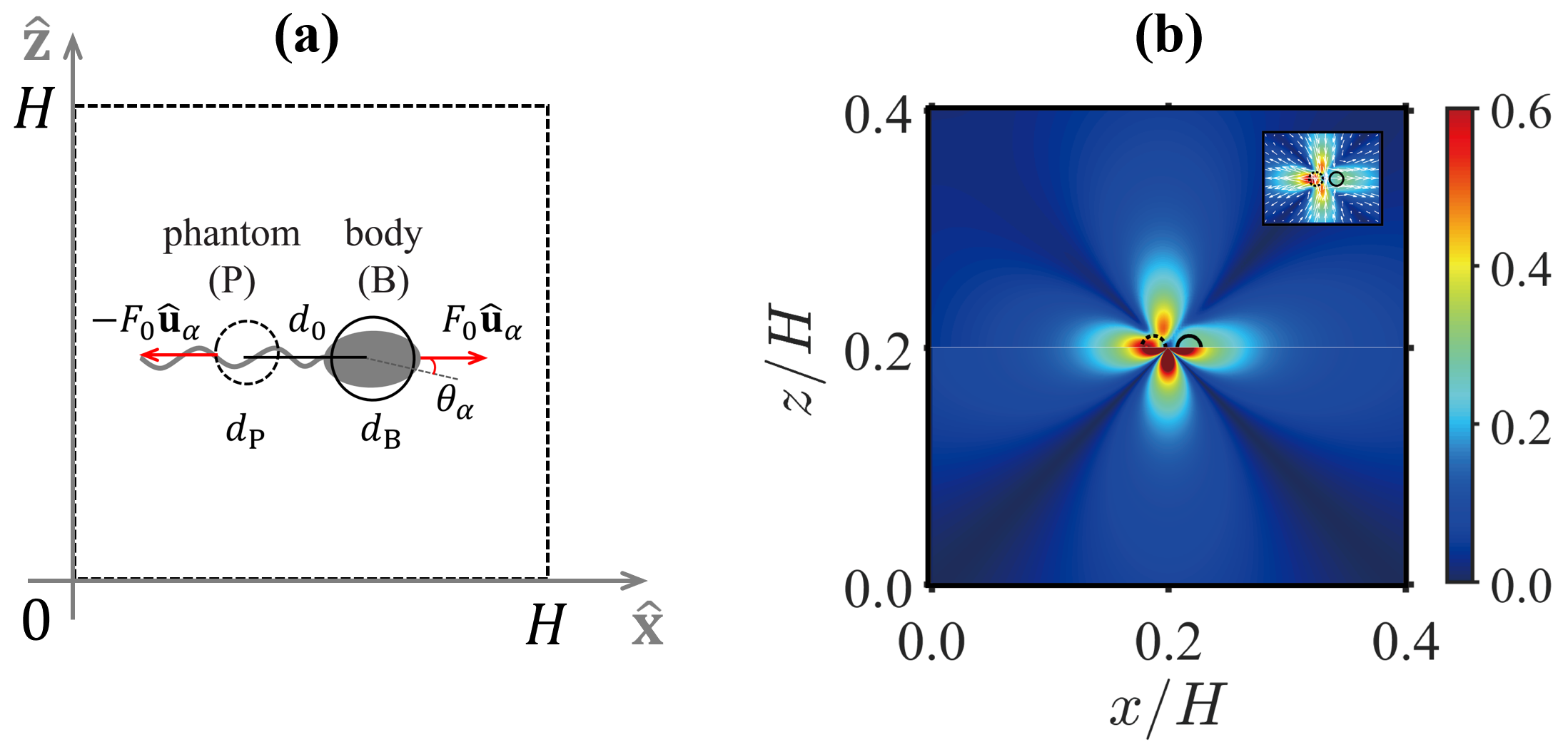}
  \caption {(Color online) (a) A dumbbell model microswimmer (no tumbling) composed of a body particle with diameter $d_{\mathrm B}$ and a ``phantom" particle with diameter $d_{\mathrm{P}}$. The center-to-center distance between the body particle and the phantom particle is fixed at $d_0$. (b) Simulated flow field (the top half part) induced by the dumbbell microswimmer (in the reference frame moving with the microswimmer) in a square computational domain of dimensions $H\times H$. Analytical far-field flow field induced by an extensible force dipole is also plotted (the bottom half part) for comparison. (Inset) The simulated near-field flow in the vicinity of (within a distance of $4d_0$ away from) the center of the microswimmer. Here, we take the physics parameters to be: ${\mathrm{Re}} \equiv \rho V_0H/\eta_{\mathrm f} = 1.0$ and hence ${\mathrm{Re_p}} \equiv \rho V_0 d_{\mathrm{B}}/\eta_{\mathrm f} = 0.05$; $d_{\mathrm{B}}/H = d_{\mathrm{P}}/H = 0.05$, and $d_0 / H = 0.075$. We take the two ``free'' parameters to be: $r_\eta = r_{\mathrm{s}}/r_{\mathrm{f}}=100.0$ and $r_{\epsilon_{\mathrm{s}}}=\epsilon_{\mathrm{s}}/H=0.004$. Periodic boundary conditions have been used in both $x-$ and $z-$ directions.} 
  \label{Fig:App2-microswimmer}
\end{figure}

To apply the DRD approach to simulate mFSI in active matter hydrodynamics, we first validated the DRD approach by simulating the flow field generated by a dumbbell microswimmer (see Fig.~\ref{Fig:App2-microswimmer}(a)), which is known as a ``pusher'' and the far-field flow field surrounding the pusher (centered at the origin) is approximated by that of a point force dipole~\cite{Furukawa2014PRE}
\begin{equation}\label{eq:App2-dipole}
\boldsymbol{v}(\boldsymbol{r})=\frac{d_0 F_0}{4 \pi \eta_{\mathrm{f}}} \frac{2\left(\hat{\boldsymbol{r}} \cdot \widehat{\boldsymbol{u}}_\alpha\right)^2-1}{|\boldsymbol{r}|} \hat{\boldsymbol{r}}, 
\end{equation}
where $\hat{\boldsymbol{r}}={\boldsymbol{r}}/{|\boldsymbol{r}|}$, and $\boldsymbol{r}$ is the distance vector relative to the center of the dipole. In Fig.~\ref{Fig:App2-microswimmer}(b), the flow field at a distance of several swimmer body-lengths is shown to match the dipolar flow field given in Eq.~(\ref{eq:App2-dipole}) very well. 

\subsection{Benchmark simulation 3: Two-phase flows on solid surfaces}\label{sec:Benchmark-3}

\subsubsection{Governing dynamic equations}\label{sec:App3-DynEqn}

\begin{figure}[htbp]  
  \centering
  \includegraphics[width=0.6\columnwidth]{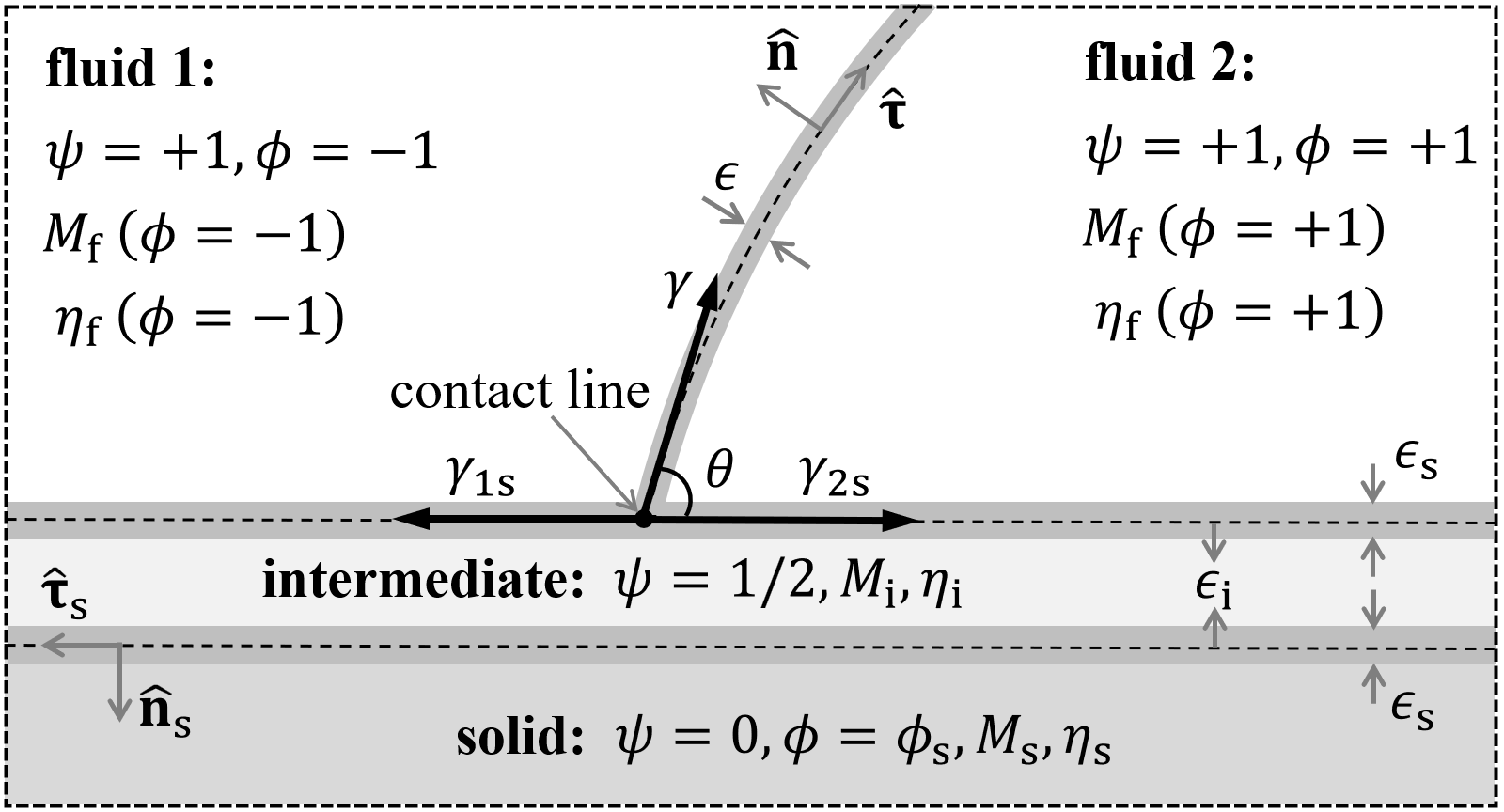}
  \caption {Schematic illustration for the moving contact line in immiscible two-phase flows on a planar solid surface. The immiscible two-phase fluid domain is distinguished by different values of the phase parameter $\phi$ while sharing the same value of the phase parameter $\psi=+1$. The solid domain is represented by $\psi=0$ and $\phi=\phi_{\mathrm{s}}$. To fully reproduce Qian-Wang-Sheng’s model~\cite{Qian2006JFM,Qian2008JFM}, we introduce an intermediate domain, which has a small thickness $\epsilon_{\mathrm{i}}$ (magnified in the figure for better illustration), a very low viscosity $\eta_{\mathrm i}$ (with $\eta_{\mathrm i}<\eta_{\mathrm{f}}\ll \eta_{\mathrm{s}}$), and a very low mobility $M_{\mathrm i}$ (with $M_{\mathrm i}<M_{\mathrm{f}}$, and noting $M_{\mathrm{s}}\ll M_{\mathrm{f}}$). } 
  \label{Fig:App3-MCL}
\end{figure}

In this section, we demonstrate the third advantage of our DRD approach -- the capability of simulating multiphase flows on solid surfaces. Specifically, we consider two-phase flows on solid surfaces: moving contact line dynamics of two immiscible two-phase flows on solid surfaces, as shown in Fig.~\ref{Fig:App3-MCL}. In this triple-phase flow, the fluid-fluid interface and the fluid-solid interface are described by two-phase parameter functions $\phi$ and $\psi$, respectively. The interfacial profile function $\psi$ of the solid boundary takes the general form of Eq.~(\ref{Eq:theory-DRD-psi}). In this case, the total free energy is given from Eq.~(\ref{Eq:theory-DRD-F}) by
\begin{equation}\label{Eq:App3-FT}
\mathcal{F}\left[\phi(\boldsymbol{r}), \psi\left(\boldsymbol{r}\right)\right]=\int d\boldsymbol {r} \left[\psi \left(f_{\mathrm b}(\phi)+\frac{1}{2}K|\nabla \phi|^2\right)+  {\epsilon_{\mathrm s}} f_{\mathrm s}(\phi)\left|\nabla \psi\right|^2\right], 
\end{equation}  
where $f_{\mathrm b}(\phi)=\frac{1}{4}a\left(\phi^2-1\right)^2$ is the bulk free energy density and $f_{\mathrm s}(\phi)=-\frac{1}{4} \gamma \cos \theta_{\mathrm{s}} (\phi^3-3\phi)$ is the surface energy densities at the fluid-particle interface with $\theta_{\mathrm{s}}$ being the static contact angle at the particle surface.  
The dissipation function is given by Eq.~(\ref{Eq:theory-DRD-Phi}) and then minimizing the Rayleighian $\mathcal {R}[\boldsymbol{v},\boldsymbol{J}]=\dot{\mathcal{F}}+\Phi-\int d \boldsymbol{r} P\nabla\cdot\boldsymbol{v}$ in Eq.~(\ref{Eq:theory-DRD-R}) with respect to $\boldsymbol{v}$ and $\boldsymbol{J}$ gives dynamic equations similar to Eqs.~(\ref{Eq:theory-DRD-Dyn}):
\begin{subequations} \label{Eq:App3-Dyn-NSCH}
\begin{equation} \label{Eq:App3-Dyn-NS}
\rho\left(\partial_t {\boldsymbol v} +{\boldsymbol v} \cdot \nabla {\boldsymbol v} \right)=-\nabla P+\nabla \cdot\left[\eta(\phi,\psi) \left(\nabla {\boldsymbol{v}}+\nabla {\boldsymbol{v}}^{\mathrm T} \right)\right]+\hat{\mu}\nabla(\psi \phi),
\end{equation}
\begin{equation}\label{Eq:App3-Dyn-CH}
\partial_t(\psi\phi)+\boldsymbol{v}\cdot \nabla(\psi\phi)=\nabla \cdot \left[M(\phi,\psi) \nabla\hat{\mu}\right],
\end{equation}
\end{subequations}
and the incompressibility condition $\nabla\cdot\boldsymbol{v}=0$, which are supplemented with proper initial and boundary conditions at the boundary of the computational domain. Here the total chemical potential, $\hat{\mu}$ defined in Eq.~(\ref{Eq:theory-DRD-muT}) reduces to
\begin{equation}\label{Eq:App3-muT}
\psi \hat{\mu} = \psi\mu_{\mathrm{b}}-\nabla \cdot (K\psi \nabla \phi)+ {\epsilon_{\mathrm s}}\mu_{\mathrm{s}}\left|\nabla\psi \right|^{2}.
\end{equation}  
Moreover, since fluid slip and relaxation dissipation across the fluid-solid interface are both critical in multiphase dynamics, particularly, in the contact line motion~\cite{Qian2006JFM}, we choose the non-monotonic smooth profiles of viscosity $\eta(\phi,\psi)$ and mobility $M(\phi,\psi)$ given in Eq.~(\ref{Eq:theory-Dprofile3}).

\subsubsection{Sharp-interface limit and generalized Navier boundary conditions (GNBC)}\label{sec:App3-SharpLimit}

Now we consider the sharp-interface limit of the general dynamics equations in Sec.~\ref{sec:App3-DynEqn} that are derived based on the diffuse fluid-solid interface, from which we show that generalized Navier boundary conditions (GNBC) at solid surface derived by Qian \emph{et al.}~\cite{Qian2006JFM,Qian2008JFM} can be recovered. For specificity and simplicity, we consider immiscible two-phase flows over a rigid solid domain with a planar surface at $z=0$ (with outward normal direction $\boldsymbol{\hat{n}}_{\mathrm{s}}$ along $-\boldsymbol{\hat{z}}$-direction) as shown in Fig.~\ref{Fig:App3-MCL}. In this case, the interfacial profile function $\psi$ in Eq.~(\ref{Eq:theory-DRD-psi}) of the solid boundary reduces to a simpler form 
\begin{equation}\label{Eq:App3-psi-sharp}
\psi(z)=\frac{1}{2}\left[1-\tanh \left(\frac{z}{\sqrt{2} \epsilon_{\mathrm {s}}}\right)\right].
\end{equation}
Moreover, to fully reproduce Qian-Wang-Sheng’s model, we introduce a third intermediate domain between the fluid and solid regions (see Fig.~\ref{Fig:App3-MCL}). This intermediate domain has a small thickness $\epsilon_{\mathrm{i}}$, a very low viscosity $\eta_{\mathrm i}$ (with $\eta_{\mathrm i}<\eta_{\mathrm{f}}\ll \eta_{\mathrm{s}}$), and a very low mobility $M_{\mathrm i}$ (with $M_{\mathrm i}<M_{\mathrm{f}}$, and noting $M_{\mathrm{s}}\ll M_{\mathrm{f}}$). Both $\eta(\psi)$ and $M(\psi)$ are interpolated using the piecewise linear form given in Eq.~(\ref{Eq:theory-Dprofile3}).

Here, instead of using traditional mathematical methods to derive sharp-interface limits of dynamic equations, we directly carry out the sharp-interface limits of the change rate of the total free energy $\dot{\mathcal{F}}_{\mathrm T}$ in Eq.~(\ref{Eq:theory-DRD-FTdot}) and the dissipation function in Eq.~(\ref{Eq:theory-DRD-Phi}). 
Firstly, integrating $\dot{\mathcal{F}}_{\mathrm T}$ in Eq.~(\ref{Eq:theory-DRD-FTdot}) cross the fluid-solid interface of small thickness $\epsilon_{\mathrm{s}}$, we obtain the rate of change of the fluid-solid interface energy in the sharp-interface limit as
\begin{align} \label{Eq:App3-FTdot-Sharp}
\dot{\mathcal{F}}_{\mathrm{surf}} \approx \int d{\mathcal A} \int_{-\epsilon_{\mathrm{s}}/2}^{+\epsilon_{\mathrm{s}}/2} dz \left[-\partial_z (K\psi \partial_z\phi)+ \epsilon_{\mathrm s}{\mu_{\mathrm{s}}} (\partial_z\psi )^{2}\right]\partial_t\phi
\approx \int d{\mathcal A} \left(-K\partial_z\phi+\frac{\partial f_{\mathrm{s}}}{\partial \phi } \right) \partial_t\phi, 
\end{align} 
where $d\mathcal A$ is the area element of the solid surface, the surface energy density is $f_{\mathrm s} (\phi) =-\frac{1}{4} \gamma \cos \theta_{\mathrm s} (\phi^3-3\phi)$, 
and we have used the identity $\int_{-\infty}^{+\infty}dz\epsilon_{\mathrm s}(\partial_z\psi )^{2}= {2\sqrt{2}}/{3}$.
Secondly, noting that the dominant component of the shear rate tensor in the fluid-solid interfacial region is $\partial_zv_x\sim v_x^{\mathrm{slip}}/\epsilon_{\mathrm{s}}$ and $\boldsymbol{J}\approx J_{z}\boldsymbol{\hat z}\sim -\dot{\phi} \epsilon_{\mathrm{s}}$, we obtain the dissipation function at the fluid-solid interface from Eq.~(\ref{Eq:theory-DRD-Phi}) as
\begin{equation}\label{Eq:App3-Phi-Sharp}
\Phi_{\mathrm{surf}}[\boldsymbol{v},\boldsymbol{J}]\approx \int d{\mathcal A} \int_{-\epsilon_{\mathrm{s}}/2}^{+\epsilon_{\mathrm{s}}/2} dz \left[\frac{1}{4}\eta_{\mathrm{i}}(\partial_zv_x)^{2}+ \frac{J_{z}^2}{2M_{\mathrm{i}}} \right]= \int d{\mathcal A}  \left[\frac{1}{2}\beta \left(v_x^{\mathrm{slip}}\right)^2+ \frac{\dot{\phi}^2}{2\Gamma}\right],
\end{equation}
with $v_x^{\mathrm{slip}}$ being the slip velocity of the fluids relative to the solid at the solid surface. Here the friction coefficient $\beta\sim \eta_{\mathrm{i}}/\epsilon_{\mathrm{s}}$ and hence the slip length $\ell_{\mathrm{s}}\equiv \eta_{\mathrm{f}}/\beta$ relative to the thickness $\epsilon_{\mathrm{i}}$ of intermediate region scales as $\ell_{\mathrm{s}}/\epsilon_{\mathrm{s}}\sim \eta_{\mathrm{f}}/\eta_{\mathrm{i}}$. On the other hand, the relaxational parameter $\Gamma \sim M_{\mathrm{i}}/\epsilon_{\mathrm{s}}^3$ and hence the relaxation time $\tau_{\mathrm{s}}\equiv \epsilon_{\mathrm{s}}/K\Gamma$ at the fluid-solid interfacial region relative to that $\tau_{\mathrm{f}}\sim \epsilon^4/M_{\mathrm{f}}K$ in the fluid-fluid interfacial region scales as $\tau_{\mathrm{s}}/\tau_{\mathrm{f}} \sim M_{\mathrm{f}}/M_{\mathrm{i}}$ (since the thicknesses of fluid-fluid and fluid-solid are of same order $\epsilon\sim \epsilon_{\mathrm{s}}$).
Thirdly, the change rate of the free energy functional and the dissipation functional in the bulk two-phase fluids (with $\psi=+1$) are given by
\begin{align} \label{Eq:App3-FTdot-sharp-bulk}
\dot{\mathcal{F}}_{\mathrm{bulk}} = \int d\boldsymbol{r} \hat{\mu}_{\mathrm{CH}} \partial_t\phi, \quad
\Phi_{\mathrm{bulk}}[\boldsymbol{v},\boldsymbol{J}]=\int d \boldsymbol{r}\left[\frac{1}{4}\eta_{\mathrm{f}}(\nabla {\boldsymbol{v}}+\nabla {\boldsymbol{v}}^{\mathrm{T}})^{2}+ \frac{1}{2}M_{\mathrm{f}}^{-1}\boldsymbol{J}^2\right],
\end{align}
respectively, with the Cahn-Hilliard chemical potential $\hat{\mu}_{\mathrm{CH}}\equiv \mu_{\mathrm b}-\nabla \cdot (K \nabla \phi)$.  

Furthermore, using $\dot{\phi}=\partial_t\phi+v_x\partial \phi$ (due to the impermeability condition, imposed by large viscosity in the solid domain in our DRD approach) at the sharp fluid-solid interface and the incompressibility condition $\nabla\cdot\boldsymbol{v}=0$, we obtain the Rayleighian in the limit of sharp fluid-solid interface: $\mathcal {R}[\boldsymbol{v},\boldsymbol{J},v_x^{\mathrm{slip}},\dot{\phi}]=\dot{\mathcal{F}}_{\mathrm{bulk}}+\dot{\mathcal{F}}_{\mathrm{surf}}+\Phi_{\mathrm{bulk}}+\Phi_{\mathrm{surf}}-\int d \boldsymbol{r} P\nabla\cdot\boldsymbol{v}$, which has been constructed by Qian \emph{et al.}~\cite{Qian2006JFM,Qian2008JFM}. That is, our general model in the DRD approach can recover all the thermodynamically consistent boundary conditions in the sharp interface limit, including the famous generalized Navier boundary
conditions (GNBC). Finally, note that if we take the linear monotonic form of $\eta$ and $M$ as shown in Eqs.~(\ref{Eq:theory-Dprofile12}), we will recover the no-slip boundary condition for velocity and equilibrium condition for $\phi$, \emph{i.e.}, $v_x^{\mathrm{slip}}=0$ and $-K\partial_z\phi+{\partial\gamma_{\mathrm{FS}}}/{\partial \phi}=0$ with $\gamma_{\mathrm{FS}} = f_{\mathrm s} (\phi)$ being the fluid-solid interfacial tension.

\begin{figure}[htbp]  
  \centering
  \includegraphics[width=0.75\columnwidth]{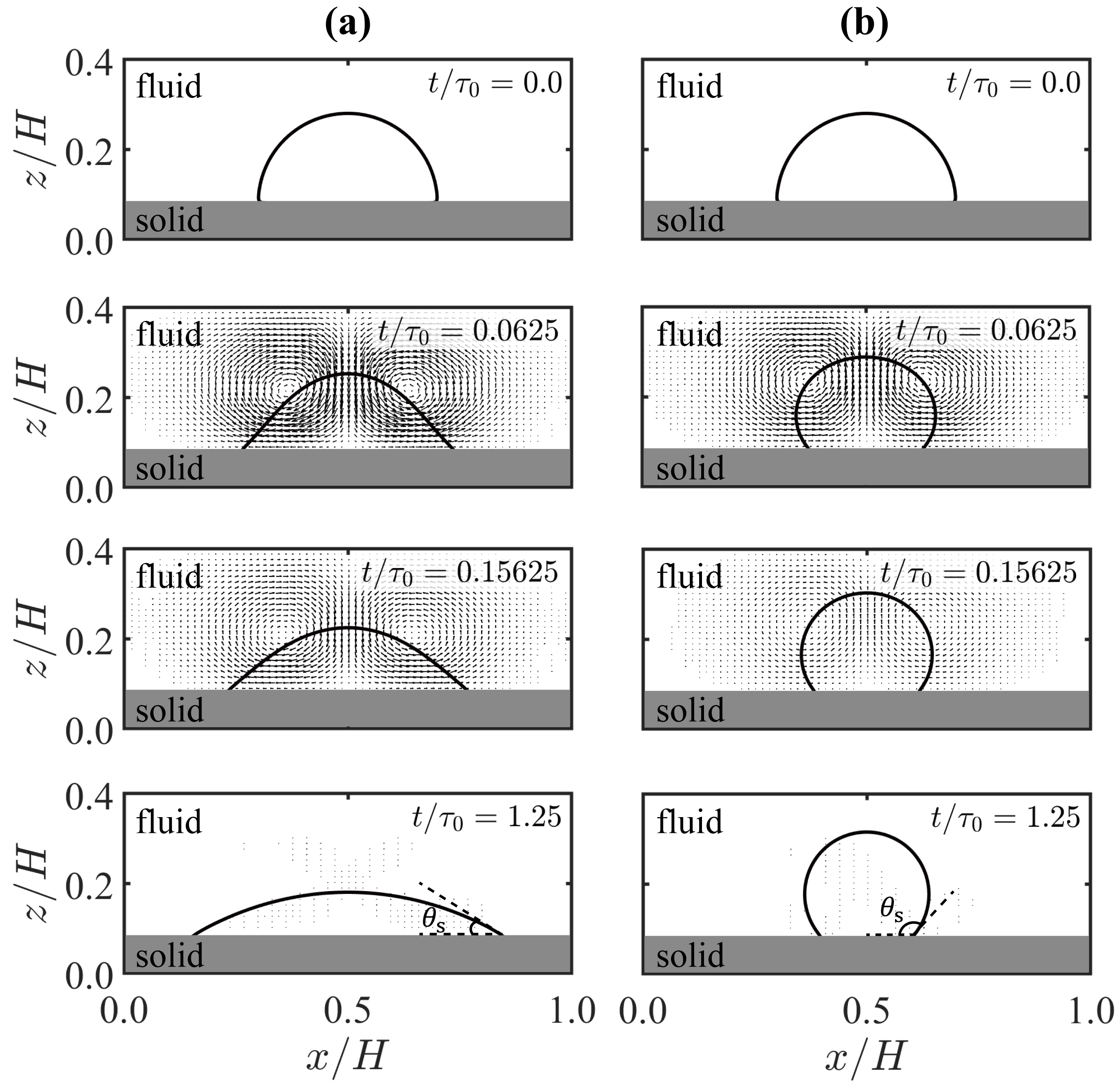}
  \caption {Equilibrating dynamics of a sessile droplet on a planar solid substrate with two different (static) contact angles: (a) $\theta_{\mathrm{s}}=30^{\rm o}$ and (b) $\theta_{\mathrm{s}}=150^{\rm o}$. The interface profiles (solid curves) of the droplet, the flow fields (arrows), and the solid phase (gray) are all shown in the snapshots of the simulations. Here, we take the physics parameters to be: ${\mathrm{Re}} \equiv \rho V_0H/\eta_{\mathrm f} = 3.0$, $\mathrm{Ca}\equiv \eta_{\mathrm{f}}V_0/\gamma = 0.08$, $H_\mathrm{s}/H = 0.2$, and $\ell_{\mathrm{s}}/H = 0.0$. We take the ``free'' parameters to be: (1) free parameters due to the use of diffuse fluid-fluid interface model, $r_{\epsilon}=\epsilon/H = 0.01$,  ${\mathcal{M}}\equiv {M_{\mathrm{f}}} a/V_0H = 0.05$; (2) free parameters due to the use of DRD method, $r_{\epsilon_{\mathrm{s}}}=\epsilon_{\mathrm{s}}/H = 0.005$, $r_\eta=\eta_{\mathrm{s}} / \eta_{\mathrm{f}} = 100.0$, and $r_M=M_{\mathrm{s}} /M_{\mathrm{f}} =0.0$. Periodic boundary conditions have been used in the $x$-direction. } 
  \label{Fig:App3-wetting}
\end{figure}

\subsubsection{Equilibrating dynamics of sessile droplets on planar surfaces}\label{sec:App3-Benchmark-Non} 

As done in Sec.~\ref{sec:App1-Benchmark}, we non-dimensionalize the above dynamic equations by scaling the length by the system width or height $H$, velocity by some characteristic velocity $V_0$, time by $\tau_0 \equiv H/V_0$, pressure and stress by $\eta_{\mathrm{f}}V_0/H$, chemical potential by $a$, viscosity by $\eta_{\mathrm{f}}$, the mobility coefficient by $M_{\mathrm{f}}$.
Fourteen dimensionless parameters appear, as shown in the following comprehensive list along with their physical interpretations. Moreover, we classify them into three general categories: 
\begin{itemize}
\item (i) Measurable physics parameters defined based on measurable quantities. (1) Reynolds number: ${\mathrm{Re}} \equiv \rho V_0H/\eta_{\mathrm f}$. (2) Capillary number: $\mathrm{Ca}\equiv \eta_{\mathrm{f}}V_0/\gamma$.
(3) Relative contact-angle relaxation time: $\tau_{\mathrm{s}} / \tau_0=\tau_{\mathrm{s}} V_0/H$. (4) Static contact angles $\theta_{\mathrm{s}} \in [0,\pi]$. (5) Solid-to-fluid dimension ratio: $H_{\mathrm{s}}/H$. (6) Relative ``slip'' length in GNBC: $\ell_{\mathrm{s}}/H$. 
\item (ii) ``Free'' parameters in diffuse-interface models for multiphase fluids. These include parameters intrinsic to diffuse-interface formulations. (1) Relative interfacial thicknesses $r_\epsilon=\epsilon / H$. (2) Interfacial mobility parameter: ${\mathcal{M}}\equiv {M_{\mathrm{f}}} a/V_0H$, the ratio of $\tau_0$ to the relaxation time $\tau_{\mathrm{f}} /r_{\epsilon}^{2}$ with $\tau_{\mathrm{f}}=\epsilon^2/aM_{\mathrm{f}}=\epsilon^4/M_{\mathrm{f}}K$ being the fluid-fluid interfacial relaxation time. 
\item (iii) Parameters arising from the DRD treatment of fluid-solid interfaces. (1,2) Viscosity ratios: $r_\eta=\eta_{\mathrm{s}} / \eta_{\mathrm{f}}$ and $r_{\eta_{\mathrm{i}}}=\eta_{\mathrm{i}} / \eta_{\mathrm{f}}$. (3,4) Mobility ratios, $r_M=M_{\mathrm{s}} / M_{\mathrm{f}}$ and $r_{M_{\mathrm{i}}}=M_{\mathrm{i}} / M_{\mathrm{f}}$.
(5,6) Relative interfacial thicknesses: $r_{\epsilon_{\mathrm{s}}}=\epsilon_{\mathrm{s}}/H$ and $r_{\epsilon_{\mathrm{i}}}=\epsilon_{\mathrm{i}}/H$. 
\end{itemize}

Moreover, in Sec.~\ref{sec:Benchmark-3}, we analyze the sharp-interface limit $\left(r_{\epsilon_{\mathrm{s}}}=\epsilon_{\mathrm{s}} / H_{\mathrm{s}} \rightarrow 0\right)$ of the DRD model for two-phase flows on solid surfaces. It reveals that some parameters from the DRD treatment are related to measurable quantities. Consequently, the effective number of free parameters is reduced. For example, the slip length $\ell_{\mathrm{s}}$ in the GNBC is found to scale as $\ell_{\mathrm{s}} / \epsilon_{\mathrm{i}} \sim \eta_{\mathrm{f}}/\eta_{\mathrm{i}}=r_{\eta_{\mathrm{i}}}^{-1}$ and the relative relaxation time of dynamic contact angle $\theta$ to static contact angle $\theta_{\mathrm{s}}$ scales as $\tau_{\mathrm{s}}/\tau_{\mathrm{f}} \sim M_{\mathrm{f}} / M_{\mathrm{i}}=$ $r_{M_{\mathrm{i}}}^{-1}$. Therefore, only \emph{six} free parameters remain: $r_\epsilon=\epsilon/H$, $\tau_{\mathrm{f}} / \tau_0$, $r_{\epsilon_{\mathrm{s}}}=\epsilon_{\mathrm{s}}/H$, $r_{\epsilon}=\epsilon/H$, $r_\eta=\eta_{\mathrm{s}} / \eta_{\mathrm{f}}$, and $r_M=M_{\mathrm{s}}/M_{\mathrm{f}}$.

\begin{figure}[htbp] 
  \centering
  \includegraphics[width=0.75\columnwidth]{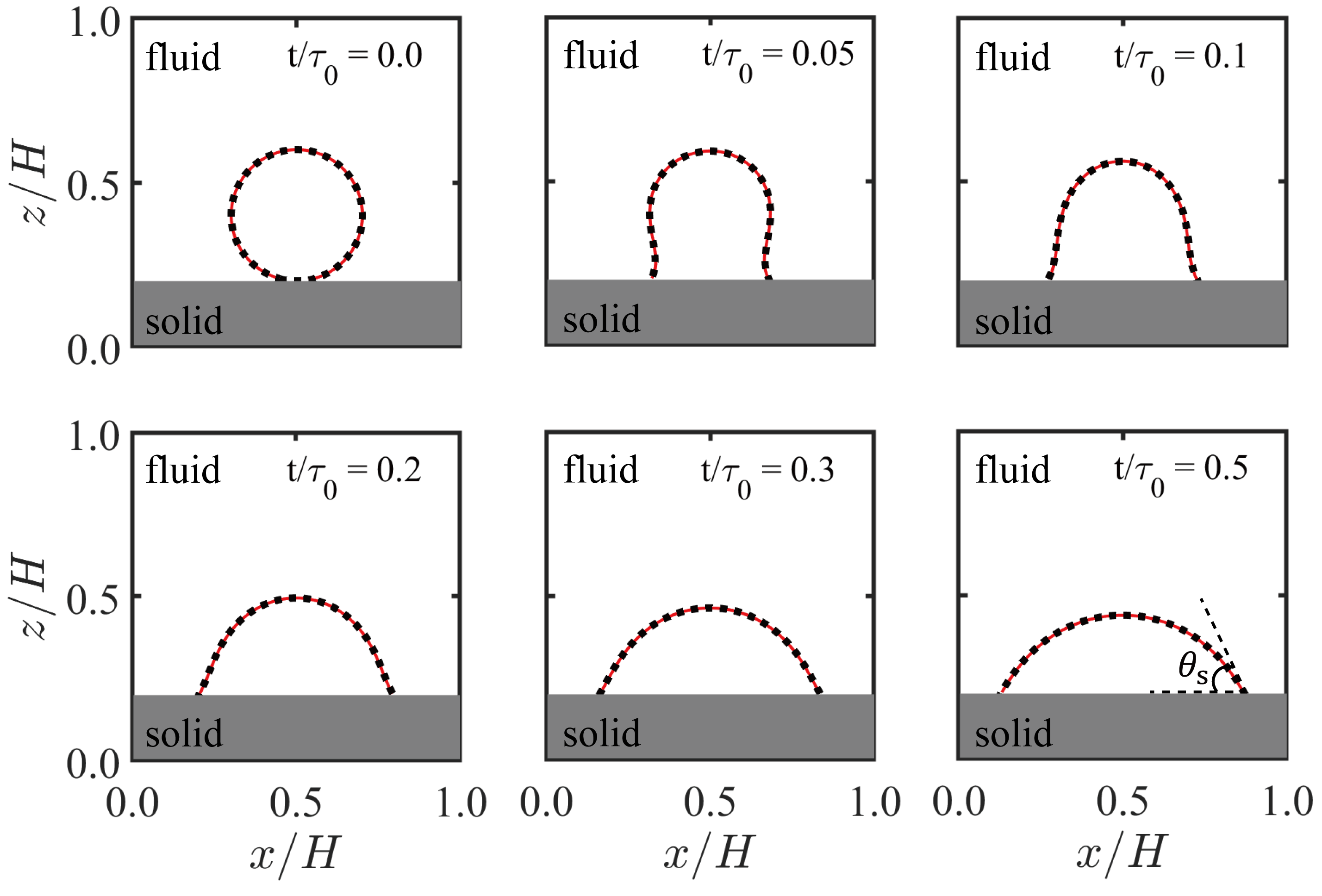}
  \caption {Spreading dynamics of a droplet on a flat solid substrate (gray region): Comparison of the simulation results using our DRD method (black dashed lines) and the standard ``sharp fluid-solid interface'' method (red solid lines)~\cite{gao_efficient_2014} at different time steps ($t/\tau_0= 0, 0.05, 0.1, 0.2, 0.3, 0.5$). 
  Here, we take the physics parameters to be: ${\mathrm{Re}} \equiv \rho V_0H/\eta_{\mathrm f} = 3.0$, $\mathrm{Ca}\equiv \eta_{\mathrm{f}}V_0/\gamma = 0.08$, $\theta_s=60^\circ$, $H_\mathrm{s}/H = 0.2$, and $\ell_{\mathrm{s}}/H = 0.0$. We take the ``free'' parameters to be: (1) free parameters due to the use of diffuse fluid-fluid interface model, $r_{\epsilon}=\epsilon/H = 0.01$ and ${\mathcal{M}}\equiv {M_{\mathrm{f}}} a/V_0H = 0.05$; (2) free parameters due to the use of DRD method, 
   $r_{\epsilon_{\mathrm{s}}}=\epsilon_{\mathrm{s}}/H = 0.005$, $r_\eta=\eta_{\mathrm{s}} / \eta_{\mathrm{f}} = 100.0$, and $r_M=M_{\mathrm{s}} /M_{\mathrm{f}} =0.0$. Periodic boundary conditions have been used in the $x$-direction.  }
  \label{Fig:diffuse-sharp-comparison}
\end{figure}

 
\begin{figure}[htbp] 
  \centering
  \includegraphics[width=1.0\columnwidth]{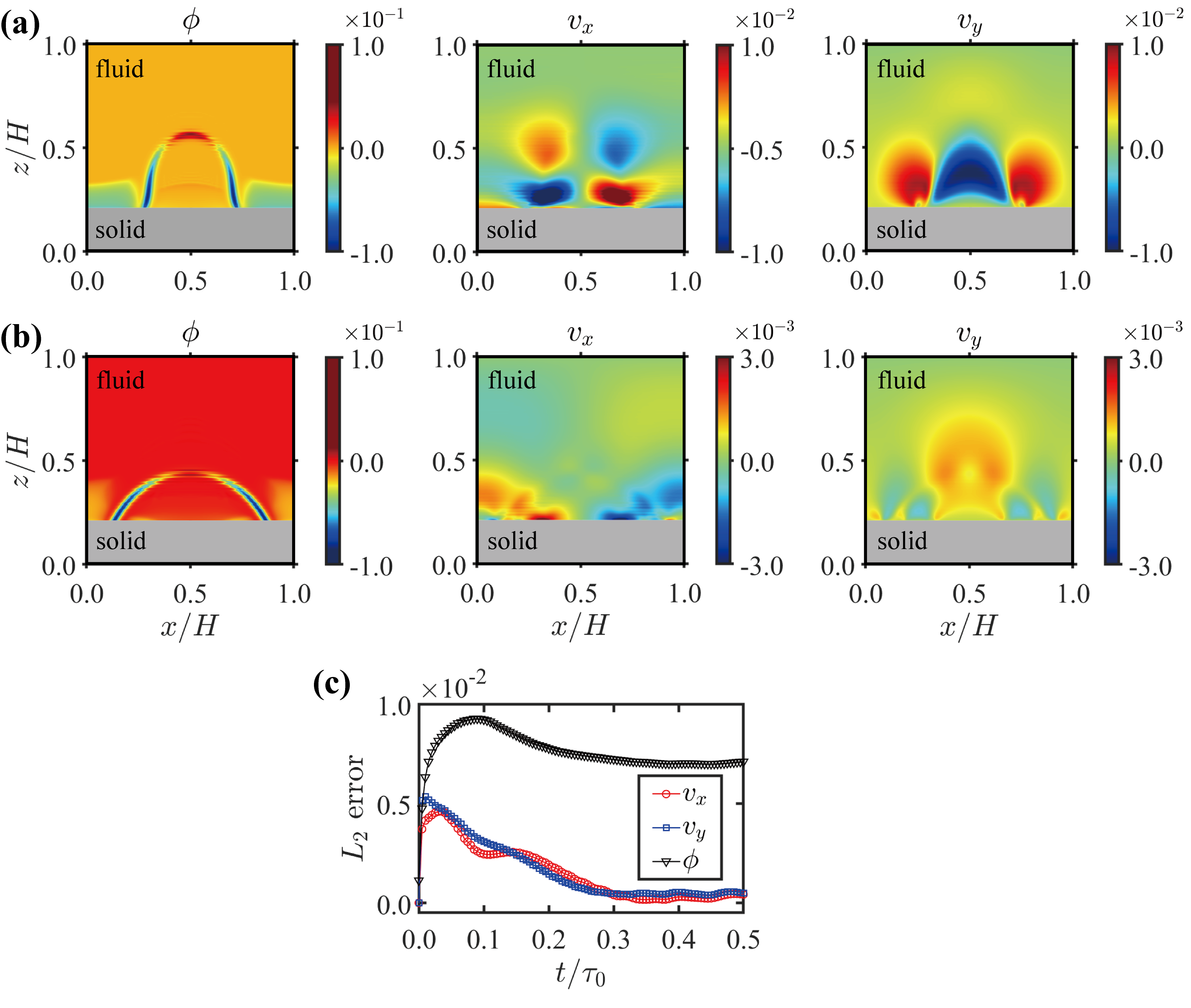}
    \caption{Analysis of simulation accuracy using DRD approach in comparison to those obtained using the standard ``sharp fluid-solid interface'' method~\cite{gao_efficient_2014} as shown in Fig.~\ref{Fig:diffuse-sharp-comparison}. (a,b) The difference map in phase variable $\phi$ field and velocity fields $\boldsymbol{v}=\{v_x,v_y\}$ between the DRD and the sharp-fluid-solid-interface methods at (a) $t/\tau_0=0.1$ and (b) at $t/\tau_0=0.5$. (c) The time evolution of the $L_2$ errors in the fields of $\phi$ and $\boldsymbol{v}=\{v_x,v_y\}$, which remain below $0.005$ and $0.01$, respectively, throughout the simulations.}
    \label{fig:example}
\end{figure} 

In Fig.~\ref{Fig:App3-wetting}, we present DRD simulation results for the equilibration dynamics of sessile droplets on a planar solid substrate with two distinct static contact angles ($30^\circ$ and $150^\circ$). Starting from an initial condition at $t=0$ with a contact angle of $90^\circ$, the droplets are observed to either spread or dewet spontaneously until they equilibrate at the preset static contact angles. This test serves as a standard benchmark for assessing spurious currents in surface tension implementations. Moreover, it provides practical guidelines for selecting the six ``free'' parameters in our model, namely: sufficiently small interface thicknesses, $r_{\epsilon}=r_{\epsilon_{\mathrm{s}}}=r_{\epsilon_{\mathrm{i}}} \leqslant 0.01$ (\emph{i.e.}, all the interfacial thicknesses are taken to be the same small value); proper values of mobility parameter ${\mathcal{M}} \sim 0.01$~\cite{yue_sharp-interface_2010}, viscosity ratio, $r_\eta=\eta_{\mathrm{s}}/\eta_{\mathrm{f}} \geqslant 100$, and mobility ratio, $r_M=M_{\mathrm{s}} / M_{\mathrm{f}}=0$.

In Fig.~\ref{Fig:diffuse-sharp-comparison}, we present the simulation results for a droplet spreading on a flat solid substrate. The evolution of the droplet profile obtained from our DRD method and the standard ``sharp fluid-solid interface'' method~\cite{gao_efficient_2014} are nearly indistinguishable(as also shown in Fig.~\ref{fig:example}(a,b)); the $L_2$ errors of the phase-field variable $\phi$ and velocity field $\boldsymbol{v}=\{v_x,v_y\}$ remain below $0.005$ and $0.01$, respectively, as shown in Fig.~\ref{fig:example}(c)), further confirming the accuracy of the proposed DRD method. Here, numerical simulations are conducted with the following parameters: interface thickness $\epsilon=0.01$, $\epsilon_{\mathrm{s}}=0.5\epsilon=0.005$, domain $[0,1]\times[0,1]$, number of grids $N_x=128$, $N_z=128$, and mesh size $\Delta x=\Delta z=1/128$.

\subsection{Cutting-edge applications across diverse fields} \label{sec:Apps}

\subsubsection{Applications in microfluidics} \label{sec:Apps-MicroFluids}

In microfluidics, researchers focus on manipulating and controlling fluids in channels at microscales~\cite{Kirby2010Book}. DNS provides detailed insights into fluid behaviors and enables researchers to study phenomena such as droplet formation, mixing, and particle transport, which are crucial for advancing microfluidic technologies and applications. However, many significant challenges have been posed to the DNS of mFSI in microfluidics due to the complex interactions between different fluid phases and the confined complex geometries involved. 

\begin{figure}[htbp]
  \centering 
  \includegraphics[width=0.6\columnwidth]{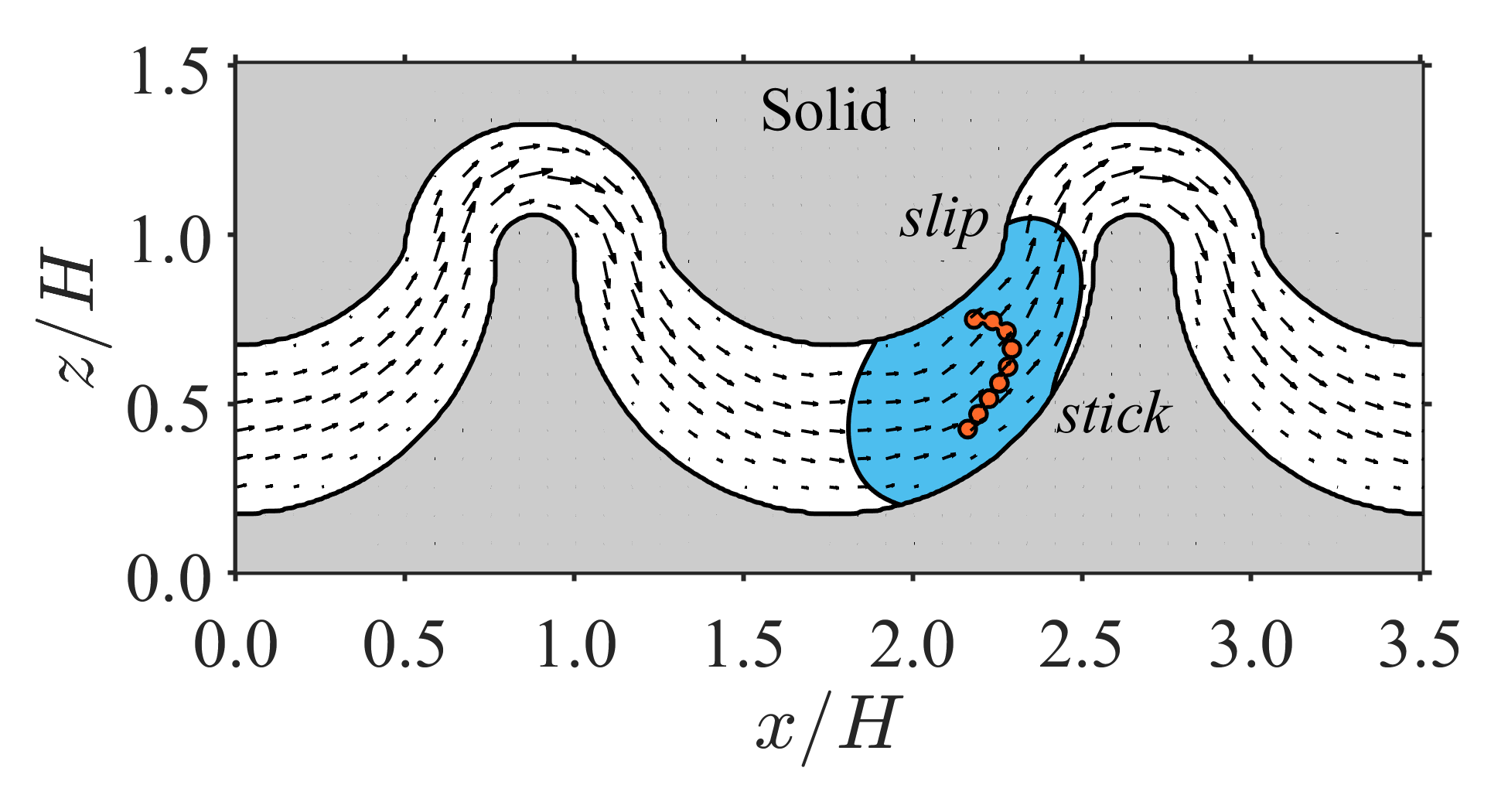}
  \caption {(Color online) Applications in microfluidics: Transport of a droplet containing a flexible fiber (composed of particles of diameter $d$) in a corrugated microchannel of inlet width $H/2$. A snapshot of the whole computational domain is shown, and the contact lines show an interesting stick-slip motion. 
    Here, we take the physics parameters to be: ${\mathrm{Re}} \equiv \rho V_0H/\eta_{\mathrm f} = 1.0$, $\mathrm{Ca}\equiv \eta_{\mathrm{f}}V_0/\gamma = 0.01$,$d/H = 0.05$, $L_x/H = 3.5$, $L_z/H = 1.5$, and $\ell_{\mathrm{s}}/H = 0.0$. The static contact angles at the channel surface and the fiber surfaces are $\theta_{\mathrm{s}}=135^{\rm o}$ and $\theta_{\mathrm{s}}=90^{\rm o}$, respectively. Particle-solid boundary interaction stiffness, $\mathcal{K}=k/H\eta_{\mathrm{f}V_0} = 25$, fiber bead-spring stiffness $\mathcal{K}_{\mathrm{s}}=k_{\mathrm{s}}/H\eta_{\mathrm{f}V_0} = 5.0$, and fiber bending stiffness, $\mathcal{K}_{\mathrm{b}}=k_{\mathrm{b}}/H\eta_{\mathrm{f}V_0} = 0.25$. We take the ``free'' parameters to be: (1) free parameters due to the use of diffuse fluid-fluid interface model, $r_{\epsilon}=\epsilon/H = 0.01$ and ${\mathcal{M}}\equiv {M_{\mathrm{f}}} a/V_0H = 0.05$; (2) free parameters due to the use of DRD method, $r_{\epsilon_{\mathrm{s}}}=\epsilon_{\mathrm{s}}/H = 0.005$, $r_\eta=\eta_{\mathrm{s}} / \eta_{\mathrm{f}} = 100.0$, and $r_M=M_{\mathrm{s}} /M_{\mathrm{f}} =0.0$. Periodic boundary conditions have been used in the $x$-direction. }
\label{Fig:Microfluidics}
\end{figure}

In Sec.~\ref{sec:App1-Benchmark}, we have demonstrated the validity and power of the DRD approach in overcoming these challenges, we carried out benchmark simulations of inertial particle focusing in 2D microchannels with planar and corrugated surfaces.  
The results agree well with those obtained from other DNS methods~\cite{Esipov2020Focusing} (see Figs.~\ref{Fig:App1-Focusing1} and \ref{Fig:App1-Focusing2}). Moreover, in Sec.~\ref{sec:App3-Benchmark-Non}, we have simulated the equilibrating dynamics of a sessile droplet on planar solid surfaces with various contact angles (see Figs.~\ref{Fig:App3-wetting}). Here, we use DRD to simulate a more advanced process in microfluidics: the transport of a droplet with a suspending flexible fiber in an asymmetrically curved microchannel (see Fig.~\ref{Fig:Microfluidics}). Interestingly, we observed a stick-slip motion of the droplet contact lines on the microchannel surfaces, which can be attributed to the strong dependence of effective slip length on the local curvature of the corrugated microchannel. It has been predicted before~\cite{einzel1990boundary} that fluid slip is enhanced (or suppressed) by concave (or convex) surfaces. Therefore, one would expect the contact lines to slip when moving from convex to concave surfaces and stick when moving in turn from concave to convex surfaces (see Fig.~\ref{Fig:Microfluidics}).

\subsubsection{Applications in active matter hydrodynamics} \label{sec:Apps-AM}

Active matter dynamics refers to the collective behaviors exhibited by self-propelled entities, such as bacteria, cells, or synthetic micro-robots, that interact with their environment~\cite{Marchetti2013}. For hydrodynamic behaviors of active matter, DNS provides valuable insights into the underlying mechanisms and enables researchers to understand emergent phenomena such as pattern formation, phase separation, and dynamic self-assembly, 
that are essential for understanding living matter systems and designing active matter-based technologies, such as micro-robots, drug delivery systems, and self-healing materials. 
However, mFSI in active matter hydrodynamics presents many significant challenges to DNS due to the complex interactions between the self-propelled entities and their surrounding fluid environment. 

\begin{figure}[htbp]
  \centering
  \includegraphics[width=0.73\columnwidth]{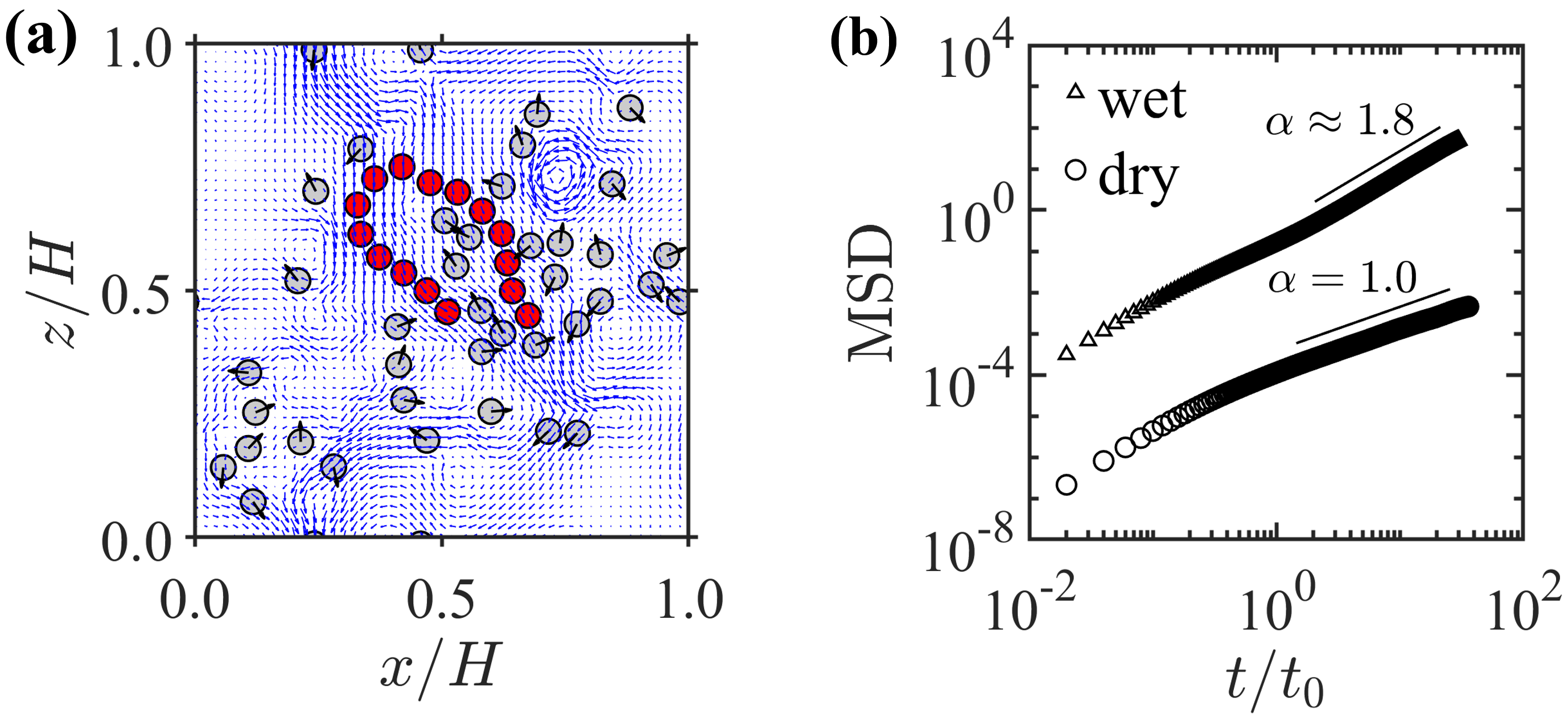}
  \caption {(Color online) Applications in active matter hydrodynamics: Dynamics of a flexible fiber (composed of rigid beads of diameter $d$) suspending in a bath of (dumbbell) microswimmer suspension (where the microswimmer is shown in Fig.~\ref{Fig:App2-microswimmer}(a)). (a) A snapshot of the whole computational domain is shown.
  (b) Comparisons of the mean-square-displacement (MSD $\propto t^\alpha$) of the center-of-mass of the fiber (superdiffusive, $\alpha>1$) in wet microswimmer baths with those (normally diffusive, $\alpha=1$) in dry active baths. 
Here, we take the physics parameters to be: ${\mathrm{Re}} \equiv \rho V_0H/\eta_{\mathrm f} = 1.0$, ${\mathrm{Pe}_{\mathrm{s}}} \equiv \tau_{\mathrm{p}}/\tau_0=0.1$, $d/H=d_{\mathrm{B}}/H = d_{\mathrm{P}}/H=0.05$; stiffness parameters of the particle-particle interaction, fiber stretching and bending: $\mathcal{K}=5.0$, $\mathcal{K}_{\mathrm s}=5.0$ and $\mathcal{K}_{\mathrm b}=0.0025$. The number of swimmers is $40$, and the number of fiber particles is $15$. We take the ``free'' parameters to be: $r_{\epsilon_{\mathrm{s}}}=\epsilon_{\mathrm{s}}/H = 0.004$, $r_\eta=\eta_{\mathrm{s}} / \eta_{\mathrm{f}} = 50.0$, and $r_M=M_{\mathrm{s}} /M_{\mathrm{f}} =0.0$. Periodic boundary conditions apply along both $x-$ and $z-$ directions.}
\label{Fig:Active}
\end{figure}

To apply the DRD approach to simulate mFSI in active matter, we have validated the DRD approach in Sec.~\ref{sec:App2-Benchmark} by simulating and confirming the far-field dipolar flow field generated by a dumbbell model microswimmer~\cite{Furukawa2014PRE} (see Figs.~\ref{Fig:App2-microswimmer}(b)). Here, we use DRD approach to simulate a much more complex active matter dynamic system: the interaction of $40$ dumbbell microswimmers (with a volume fraction $\sim 8\%$) with a fairly flexible (passive) fiber composed of $15$ rigid beads (see Fig.~\ref{Fig:Active}). 
Two types of simulations have been carried out: (i) DRD (wet) simulations with hydrodynamic interactions (HIs) and (ii) Brownian (dry) dynamics simulations without HIs, to unveil the effects of HIs on the emergent dynamics of the fiber. 
In Fig.~\ref{Fig:Active}(a), we observed the emergence of a turbulent-like flow pattern in the suspension of self-propelling microswimmers; the fiber is deformed significantly but quite differently from that in dry active baths, highlighting the potential significance of HIs. Moreover, the effects of HIs can be further recognized by tracking the center-of-mass of the fiber and comparing its mean-square-displacement (MSD) in wet microswimmer baths with those in dry active baths, as shown in Fig.~\ref{Fig:Active}(b). The diffusion of the fiber is super-diffusive in the wet bath of microswimmers with HIs, while it is normally diffusive (Brownian) for the fiber immersed in the dry active bath without HIs~\cite{Yariv2016}. 



\section{Applications to reactive porous-media flows with evolving solid surfaces} \label{sec:Apps-PorousFlows}

The DRD approach can also be applied to the study of mFSIs, which involve the evolution of solid surfaces due to phase transitions or chemical reactions, such as those occurring during crystal growth and melting~\cite{onuki2011}, as well as precipitation and dissolution in porous media~\cite{Ladd2021}, \emph{etc}. In particular, direct numerical simulation (DNS) at microscopic pore scales for flows involving precipitation and dissolution can yield detailed insights into the complex interplay among multiphase flow patterns, capillary forces, transport processes, pore-scale heterogeneities, and fluid–solid phase transitions. This knowledge is crucial for various applications~\cite{Almajed2021}, including filtration processes and microbially/enzyme-induced calcite precipitation (MICP/EICP) for soil stabilization, ultimately facilitating the design of sustainable porous-media systems and the optimization of MICP/EICP performance.  

\subsubsection{Governing dynamic equations}\label{sec:App4-Dyns}

Here, as an example, we consider the precipitation \& dissolution in an array of solid particles (nucleates, see Fig.~\ref{Fig:MICP}), where viscous flows are coupled with solute diffusion in the bulk and precipitation \& dissolution at the solid surfaces. Such precipitation \& dissolution can be regarded as a chemical reaction where the solute molecules in the fluid bulk and those in the solid are two different species transforming through the elementary reaction
\begin{equation}
\text{Fluid Solute} \rightleftharpoons \text{Solid Solute}.    
\end{equation}
In this case, the fluid is a single-phase solution with diffusing solute concentration $c(\boldsymbol{r},t)$, while in the solid phase, the solute concentration is always a constant $c_{\mathrm s}$, and the position, structure or shape of the solid evolves with time. 


In this fluid-solid two-phase flow, the fluid-fluid interface is not present, so the phase parameter $\phi$ is not needed, and the only interfacial phase parameter is $\psi$ for the solid boundary, taking the form of Eq.~(\ref{Eq:theory-DRD-psi}). 
The conservation of solutes is ensured by the balance equation of $\psi c$:
\begin{equation} \label{Eq:App4-phi} 
\partial_t (\psi c)+\boldsymbol{v}\cdot \nabla (\psi c)=-\nabla \cdot\boldsymbol{J}-r_{\mathrm p}(1-c/c_{\mathrm s})|\nabla \psi(\boldsymbol{r},t)|,
\end{equation}
where $\boldsymbol{v}$ is the velocity field following the incompressibility condition $\nabla\cdot\boldsymbol{v}=0$, $\boldsymbol{J}$ is the diffusion flux, and $r_{\mathrm p}$ denotes the net rate of precipitation ($r_{\mathrm p}>0$) or dissolution ($r_{\mathrm p}<0$) that occurs only at the fluid-solid interface (as indicated by the factor of $|\nabla \psi(\boldsymbol{r},t)|$).

To model the evolution of $\psi$ and solid shapes or structures, we have several different choices~\cite{Lowengrub2009DD,Lowengrub2021DD}. We can prescribe the fluid-solid phase parameter $\psi$ based on the interface positions obtained by the local evolving velocity of the solid surface, $\boldsymbol{V}_{\mathrm{s}}=-\boldsymbol{\hat{n}}_{\mathrm{s}}{r_{\mathrm p}}/{c_{\mathrm s}}$, determined by local precipitation kinetics, with the unit normal vector $\boldsymbol{\hat{n}}_{\mathrm{s}}$ pointing along $-\boldsymbol{\hat{z}}$-direction. 
However, this method becomes problematic when interfaces break or coalesce. To address this issue, we can use the level-set method to solve the Hamilton-Jacobi equation for the signed distance function $\mathcal{D}(\boldsymbol{r},t)$ or $\psi$:
\begin{equation}\label{Eq:App4-Dt}
\partial_t \mathcal{D}(\boldsymbol{r},t) + \frac{r_{\mathrm p}}{c_{\mathrm s}}  |\nabla \mathcal{D}(\boldsymbol{r},t)|=0, \quad \mathrm{or}, \quad 
\partial_t \psi(\boldsymbol{r},t) + \frac{r_{\mathrm p}}{c_{\mathrm s}} |\nabla \psi(\boldsymbol{r},t)|=0.
\end{equation}
In this way, the interface breakup or coalescence can be simulated without additional difficulties. Alternatively, we can also use the phase-field method through the Cahn-Hilliard equation for $\psi$~\cite{Lowengrub2009DD}:
\begin{equation}\label{Eq:App4-Psit}
\partial_t \psi(\boldsymbol{r},t) -\tau_{\mathrm{sn}}^{-1} \epsilon_{\mathrm s}^2 \nabla^2 \mu_{\psi}+ \frac{r_{\mathrm p}}{c_{\mathrm s}} |\nabla \psi(\boldsymbol{r},t)|=0,
\end{equation}
with $\mu_{\psi}=\left(\psi-3 \psi^2+2\psi^3 \right)-\epsilon_{\mathrm s}^2 \nabla^2 \psi$ being the generalized chemical potential and $\tau_{\mathrm{sn}}$ is a timescale governing the relaxational dynamics of interface profile that should be chosen to be smaller than $\epsilon_{\mathrm s}c_{\mathrm s}/r_{\mathrm p}$.    

Next, we employ the OVP to derive the governing dynamic equations. Firstly, the free energy of the two-phase system is given by
\begin{equation}\label{Eq:App4-F}
\mathcal{F}\left[c(\boldsymbol{r}), \psi (\boldsymbol{r})\right]
=\int d\boldsymbol {r} \left[\psi f(c)+ (1-\psi) f_{\mathrm{ss}}(c_{\mathrm s})\right], 
\end{equation} 
with $f(c)=k_{\mathrm B}Tc\ln c$ and $f_{\mathrm{ss}}(c_{\mathrm s})$ being the free energy density in the fluid and the solid phases, respectively. Secondly, the dissipation function includes contributions from three dissipation mechanisms (viscous dissipation, diffusion, and reaction) as
\begin{equation}\label{Eq:App4-Phi}
\Phi[\boldsymbol{v},\boldsymbol{J},r_{\mathrm p}]=\int d \boldsymbol{r}\left[\frac{1}{4}\eta(\psi)(\nabla {\boldsymbol{v}}+\nabla {\boldsymbol{v}}^{\mathrm{T}})^{2}+\frac{k_{\mathrm B}T}{2cD(\psi)}\boldsymbol{J}^2+\frac{1}{2\Gamma} r_{\mathrm p}^2|\nabla \psi(\boldsymbol{r},t)|\right].
\end{equation}
Thirdly, minimizing the Rayleighian $\mathcal{R}[\boldsymbol{v},\boldsymbol{J},r_{\mathrm p}]=\dot{\mathcal{F}}+\Phi-\int d \boldsymbol{r} P\nabla\cdot\boldsymbol{v}$ gives the following complete set of dynamic equations
\begin{subequations}\label{Eq:App4-Dyn}
\begin{equation} \label{Eq:App4-Dyn-Stokes}
\rho\left(\partial_t {\boldsymbol v} +{\boldsymbol v} \cdot \nabla {\boldsymbol v} \right)=-\nabla P+\nabla \cdot\left[\eta(\psi)\left(\nabla {\boldsymbol{v}}+\nabla {\boldsymbol{v}}^{\mathrm T} \right)\right],
\end{equation}
\begin{equation}\label{Eq:App4-Dyn-CH}
\partial_t (\psi c)+\boldsymbol{v}\cdot \nabla (\psi c)=\nabla \cdot \left(D(\psi)\nabla c\right)-r_{\mathrm p}(1-c/c_{\mathrm s}) |\nabla \psi(\boldsymbol{r},t)|,
\end{equation}
\begin{equation}\label{Eq:App4-Dyn-Dt}
\partial_t \psi(\boldsymbol{r},t) -\tau_{\mathrm{sn}}^{-1} \epsilon_{\mathrm s}^2 \nabla^2 \mu_{\psi}+ \frac{r_{\mathrm p}}{c_{\mathrm s}} |\nabla \psi(\boldsymbol{r},t)|=0,
\end{equation}  
\begin{equation}\label{Eq:App4-Dyn-rp}
r_{\mathrm{p}}=k_{\mathrm{r}}(c^2/c_{\mathrm{eq}}^2-1),
\end{equation}
where we have assumed the first-order reaction kinetics~\cite{van2008stefan} for surface precipitation
with $k_{\mathrm{r}}$ being a (positive) constant reaction rate and $c_\mathrm{eq}$ ($<c_{\mathrm s}$) being some equilibrium concentration when precipitation or dissolution stops~\cite{van2008stefan}. The smooth profiles of viscosity $\eta(\boldsymbol r)$ and diffusivity $D(\boldsymbol r)$ are taken to be the linear monotonic interpolation form given in Eq.~(\ref{Eq:theory-Dprofile12}) as
\begin{equation}\label{Eq:App4-etaSlip}
\eta(\psi)=\eta_{\mathrm{s}}+\left(\eta_{\mathrm{f}}-\eta_{\mathrm{s}}\right)\psi, \quad
D(\psi)= D_{\mathrm{s}}+\left(D_{\mathrm{f}}-D_{\mathrm{s}}\right)\psi. 
\end{equation}
\end{subequations} 
A very large viscosity $\eta_{\mathrm{s}}$ ($\gg \eta_{\mathrm{f}}$) and very small diffusivity $D_{\mathrm{s}}$ ($\ll D_{\mathrm{f}}$) in the solid domain can reproduce the following boundary conditions in the sharp interface limit: $\boldsymbol{v} =\boldsymbol{0}$ and $-\boldsymbol{\hat n}_{\mathrm{s}} \cdot D \nabla c=r_{\mathrm p}\left(1-c/c_{\mathrm{s}}\right)$. 
Four major dimensionless parameters arise: (1) Reynolds number, ${\mathrm{Re}} \equiv \rho Hk_{\mathrm{r}}/\eta_{\mathrm f}c_{\mathrm{s}}$, (2) P\'eclet number, ${\mathrm{Pe}} \equiv V_0H/D_{\mathrm f}$ with $V_0$ being the maximum inlet velocity, (3) Damk\"ohler number, ${\mathrm{Da}} \equiv Hk_{\mathrm{r}}/D_{\mathrm f}c_{\mathrm{s}}$, and (4) Interfacial mobility parameter: $\mathcal{M}\equiv\tau_{\mathrm{pr}}/\tau_{\mathrm{sn}}$, the ratio of the precipitation time $\tau_{\mathrm{pr}}\equiv H c_{\mathrm{s}}/k_{\mathrm{r}}$ to the fluid-fluid interfacial relaxation time $\tau_{\mathrm{sn}}$. Using the same algorithm as before, we integrated the dynamic equations (\ref{Eq:App4-Dyn}) in 2D using the finite difference method on staggered grids~\cite{gao_efficient_2014}. 












\subsubsection{Precipitation dynamics in 1D and 2D}\label{sec:App4-Benchmark}

\begin{figure}[htbp]  
  \centering
  \includegraphics[width=0.75\columnwidth]{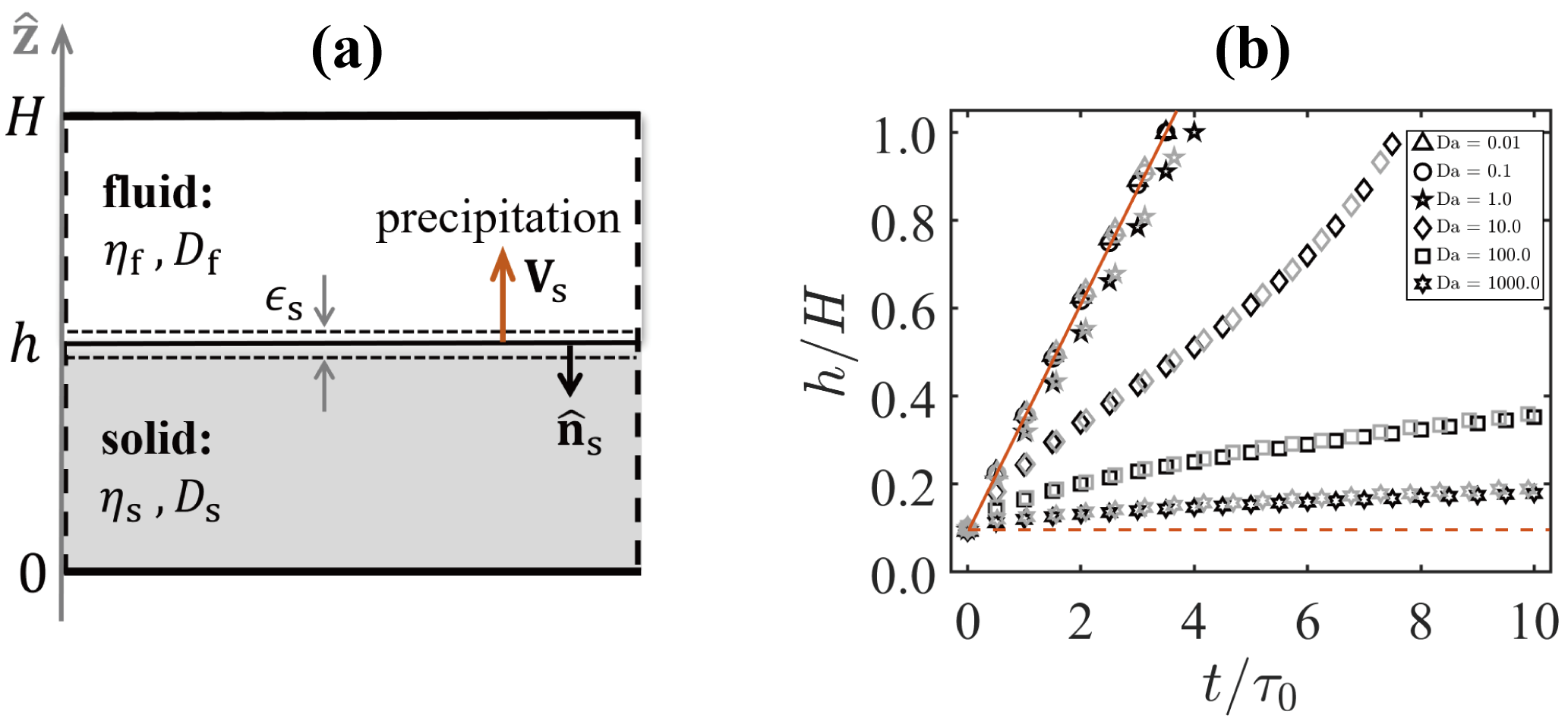}
  \caption {(a) Schematic illustration for the precipitation or dissolution at the fluid-solid interface in 1D. (b) Comparisons between the numerical solutions (gray open markers) obtained from the DRD approach with those (black open markers) from sharp interface models in a large range of Damk\"ohler number: ${\mathrm{Da}} \equiv k_{\mathrm r}H/D_{\mathrm{f}} c_{\mathrm s} \in [0.01, 1000]$. 
  Here, we take the other physics parameters to be:  equilibrium concentration $c_{\mathrm{eq}}/c_{\mathrm{s}}=0.8$, and environment concentration $c_0/c_{\mathrm{s}}=0.85$. We take the ``free'' parameters to be: $r_{\epsilon_{\mathrm{s}}}=\epsilon_{\mathrm{s}}/H = 0.003$, $r_\eta=\eta_{\mathrm{s}} / \eta_{\mathrm{f}} = 100.0$, $\mathcal{M}\equiv\tau_{\mathrm{pr}}/\tau_{\mathrm{sn}}=0.05$, and $r_D = D_{\mathrm{s}}/D_{\mathrm{f}}=0.0$. Periodic boundary conditions have been used in the $x$-direction.}
\label{Fig:App4-MICP}
\end{figure}

To validate the DRD approach for simulating mFSI in porous-media flows involving solute precipitations, we firstly applied it to model the precipitation dynamics in a 1D fluid–solid system, as schematically illustrated in Fig.~\ref{Fig:App4-MICP}(a). We then compared the numerical solutions for $h(t)$ obtained using the DRD approach with both the analytical solutions and the numerical solutions from the sharp interface model (see appendix~\ref{sec:App-MICP-1DSharp}). We observed a strong quantitative agreement between the two solutions across a wide range of Damk\"ohler number or Karlovitz number: ${\mathrm{Da}} \in [0.01, 1000]$ or ${\mathrm{Ka}} \in [0.001, 100]$, as shown in Fig.~\ref{Fig:App4-MICP}(b). 

\begin{figure}[htbp]
  \centering
  \includegraphics[width=0.68\columnwidth]{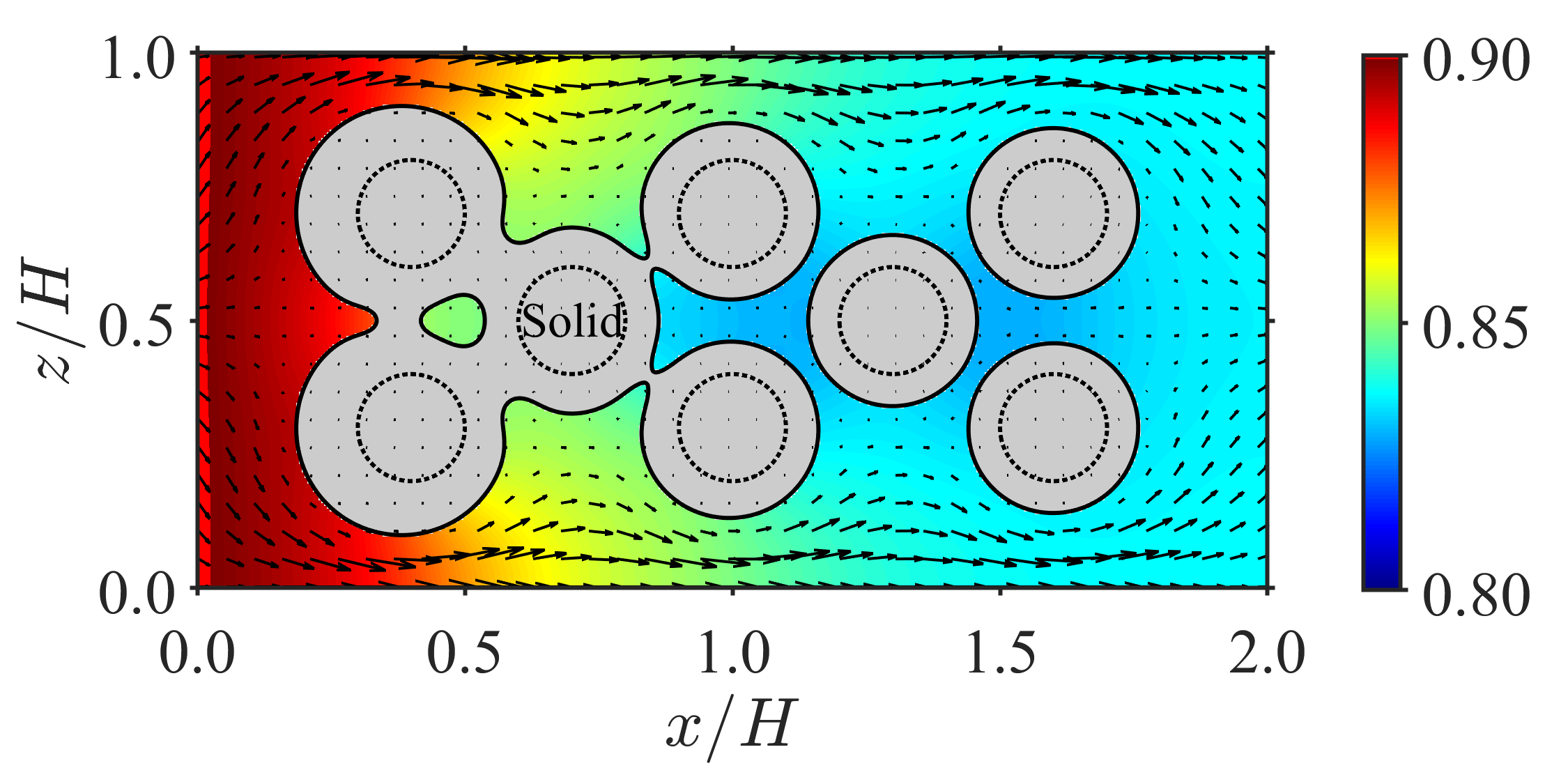}
  \caption {(Color online) Applications in reactive porous-media flows with precipitation at solid surfaces (of initial solid volume fraction $\approx 13\%$). A snapshot is taken for the whole computational domain, including the flow field (arrows), the concentration field (color), and the evolved porous structure (gray) that is composed initially of a hexagonal lattice of circular nucleates (dashed lines, of diameter $0.2H$). 
    Here, we take the physics parameters to be:  ${\mathrm{Re}}=2.72$, ${\mathrm{Pe}}={\mathrm{Da}}=1.0$, $c_{\mathrm{eq}}/c_{\mathrm{s}}=0.8$, and the inlet concentration $c_{\mathrm{0}}/c_{\mathrm{s}}=0.9$. We take the ``free'' parameters to be:  $r_{\epsilon_{\mathrm{s}}}=\epsilon_{\mathrm{s}}/H = 0.01$, $r_\eta=\eta_{\mathrm{s}} / \eta_{\mathrm{f}} = 100.0$, $\mathcal{M}\equiv\tau_{\mathrm{pr}}/\tau_{\mathrm{sn}}=0.05$, and $r_D = D_{\mathrm{s}}/D_{\mathrm{f}}=0.0$. Inlet/outlet conditions apply in the $x-$  direction, and periodic boundary conditions and inlet/outlet conditions apply in the $z-$ direction.}
  \label{Fig:MICP}
\end{figure}

Secondly, we use the DRD to simulate a single-phase flow passing through a porous media initially composed of a hexagonal lattice of circular nucleates where viscous flows are coupled with solute diffusion in the bulk and precipitation at the solid surfaces (see Fig.~\ref{Fig:MICP}). We found that some holes are formed and the highly heterogeneous porous structure can be controlled and manipulated by tuning the competing transport phenomena through dimensionless parameters. Further DNS explorations of this system would help to optimize important engineering processes such as MICP/EICP. 

\section{Conclusion}\label{sec:summary}

In summary, we have proposed and validated a generic monolithic direct numerical simulation (DNS) approach--the diffuse resistance domain (DRD) approach--for microscale fluid-structure interactions (mFSI) in multicomponent multiphase flows, which are ubiquitous in both nature and modern engineering processes. In contrast to other DNS methods for mFSI problems~\cite{Kirby2010Book,Subramaniam2020,Hou2012}, the DRD approach naturally overcomes several major challenges: (i) thermodynamic consistency is ensured by employing Onsager's principle in the construction of the model (including both the bulk equations and boundary conditions); (ii) the complex interfacial dynamics at fluid--solid interfaces are accurately captured by smoothly interpolating dynamic resistance coefficients across the fluid and solid domains. Importantly, classical boundary conditions (\emph{e.g.}, the fluid-impermeability, no-slip, Navier-slip, generalized Navier-slip, and dynamic contact-angle hysteresis conditions) are all reproduced by the DRD approach, as demonstrated by our sharp-interface limit analysis. Moreover, in contrast to the diffuse-domain method~\cite{Lowengrub2009DD}, the governing equations in the DRD approach for systems involving both multiphase fluids and solid structures assume the same form as those for multiphase fluids alone, thereby significantly simplifying the development of numerical algorithms. Finally, unlike multiphase-field methods for mFSI---which evolve phase parameter fields dynamically (\emph{e.g.}, via Cahn--Hilliard or Allen--Cahn equations)---the DRD approach delineates fluid--solid interfaces through prescribed functions that vary smoothly across the interface. This not only simplifies the energy construction for solids (especially in the rigid limit) but also substantially reduces computational costs, particularly when the solids move, deform, or evolve.
Interestingly, the advanced applications of the DRD approach presented in this work yield preliminary simulation results that capture complex mFSI phenomena in microfluidics, active matter hydrodynamics, and precipitation dynamics in porous media, qualitatively aligning with recent experimental observations. Nevertheless, a comprehensive exploration of all relevant dimensionless parameters, further quantitative analyses in 3D simulations (\emph{e.g.}, in ongoing DRD simulations of particle focusing in 3D microchannels), and direct comparisons with experimental measurements will be essential in future studies.

In addition, here we adopt diffuse-interface two-phase flows—governed by the field parameter $\phi$—as a representative model for structured complex fluids, where order parameters are required to capture microstructural evolution. The DRD approach can be further extended (\emph{e.g.}, by combining with deformable particle model~\cite{MarkD2018deformableparticle} and dynamic van der Waals theory~\cite{onuki2005dynamic}) to simulate more complex mFSI that involves multiphysics multi-field couplings, \emph{e.g.}, viscoelastic fluids, deformable elastic continuous solids, external electric or magnetic fields, two-phase fluids in heat flow and/or involving phase transitions, thermal fluctuations, and biological activities. 
Such a generic efficient DNS approach for mFSI provides a promising tool to unveil fundamental physical mechanisms in complex mFSI, perform precise control of microscale fluids, and optimize engineering processes in applications across diverse fields such as microfluidics, additive manufacturing, and biomedical engineering where the technology bottleneck usually arises from ultra-high-dimensional parameter space and high trial-and-error costs. 



\section*{Acknowledgements}
We thank Tiezheng Qian, Xiao-Ping Wang, Zhenlin Guo, Dong Wang, Haiqin Wang, Yanan Liu, and Guangyin Jing for some useful discussions and comments. We thank J. Su for the literature survey and for helping to plot some figures. 
X. Xu is supported partly by the National Natural Science Foundation of China (NSFC, No.~12131010). The numerical computations were performed on TianHe-2 through the Shanxi Supercomputing Center of China. M. Gao and Z. Li contributed equally to this work. 

\section*{Code availability}
The code for the simulations is available on the GitHub data repository at \url{https://github.com/marcusgao11/DRD_application.git.}

\appendix

\section{Sharp-interface model for 1D diffusion: Understand the validity of the DRD approach}\label{sec:App-1DDiffusion}

In Sec.~\ref{sec:theory-nutshell}, we demonstrated that the DRD approach can accurately solve the 1D diffusion equation under various boundary conditions when the interface thickness is sufficiently small $r_{\epsilon_{\mathrm{s}}} = \epsilon_{\mathrm{s}}/H \to 0$. This is achieved by selecting appropriate interpolations of diffusivity $D(\psi)$ and optimizing the limits of the diffusivity ratios ($r_{\mathrm D} \equiv D_{\mathrm{s}}/D_{\mathrm{f}}$ and $r_{\mathrm{D}_{\mathrm i}}\equiv D_{\mathrm{i}}/D_{\mathrm{f}}$) for different cases. To understand the validity of the DRD approach in accurately reproducing various boundary conditions, we analyze the sharp-interface model for the one-dimensional (1D) diffusion equation (\ref{Eq:theory-ct}) and its solutions in a triple-domain system (for $-H_{\mathrm{s}}\leqslant z\leqslant 2H$) as shown in Fig.~\ref{Fig:theory-1DSchematic}(b), where the two ``sharp'' interfaces are located at $z=\epsilon_{\mathrm i}/2$ and $z=-\epsilon_{\mathrm i}/2$, and $D(z)$ is the diffusivity that varies piecewisely from $D_{\mathrm{f}}$ in the fluid domain ($\epsilon_{\mathrm i}/2\leqslant z\leqslant 2H$), to $D_{\mathrm i}$ in the fluid-solid interfacial (intermediate) domain ($-\epsilon_{\mathrm i}/2\leqslant z < \epsilon_{\mathrm i}/2$), and to $D_{\mathrm{s}}$ in the solid domain ($-H_{\mathrm{s}}\leqslant z < -\epsilon_{\mathrm i}/2$). Under the boundary conditions $c(2H,t)=c_1$ and $c(-H_{\mathrm{s}},t)=c_{\mathrm 2}$, we obtain the steady-state solution of the 1D diffusion equation (\ref{Eq:theory-ct}) as
\begin{equation}\label{Eq:theory-3domain-c}
c(z,t\to \infty)= 
\begin{cases}  
c_1+(J_0/D_{\mathrm{f}})(2H-z) & \text { if } \quad \quad \, \, \epsilon_{\mathrm i}/2\leqslant z\leqslant 2H, \\
c_1+(J_0/D_{\mathrm{f}})\left(2H-\epsilon_{\mathrm i}/2\right) + (J_0/D_{\mathrm i})\left(\epsilon_{\mathrm i}/2-z\right) & \text { if } \quad -\epsilon_{\mathrm i}/2\leqslant z < \epsilon_{\mathrm i}/2, \\ 
c_2-(J_0/D_{\mathrm{s}})(z+H_{\mathrm{s}}) & \text { if } \quad -H_{\mathrm{s}} \leqslant z < -\epsilon_{\mathrm i}/2, 
\end{cases}
\end{equation}
with the steady state flux $J_0$ given by $J_0=-D_{\mathrm{f}}(c_1-c_2)/\left[D_{\mathrm{f}}\epsilon_{\mathrm i}/D_{\mathrm i}+2H-\epsilon_{\mathrm i}/2+(D_{\mathrm{f}}/D_{\mathrm{s}})(H_{\mathrm{s}}-\epsilon_{\mathrm i}/2)\right]$. We consider several special cases and pay attention to the effects of the diffusivity ratios $r_{\mathrm D} = D_{\mathrm{s}}/D_{\mathrm{f}}$ and $r_{\mathrm{D}_{\mathrm i}}= D_{\mathrm{i}}/D_{\mathrm{f}}$ on the solution in the fluid domain (domain-1) as follows. 
\begin{itemize}
\item Double-domain case 1 (with $\epsilon_{\mathrm i}= 0$): the solid domain-2 has a very large diffusivity, \emph{i.e.}, $r_{\mathrm D} = D_{\mathrm{s}}/D_{\mathrm{f}} \gg 1$. In this case, we find the solution (\ref{Eq:theory-3domain-c}) in the fluid domain ($0\leqslant z\leqslant 2H$) reduces to be $c\approx c_1-(c_1-c_2)(2H-z)/H$. That is, $r_{\mathrm D} = D_{\mathrm{s}}/D_{\mathrm{f}} \gg 1$ sets an effective Dirichlet boundary condition at the interface $z=0$ for the diffusion in the fluid domain-1 (with $0\leqslant z\leqslant 2H$): 
\begin{subequations}\label{Eq:DRD-BCD}
\begin{equation}\label{Eq:theory-BCD-Dirichlet}
c(z=0)=c_2.
\end{equation}
Alternatively, we can understand the result in another way. In the steady state, the diffusion flux is a constant: $J=-D(z)\partial_z c=J_0=\mathrm{const.}$. Then, in solid domain-2 with $D(z)=D_{\mathrm{s}} \gg D_{\mathrm{f}}$, we have $\partial_z c \to 0$ and hence $c(z<0)\approx c_2$ in the solid domain, yielding the effective Dirichlet boundary condition $c(z=0)=c_2$. 

Such an analytical double-domain result in the limit of $r_{\mathrm D}\gg 1$ explains the good agreement of the numerical DRD solution (implemented using the linear interpolation of diffusivity $D(\psi)$ in Eq.~(\ref{Eq:theory-Dprofile12})) with the analytical steady-state solution of 1D diffusion equation (\ref{Eq:theory-ct}) in the fluid domain ($0\leqslant z\leqslant 2H$) under Dirichlet boundary condition $c(z=0)=c_2$, as shown in Figs.~\ref{Fig:theory-BC12}(a,b). 

\item Double-domain case 2 (with $\epsilon_{\mathrm i}= 0$): the solid domain-2 has a very small diffusivity, \emph{i.e.}, $r_{\mathrm D} = D_{\mathrm{s}}/D_{\mathrm{f}} \ll 1$. In this case, the solution (\ref{Eq:theory-3domain-c}) in the fluid domain-1 ($0\leqslant z\leqslant 2H$) reduces to be $c\approx c_1$. That is, $r_{\mathrm D} = D_{\mathrm{s}}/D_{\mathrm{f}} \to 0$ sets an effective Neumann (impermeable zero-flux) boundary condition at the interface $z=0$ for the diffusion in the fluid domain-1 (with $0\leqslant z\leqslant 2H$):  
\begin{equation}\label{Eq:theory-BCD-Neumann}
D_{\mathrm{f}}\partial_{z} c|_{z=0}=0,
\end{equation}
Alternatively, we can understand the result in another way. In the steady state and in the solid domain-2 with $D(z)= D_{\mathrm{s}} \to 0$, we have $J=-D_{\mathrm{s}}\partial_z c \to 0$, yielding the effective Neumann boundary condition $D_{\mathrm{f}}\partial_{z} c|_{z=0}=0$.   

Such an analytical double-domain result in the limit of $r_{\mathrm D}\to 0$ explains the good agreement of the numerical DRD solution (implemented also using the linear interpolation of diffusivity $D(\psi)$ in Eq.~(\ref{Eq:theory-Dprofile12})) with the analytical steady-state solution of 1D diffusion equation (\ref{Eq:theory-ct}) in the fluid domain ($0\leqslant z\leqslant 2H$) under the Neumann boundary condition $\partial_{z} c|_{z=0}=0$, as shown in Figs.~\ref{Fig:theory-BC12}(c,d). 

\item Triple-domain case ($\epsilon_{\mathrm i}\ll H$ but $\epsilon_{\mathrm i}\neq 0$): the solid domain-2 has a very large diffusivity but the fluid-solid interfacial (intermediate) domain has a small diffusivity, \emph{i.e.}, $D_{\mathrm i}<D_{\mathrm{f}}\ll D_{\mathrm{s}}$ or $r_{\mathrm D} =D_{\mathrm{s}}/D_{\mathrm{f}} \gg 1$ and $r_{\mathrm{D}_{\mathrm i}}=D_{\mathrm{i}}/D_{\mathrm{f}}<1$. In this case, the solution (\ref{Eq:theory-3domain-c}) in the fluid domain-1 ($\epsilon_{\mathrm i}/2\leqslant z\leqslant 2H$) reduces to be $c\approx c_1-(c_1-c_2)(2H-z)/(2H+\ell_{\mathrm s}-\epsilon_{\mathrm i}/2)$ with $\ell_{\mathrm s} \equiv D_{\mathrm{f}} \epsilon_{\mathrm i}/D_{\mathrm i}$ and hence $\ell_{\mathrm s}/\epsilon_{\mathrm i}=D_{\mathrm{f}}/D_{\mathrm i}>1$. That is, $r_{\mathrm D} =D_{\mathrm{s}}/D_{\mathrm{f}}\gg 1$ and $r_{\mathrm{D}_{\mathrm i}}=D_{\mathrm{i}}/D_{\mathrm{f}}<1$ together set an effective Robin (concentration ``slip'') boundary condition at the interface $z=0$ (as $\epsilon_{\mathrm i} \to 0$) for the diffusion in the fluid domain-1 (with $\epsilon_{\mathrm i}/2\leqslant z\leqslant 2H$): 
\begin{equation*}\label{Eq:DRD-BCD-Robin1}
\ell_{\mathrm s}\partial_z c|_{z=\epsilon_{\mathrm i}/2}=c(z=\epsilon_{\mathrm i}/2)-c_2.
\end{equation*}

A similar alternative way of understanding the effective Robin boundary condition can be made as follows. Firstly, as discussed in the above double-domain case 1, $r_{\mathrm D} \gg 1$ sets $c(z\leqslant -\epsilon_{\mathrm i}/2)\approx c_2$. Secondly, in the steady state, the diffusion flux is a constant: $J=-D(z)\partial_z c=J_0$ and hence at the interface $z=\epsilon_{\mathrm i}/2$, we have $D_{\mathrm{f}}\partial_z c|_{z=\epsilon_{\mathrm i}/2^+}=D_{\mathrm{i}}\partial_z c|_{z=\epsilon_{\mathrm i}/2^-}=D_{\mathrm{i}}\left[c(z=\epsilon_{\mathrm i}/2^+)-c_2\right]/\epsilon_{\mathrm{i}}$, which gives the effective Robin boundary condition at $z=\epsilon_{\mathrm i}/2$: $\ell_{\mathrm s}\partial_z c|_{z=\epsilon_{\mathrm i}/2^+}=c(z=\epsilon_{\mathrm i}/2^+)-c_2$, and as $\epsilon_{\mathrm i} \to 0$ yields the required Robin boundary condition at $z=0$:
\begin{equation}\label{Eq:DRD-BCD-Robin}
\ell_{\mathrm s}\partial_{z} c|_{z=0} =c(z=0)-c_2.
\end{equation} 
\end{subequations}

Again, such an analytical triple-domain result explains the good agreement of the numerical DRD solution (implemented using the piecewise linear interpolation of diffusivity $D(\psi)$ in Eq.~(\ref{Eq:theory-Dprofile3})) with the analytical steady-state solution of 1D diffusion equation (\ref{Eq:theory-ct}) in the fluid domain ($0\leqslant z\leqslant 2H$) under the Robin boundary condition (\ref{Eq:DRD-BCD-Robin}), as shown in Figs.~\ref{Fig:theory-BC3}(b). Moreover, the linear dependence of the slip length $\ell_{\mathrm s}/\epsilon_{\mathrm i}=D_{\mathrm{f}}/D_{\mathrm i}>1$ on $r_{\mathrm{D}_{\mathrm i}}^{-1}=D_{\mathrm{f}}/D_{\mathrm i}$ has also been confirmed numerically as shown in Figs.~\ref{Fig:theory-BC3}(c).
\end{itemize}

\section{Some relations for deriving general dynamic equations of mFSI in two-phase flows\label{sec:App-DRD-theory}}

We present some more details for the derivation of the general dynamic equations of mFSI in two-phase flows in Sec.~\ref{sec:theory-Eqns}. Firstly, from Eqs.~(\ref{Eq:theory-DRD-F}) and (\ref{Eq:theory-DRD-FbFs}), we obtain the rate of change of free energy as
\begin{equation}\label{Eq:App-DRD-theory-Fdot1}
\begin{aligned}
\dot{\mathcal{F}} &=\int d\boldsymbol {r} \left[\hat{f}_{\mathrm b}\partial_{t}\Psi+ \Psi\mu_{\mathrm{b}}\partial_{t}\phi + K\Psi\nabla\phi\cdot\nabla\left(\partial_{t}\phi\right)\right]+\int d\boldsymbol{r} \left(\sum_{\alpha=1}^N\left|\nabla\psi_{\alpha}\right|^{2}+\left|\nabla\psi\right|^{2}\right)\epsilon_{\mathrm{s}}\mu_{\mathrm{s}}\partial_{t}\phi\\
&+\int d\boldsymbol{r}2\epsilon_{\mathrm{s}}f_{\mathrm{s}}\left[\sum_{\alpha=1}^N\nabla\psi_{\alpha}\cdot\nabla\left(\partial_{t}\psi_{\alpha}\right) +\nabla\psi\cdot\nabla\left(\partial_{t}\psi\right)\right] ,
\end{aligned}
\end{equation} 
with $\mu_{\mathrm b}=\partial f_{\mathrm b}(\phi)/\partial \phi$ 
and $\mu_{\mathrm{s}}=\partial f_{\mathrm{s}}(\phi)/\partial \phi$. 
Doing integration by parts in Eq.~(\ref{Eq:App-DRD-theory-Fdot1}), the rate of change of free energy becomes
\begin{equation}\label{Eq:App-DRD-theory-Fdot2}
\begin{aligned}
\dot{\mathcal{F}} =\int d\boldsymbol{r}\left[\hat{\mu}\partial_{t}(\Psi\phi)-\hat{p}\partial_{t}\Psi-\nabla\cdot\left(2\epsilon_{\mathrm{s}}f_{\mathrm{s}}\nabla\psi_{\alpha}\right)\partial_{t}\psi_{\alpha}-\nabla\cdot\left(2\epsilon_{\mathrm{s}}f_{\mathrm{s}}\nabla\psi\right)\partial_{t}\psi\right],
\end{aligned}
\end{equation} 
where the total chemical potential $\hat{\mu}$ is defined in Eq.~(\ref{Eq:theory-DRD-muT}) and the total pressure $\hat{p}=-\hat{f}_{\mathrm b} +\hat{\mu}\phi$ is defined in Eq.~(\ref{Eq:theory-DRD-phat}); we have used the natural condition, $K\Psi\hat{\boldsymbol{n}}_{\mathrm{d}}\cdot \nabla\phi =0$ and the conditions of constant $\psi_{\alpha}$ and constant $\psi$ (and hence $\partial_{t}\psi_{\alpha}=\partial_{t}\psi=0$) at the boundaries of the whole computational domain with $\hat{\boldsymbol{n}}_{\mathrm{d}}$ being their outward unit vector.  

Moreover, since the fluid-solid interfacial functions, $\psi_{\alpha}$ and $\psi$, both take the form of Eq.~(\ref{Eq:theory-DRD-psi}), we obtain the following two identities
\begin{equation}\label{Eq:App-DRD-theory-dtpsi}
\partial_t \psi_\alpha =-\boldsymbol{V}_\alpha \cdot \nabla \psi_\alpha, \quad \quad 
\partial_t \psi =-\boldsymbol{V}_{\mathrm{w}} \cdot \nabla \psi,
\end{equation}
with $\boldsymbol{V}_\alpha=\dot{\boldsymbol{R}}_{\alpha}$ (defined in Eq.~(\ref{Eq:theory-DRD-Valpha})) and $\boldsymbol{V}_{\mathrm{w}}$ being the velocity of the $\alpha$-th particle and the boundary wall, respectively.
Substituting Eq.~(\ref{Eq:App-DRD-theory-dtpsi}) into Eq.~(\ref{Eq:App-DRD-theory-Fdot2}) and using the definition of $\Psi(\boldsymbol{r},t) \equiv \psi\prod_{\alpha=1}^N \psi_\alpha$, we obtain the rate of change of the total free energy $\dot{\mathcal{F}}_{\mathrm T}=\dot{\mathcal{F}}-\dot{\mathcal{W}}_{\mathrm{ext}}=\dot{\mathcal{F}}-\sum_{\alpha}\boldsymbol{F}_{\mathrm{ext},\alpha}\cdot\boldsymbol{V}_{\alpha}$ as
\begin{equation}\label{Eq:App-DRD-theory-FTdot}
\dot{\mathcal{F}}_{\mathrm{T}} =\int d\boldsymbol{r}\left[\hat{\mu}\partial_{t}(\Psi\phi)-\sum_{\alpha}\left(\boldsymbol{F}_{\mathrm{ext},\alpha}-\hat{p}_{\alpha}\nabla\psi_{\alpha} \right)\cdot\boldsymbol{V}_{\alpha} \right],
\end{equation}
where $\hat{p}_{\alpha} \equiv \hat{p}\psi \prod_{\beta\neq \alpha} \psi_\beta+\nabla \cdot (2\epsilon_{\mathrm{s}}f_{\mathrm{s}}\nabla\psi_\alpha)$ and we have taken $\boldsymbol{V}_{\mathrm{w}}=0$ for simplicity.   
Substituting Eqs.~(\ref{Eq:theory-DRD-Valpha}) and (\ref{Eq:theory-DRD-phi}) into Eq.~(\ref{Eq:App-DRD-theory-FTdot}) and using the incompressibility condition $\nabla\cdot \boldsymbol{v}=0$, we obtain the rate of change of the total free energy $\dot{\mathcal{F}}_{\mathrm T}$ in Eq.~(\ref{Eq:theory-DRD-FTdot}) as
\begin{equation*} 
\dot{\mathcal{F}}_{\mathrm T} [\boldsymbol{v},\boldsymbol{J}]=\int d\boldsymbol{r}\left(-\boldsymbol{v}\cdot \hat{\mu}\nabla(\Psi\phi) +\boldsymbol{J}\cdot \nabla\hat{\mu}  -\boldsymbol{v} \cdot\boldsymbol{f}_{\mathrm{tot}}\right),
\end{equation*} 
where we have used the natural condition, $\hat{\boldsymbol{n}}_{\mathrm{d}}\cdot \boldsymbol{J} =0$.

\section{Sharp-interface model for precipitation in 1D system}\label{sec:App-MICP-1DSharp}

We formulate a sharp interface model for the homogeneous precipitation of an ionic liquid solution onto a solid surface in 1D~\cite{van2008stefan} with $0\leqslant z \leqslant H$ and the liquid-solution interface located at $z=h$, as shown in Fig.~\ref{Fig:App4-MICP}. For simplicity, we consider a one-component-solute theory where the ion concentration follows the 1D diffusion equation
\begin{equation}\label{Eq:1D-Sharp-DynEqn0}
\partial_t c=D\partial_z^2c,
\end{equation} 
supplemented with $c(H,t)=c_0$ and a Robin-type boundary condition at $z=h$:
\begin{equation}\label{Eq:1D-Sharp-ji0}
D\partial_z c= {r_{\mathrm{p}}}(1-c/c_{\mathrm{s}}),
\end{equation}
\begin{align}\label{Eq:1D-Sharp-jpLaw0}
r_{\mathrm{p}}=c_{\mathrm{s}}dh/dt=k_{\mathrm{r}}(c^2/c_{\mathrm{eq}}^2-1).
\end{align}  
In such a 1D sharp-interface precipitation model, there exist two major time scales: 
(i) $\tau_{\mathrm{D}}\sim H^2/D_{\mathrm f}$, the characteristic diffusion time of solute in the liquid solution; 
(ii) $\tau_{\mathrm{pr}}\sim H c_{\mathrm{s}}/k_{\mathrm{r}}$,  the characteristic precipitating time for the solution. The dimensionless {Damk{\"o}hler number}, $\mathrm{Da} \equiv \tau_{\mathrm{D}}/\tau_{\mathrm{p}}$ or the {Karlovitz number}, $\mathrm{Ka} \equiv \mathrm{Da}^{-1}$ are the ratios between the diffusion time and the interfacial precipitation time, which characterize their relative importance in determining a steady-state ion mass distribution and precipitation rate over the length and time scales of interest. 
Two limiting regimes can be identified according to the magnitude of $\mathrm{Da}$ as follows. 
\begin{itemize}
\item For $\mathrm{Da}\gg 1$, or, $\mathrm{Ka}\ll 1$, or equivalently, $\tau_{\mathrm{pr}}\ll \tau_{\mathrm{D}}$, the interface-equilibration rate is much larger than the diffusion rate, and the precipitation is said to be \emph{diffusion-limited}. In this case, diffusion is the slowest process so diffusion characteristics dominate and the liquid-solid interface is assumed to be instantaneously in equilibrium. The boundary condition in Eq.~(\ref{Eq:1D-Sharp-jpLaw0}) should be replaced by 
\begin{equation}\label{Eq:1D-Sharp-BCceq}
c(h,t)=c_{\mathrm{eq}}. 
\end{equation}
The steady-state solution $c(z,t\to \infty)$ and $h(t)$ (where the timescale $t$ is much larger than $\tau_{\mathrm{D}}$ and $\tau_{\mathrm{pr}}$) of the above equation system can be found to be
\begin{subequations}\label{Eq:1D-Sharp-Solution}
\begin{equation}\label{Eq:1D-Sharp-Solution-rhos}
c(z)=c_0 -\left(c_0-c_{\mathrm{eq}}\right) \frac{H-z}{H-h},
\end{equation}
\begin{equation}\label{Eq:1D-Sharp-Solution-h}
(H-h(t))^2=(H-h_0)^2- {2D} \frac{c_0-c_{\mathrm{eq}}}{\rho_{\mathrm{s}}-c_{\mathrm{eq}}}t.
\end{equation}
\end{subequations} 

\item For $\mathrm{Da} \ll 1$, or, $\mathrm{Ka}\gg 1$, or equivalently, $\tau_{\mathrm{pr}}\gg \tau_{\mathrm{D}}$, diffusion occurs much faster than the interface equilibration, and the precipitation is said to be \emph{reaction-limited}. In this case, diffusion reaches an ``equilibrium" well before the interface is at equilibrium. The steady-state solution $c(z,t\to \infty)$ and $h(t)$ (where the timescale $t$ is much larger than $\tau_{\mathrm{D}}$ and $\tau_{\mathrm{pr}}$) of the above equation system can be expanded in terms of $\mathrm{Da}$ to the first order as
\begin{subequations}\label{Eq:1D-Sharp-Solution2}
\begin{equation}\label{Eq:1D-Sharp-Solution2-rhos}
c(z)\approx c_0 - \mathrm{Da} c_{\mathrm{eq}} \left(\frac{c_0^2}{c_{\mathrm{eq}}^2}-1\right) \left(1-\frac{c_0}{\rho_{\mathrm{s}}}\right)\left(1-\frac{z}{H}\right),
\end{equation}
\begin{equation}\label{Eq:1D-Sharp-Solution2-h}
h(t)\approx h_0+\mathrm{Da} \frac{c_{\mathrm{eq}}}{\rho_{\mathrm{s}}}(\frac{c_0^2}{c_{\mathrm{eq}}^2}-1)\frac{Dt}{H}.
\end{equation}
\end{subequations} 
\end{itemize}


\bibliography{references}

\end{document}